\documentclass[article]{aa}
\usepackage{graphicx}
\usepackage{float}
\usepackage[usenames,dvipsnames]{color}
\usepackage[]{subfig}
\usepackage{ulem}
\usepackage{txfonts}
\usepackage{amsmath,mathtools}
\usepackage{here}
\usepackage{array}
\newcolumntype{L}[1]{>{\raggedright\let\newline\\\arraybackslash\hspace{0pt}}m{#1}}

\newcommand{\Rsun}{\ensuremath{\,\mathrm{R_\odot}}\xspace}
\newcommand{\Msun}{\ensuremath{\,\mathrm{M_\odot}}\xspace}

\definecolor{wildstrawberry}{rgb}{1.0, 0.26, 0.64}
\definecolor{electricviolet}{rgb}{0.56, 0.0, 1.0}
 
\definecolor{changes}{rgb}{0.0, 0.44, 1.0}
\definecolor{darkorange}{rgb}{1.0, 0.55, 0.0}

\usepackage{natbib,twoopt}
\definecolor{pinegreen}{RGB}{1, 121, 111}
\definecolor{salmon}{RGB}{255,160,122}

\usepackage[breaklinks=true,colorlinks,linkcolor=blue,citecolor=pinegreen]{hyperref}         

\usepackage{etoolbox}
\makeatletter
\patchcmd\@combinedblfloats{\box\@outputbox}{\unvbox\@outputbox}{}{\errmessage{\noexpand patch failed}}
\makeatother

\bibpunct{(}{)}{;}{a}{}{,}             
\makeatletter
\newcommandtwoopt{\citeads}[3][][]{\href{http://adsabs.harvard.edu/abs/#3}%
	{\def\hyper@linkstart##1##2{}%
		\let\hyper@linkend\@empty\citealp[#1][#2]{#3}}}
\newcommandtwoopt{\citepads}[3][][]{\href{http://adsabs.harvard.edu/abs/#3}%
	{\def\hyper@linkstart##1##2{}%
		\let\hyper@linkend\@empty\citep[#1][#2]{#3}}}
\newcommandtwoopt{\citetads}[3][][]{\href{http://adsabs.harvard.edu/abs/#3}%
	{\def\hyper@linkstart##1##2{}%
		\let\hyper@linkend\@empty\citet[#1][#2]{#3}}}
\newcommandtwoopt{\citeyearads}[3][][]%
{\href{http://adsabs.harvard.edu/abs/#3}
	{\def\hyper@linkstart##1##2{}%
		\let\hyper@linkend\@empty\citeyear[#1][#2]{#3}}}
\makeatother

\begin{document}

   \title{The expansion of stripped-envelope stars: \\ consequences for supernovae and gravitational-wave progenitors\thanks{The models are available in electronic form at the CDS via anonymous ftp to \href{http://cdsarc.u-strasbg.fr}{cdsarc.u-strasbg.fr} (130.79.128.5) or via \href{http://cdsarc.u-strasbg.fr/viz-bin/qcat?J/A+A/}{http://cdsarc.u-strasbg.fr/viz-bin/qcat?J/A+A/}}
   }

   \author{
          E. Laplace
          \inst{1, *}
          \and
	      Y. G\"{o}tberg
	      \inst{2}
	      \and
	      S. E. de Mink
      \inst{3,1}
	      \and 
	      S. Justham
	      \inst{1,4,3}
	      \and
	      R. Farmer
	      \inst{1,3}
          }

   \institute{Anton Pannekoek Institute of Astronomy and GRAPPA, University of Amsterdam, 1090 GE Amsterdam, the Netherlands.
   \and
   	The Observatories of the Carnegie Institution for Science, 813 Santa Barbara Street, Pasadena, CA 91101, USA.
   	\and 
   	Center for Astrophysics, Harvard-Smithsonian, 60 Garden Street, Cambridge, MA 02138, USA.
   	\and
   	School of Astronomy and Space Science, University of the Chinese Academy of Sciences, Beijing 100012, China.\\
    $^{*}$\email{e.c.laplace@uva.nl}
             }

   \date{Received ...; accepted ...}

 
  \abstract
  { 
Massive binaries that merge as compact objects are the progenitors of  gravitational-wave sources. Most of these binaries experience one or more phases of mass transfer, during which one of the stars loses part or all of its outer envelope and becomes a stripped-envelope star. The evolution of the size of these stripped stars is crucial in determining whether they experience further interactions and their final fate.  We present new calculations of stripped-envelope stars based on binary evolution models computed with MESA. We use these to investigate their radius evolution as a function of mass and metallicity. We further discuss their pre-supernova observable characteristics and potential consequences of their evolution on the properties of supernovae from stripped stars.  At high metallicity we find that practically all of the hydrogen-rich envelope is removed, in agreement with earlier findings. Only progenitors with initial masses below 10\Msun expand to large radii (up to 100\Rsun), while more massive progenitors stay compact. At low metallicity, a substantial amount of hydrogen remains and the progenitors can, in principle, expand to giant sizes (> 400\Rsun), for all masses we consider. This implies that they can fill their Roche lobe anew. We show that the prescriptions commonly used in population synthesis models underestimate the stellar radii by up to two orders of magnitude. We expect that this has consequences for the predictions for gravitational-wave sources from double neutron star mergers, in particular for their metallicity dependence.
}
   \keywords{binaries: close -- gravitational waves -- stars: massive stars: evolution -- stars: Wolf-Rayet -- supernovae: general}

\maketitle
   
%

\section{Introduction}
The formation of gravitational-wave (GWs) sources is a key problem that is becoming increasingly important to discuss in the new era of gravitational-wave detections \citep{abbott_observation_2016,abbott_gw170817:_2017,ligo_&_virgo_collaboration_gwtc-1:_2019}. The compact objects, neutron stars (NSs) or black holes (BHs), whose mergers give rise to the gravitational wave chirp, represent the end products of massive stars (above about $8 \, \rm{M}_{\odot}$). 

How the remnants of two stars can eventually be in an orbit close enough for them to merge by emission of gravitational waves within a Hubble time is a process of which many aspects are still poorly understood. Not only does it require a detailed understanding of the evolution and fate of massive stars, but also of their binary interaction. Moreover, with the reach of present-day GW detectors we probe nearby mergers of compact objects, but we expect their progenitors to have formed at appreciable or even large redshifts. This is because of the time delay between the formation of a double compact objects and the final merger. We thus expect that many of the progenitors formed out of more pristine gas, i.e.\ gas that is less enriched in heavy elements contributed by previous generations of stars. This means we must carefully understand the effect that metallicity has on the evolution of massive stars in binaries.

Several scenarios for the formation of gravitational-wave sources have been proposed. Many of these involve the interaction between two stars in a close binary system through one or multiple phases of Roche-lobe overflow \citep[e.g.][and references therein]{kippenhahn_entwicklung_1967, tauris_formation_2017} that strip one star, or eventually both stars, of most of their hydrogen-rich envelopes, which is about two thirds of their initial mass. The stars that result from this process is what we will refer to as a stripped star, hereafter. 

Stripped stars are largely composed of helium and, later, heavier elements. As a result, one may naively expect these stars to be very compact. This is largely the case, at least during the long-lived phase of core helium fusion. However, it has been shown that stripped stars can swell and reach giant dimensions in the late stages of their evolution, depending on their mass. Some of the early numerical calculations already demonstrated this phenomenon, approximating stripped stars as pure helium stars \citep[e.g.,][]{divine_structure_1965,habets_evolution_1986}. More recent calculations confirmed this either considering pure helium stars \citep{dewi_evolution_2002, dewi_late_2003, Ivanova_The_Role_2003} or by fully following the evolution of the massive star progenitor through the stripping process in a binary system 
\citep[e.g.][]{yoon_type_2010, eldridge_death_2013, sravan_progenitors_2018}

Whether or not a stripped star expands and by how much is very relevant for understanding their fate as the progenitors of core collapse supernovae and possibly gravitational-wave sources. The large sizes of stripped stars imply that they can fill their Roche lobe anew and undergo an additional phase of mass transfer \citep{dewi_evolution_2002, dewi_late_2003, Ivanova_The_Role_2003}. Additional phases of mass transfer can produce stars with even lower envelope masses, known as ultra-stripped stars \citep{tauris_ultra-stripped_2015}. When these stars end their lives it is believed that the resulting low ejecta masses lead to very small supernova kicks \citep[e.g.,][]{podsiadlowski_effects_2004}, which can prevent disruption of the binary system at the moment of explosion. This favors the formation of close NS binaries, some of which will be tight enough to merge as a result of GWs within a Hubble time \citep{tauris_formation_2017}. The occurrence and outcome of additional phases of mass transfer thus directly impact predictions for the formation of gravitational-wave sources, in particular double neutron star mergers. 

Improving our understanding of the expansion of stripped stars is also very relevant in the light of upcoming electromagnetic transient surveys, because they are responsible for about a third of all core-collapse supernovae \citep{Graur+2017a,Graur+2017}. The radius and the mass of their envelopes impact the light-curves predicted for stripped-envelope supernovae \citep{Kleiser+2014, dessart_supernovae_2018,kleiser_helium_2018} together with the mass-loss rate expected at late times and thus the circumstellar material around the progenitor at the moment of explosion \citep[][]{ouchi_radii_2017}. Of particular interest is the case of type Ib supernova iPTF13bvn \citep{Cao+2013}, which appears to provide the most direct evidence we have that stripped stars can end their lives as helium giants 
\citep[e.g.,][]{Fremling+2014,Bersten+2014,Eldridge+2016}

Recently, several studies have drawn attention to the fact that metallicity can have a large impact on the properties of stripped stars. \citet{gotberg_ionizing_2017} and \citet{yoon_type_2017} find that metallicity strongly affects the radial extent of the hydrogen layer that is left at the surface of the star after stripping. At low metallicity the reduced internal opacity makes it possible for stars to retract within their Roche lobe before the hydrogen envelope has been fully removed, see also \citet[][]{gotberg_ionizing_2017}. In addition, metallicity affects the stellar winds, which can strip the stars even further \citep[see also][]{vink_mass-loss_2001,gilkis_effects_2019}. 

Calculations of the formation of gravitational-wave progenitors are typically made with binary population synthesis simulations, which rely on simplified assumptions for the stellar structure and interaction phases. Such simplifications are necessary because the simulations typically involve following the evolution of millions of stars in binary systems through complex phases of interaction. In the vast majority of binary population synthesis codes stars are treated with analytic fits by \citet{hurley_comprehensive_2000} against evolutionary models by \citet{pols_stellar_1998}. A list of examples is given in Section~\ref{sec:comparison_hurley}. An alternative is to interpolate in pre-computed grids of models. Two examples of recent studies that use grids of pre-computed single star models are \citet{kruckow_progenitors_2018} and \citet{giacobbo_progenitors_2018}, see also Section~\ref{sec:comparison_SEVN}. A third alternative is to post-process extended grids of binary evolutionary models. The most prominent example of this is the BPASS code \citep{eldridge_BPASS_2016, eldridge_binary_2017}. 

For the treatment of stripped stars, all population synthesis studies listed above (with the exception of BPASS) use fits against or interpolation in grids of models of single helium stars computed at solar metallicity. They do not make use of models where the stripped star has been computed self consistently through the Roche stripping process. This has two drawbacks: (1) they do not account for the effect that a left-over layer of hydrogen on the surface has on the properties of stripped stars; (2) they do not fully account for the effect that metallicity has on the properties of stripped stars. 
 
In this paper, we present a study of the radius evolution of stars stripped in massive binaries considering solar and low metallicity. For this purpose, we compute a grid of representative progenitor models for different masses that are relevant as supernova and possible neutron star progenitors. We follow their evolution through Roche-lobe overflow with a detailed binary evolution code. We focus on the expansion phases after central helium depletion and discuss how this is linked with their interior evolution and in particular the burning phases. We then show how the radii compare to sizes usually assumed in binary population synthesis models, and estimate how the differences impact the number of systems that can interact a second time through Roche-lobe overflow. We estimate and discuss the implications for core-collapse supernovae and gravitational-waves progenitors. 

This paper can be considered as a companion of the paper by \citet{gotberg_ionizing_2017}, which presents an extensive discussion of the effect of metallicity on the long-lived phase of central helium burning. In this work we extend to the late evolutionary phases. 

The paper is structured as follows: we summarize our model assumptions in section \ref{sec:methods}, before discussing the effect of metallicity using two representative stellar models in section \ref{sec:ev}. In section \ref{sec:grid}, we present our full grids of evolutionary stellar models. We compare the radii obtained to those commonly used in population synthesis models in section \ref{sec:comparison}. We then discuss the impact of these large radii on the progenitors of supernova and GW sources in section \ref{sec:discussion} together with a discussion of the uncertainties. A summary with our conclusions is provided in section \ref{sec:conclusion}.

\section{Binary evolution models}
\label{sec:methods} 
For our calculation of the interacting binary stars, we use the open-source 1D stellar evolution code \emph{MESA} \citep[version 10398,][]{paxton_modules_2011,paxton_modules_2013, paxton_modules_2015, paxton_modules_2018, paxton_modules_2019}. The models are computed at solar metallicity (initial metal fraction of $Z_{\odot} \equiv 0.0142$, based on values from \citet{asplund_chemical_2009} and sub-solar metallicity ($Z = 0.001$, representative of nearby low-metallicity environments such as the Small Magellanic Cloud, $Z_{\rm{SMC}} \approxeq Z_{\odot} / 5$).\\
Our zero age main sequence (ZAMS) models are computed by following their pre-main sequence evolution until the central helium abundance has increased by 5\%. Following \citet{tout_zero-age_1996}, which is consistent with \citet{pols_stellar_1998}, we assume an initial hydrogen mass fraction of $X=1-Z-Y$, where $Y = 0.24 + 2Z$ is the helium mass fraction. Strictly speaking, given the updated abundances from \citet{asplund_chemical_2009}, the helium abundances should be adapted, even though the difference is small. We compute the evolution of the stars until core carbon depletion (defined as the moment when the core carbon abundance decreases below $10^{-4}$). This is sufficient for the purpose of this work since the remaining evolutionary time is so short, less than 100 yr in these models, that the outer layers do not have time to react to quasi-hydrostatic changes in the core. 

We use a nuclear network comprising 21 isotopes which follows the most prominent nuclear processes that influence the life of massive stars from hydrogen burning through the CNO cycle until silicon burning with sufficient accuracy \citep[approx21,][]{paxton_modules_2011}. It has the advantage of enabling fast calculations while containing the most important isotopes relevant for our study. We use default opacity tables of \emph{MESA} \citep{iglesias_radiative_1993,iglesias_updated_1996,buchler_compton_1976,cassisi_updated_2007}.\\
Convective mixing is accounted for by using mixing-length theory \citep{bohm-vitense_uber_1958} with a mixing length parameter of $\alpha=1.5$ commonly used in stellar evolution models \citep[e.g., ][]{pols_stellar_1998}. We employ values for step overshooting above convective regions of 0.335 pressure scale heights, based on the calibration of stellar models for early B-type stars against observations valid in a comparable mass range as our models, (${10-20 \,\textrm{M}_{\odot}}$; \citealt{brott_rotating_2011}). 
We also take into account rotational mixing \citep{paxton_modules_2013}, semi-convection, and thermohaline mixing \citep{kippenhahn_time_1980} until the end of core helium burning, even though earlier studies have shown that these have little impact on the stellar structures of stripped stars \citep{yoon_type_2010,gotberg_ionizing_2017}.\\

We use the theoretical wind mass-loss algorithm from \citet{vink_mass-loss_2001} for the main sequence evolution and the \cite{de_jager_mass_1988} prescription for stars with hydrogen mass fractions lower than $X_{\text{H, s}} = 0.4$ and effective temperatures lower than $10^{4}$K. Because of the scarcity of observational constraints for the wind mass-loss from stripped stars, we employ the empirically derived wind mass-loss prescription from \citet{nugis_mass-loss_2000} based on the winds of Wolf-Rayet (WR) stars. This mass-loss rate is very close to the value derived for the observed intermediate mass stripped star \citep[the qWR star HD 45166,][]{groh_qwr_2008}. It is possible that stripped stars have lower wind mass-loss rates than what is predicted from the extrapolated wind mass-loss scheme of \citet{nugis_mass-loss_2000} since they are not close to the Eddington limit \citep{bestenlehner_vlt-flames_2014}. Indeed, \citet{vink_winds_2017} suggested that the wind mass loss rate from stripped stars is about 10 times lower than what is predicted from the \citet{nugis_mass-loss_2000} prescription. However, the models from \citet{vink_winds_2017} assume an effective temperature for stripped stars of $T_{\rm{eff}} = 50,000$ K, which is much lower than what our models imply for the long-lived phase of central helium burning. If stripped stars have lower wind mass-loss rates than what we employ in this study, they may expand more because less of the hydrogen is removed from the surface \citep[cf.][]{gilkis_effects_2019}. \\
We compute our models by employing the default spatial and temporal resolution of \emph{MESA} (\texttt{varcontrol\_target = $10^{-4}$} and\texttt{ mesh\_delta\_coeff = 1.0}). We increase the temporal resolution for phases involving the depletion of fuel in the core (e.g., core hydrogen depletion). Due to numerical issues, we lower the temporal resolution to a maximum of $10^{-3}$ after core helium depletion and lower the sensitivity of the models to changes in abundances of elements in the core after the formation of an oxygen core.\\
We compute the interaction with a binary companion using the approach described in \citet{paxton_modules_2015}. We take the effect of tides into account using \citet{hut_tidal_1981} and use the implicit mass-transfer scheme of \citet{ritter_turning_1988} for the Roche-lobe overflow. As a representative case for stable mass transfer, we treat the secondary component of the binary system as a point mass that has 80\% of the primary's mass. We assume conservative mass transfer. We only consider stripped-envelope stars created by the transfer of mass from stars that fill their Roche lobe due to a rapid expansion during their hydrogen shell burning phase after leaving the main sequence \citep[case B mass-transfer, see][]{kippenhahn_entwicklung_1967,podsiadlowski_presupernova_1992}. To this end, we adopt an initial orbital period of $25\,\rm{d}$ for all models. Stripped stars created through this channel at solar metallicity have very similar properties regardless of the exact choice of the initial orbital period and companion mass \citep{gotberg_ionizing_2017}. At low metallicity the efficiency of stripping of the envelope is dependent on the initial orbital period \citep[see][]{yoon_type_2017,ouchi_radii_2017}. Therefore, at low metallicity our results should be regarded as a representative approximation. 

Stars may alternatively be stripped by common-envelope evolution. It is currently not known whether stars which have their envelopes ejected in this manner have different post-envelope-ejection properties than stars which are stripped by stable mass transfer \citep[see, e.g.,][]{ivanova_common_2013}, or even whether there is a metallicity dependence.

Stripped stars have relatively small radii during central helium burning \citep[around 1\Rsun][]{habets_evolution_1986,gotberg_ionizing_2017,yoon_type_2010}, but they are expected to swell up once helium is depleted in the core. In this paper, we investigate the impact of the radius expansion at the end of core helium burning. If models expand enough to fill their Roche lobe anew, they are expected to start an additional mass-transfer phase \citep[sometimes referred to as case BB and BC mass-transfer, see e.g,][]{dewi_evolution_2002}. The size of Roche lobe varies depending on the size of the orbit, which in turn depends on how angular momentum and mass is transferred. This is still a major uncertainty in binary evolution \citep[e.g.,][]{de_mink_efficiency_2007}. To avoid these complications and derive results that are of more generic use, we follow the late evolution of stripped stars by letting them expand as much as their internal structure dictates, ignoring any limitation imposed by the finite but highly uncertain size of the Roche lobe at this stage. In other words, we effectively treat the stripped stars as single stars during their late evolutionary phases. This allows us to investigate their full expansion and simplifies the interpretation of the physical processes involved in the radial expansion. It also makes the results of our simulations more suitable for inclusion in future populations synthesis simulations.

\section{Evolution of two representative models}

\label{sec:ev}
\begin{figure*}[ht!]
    \vspace{-0.25cm}
	\resizebox{\hsize}{!}{
		\includegraphics[width=\textwidth]{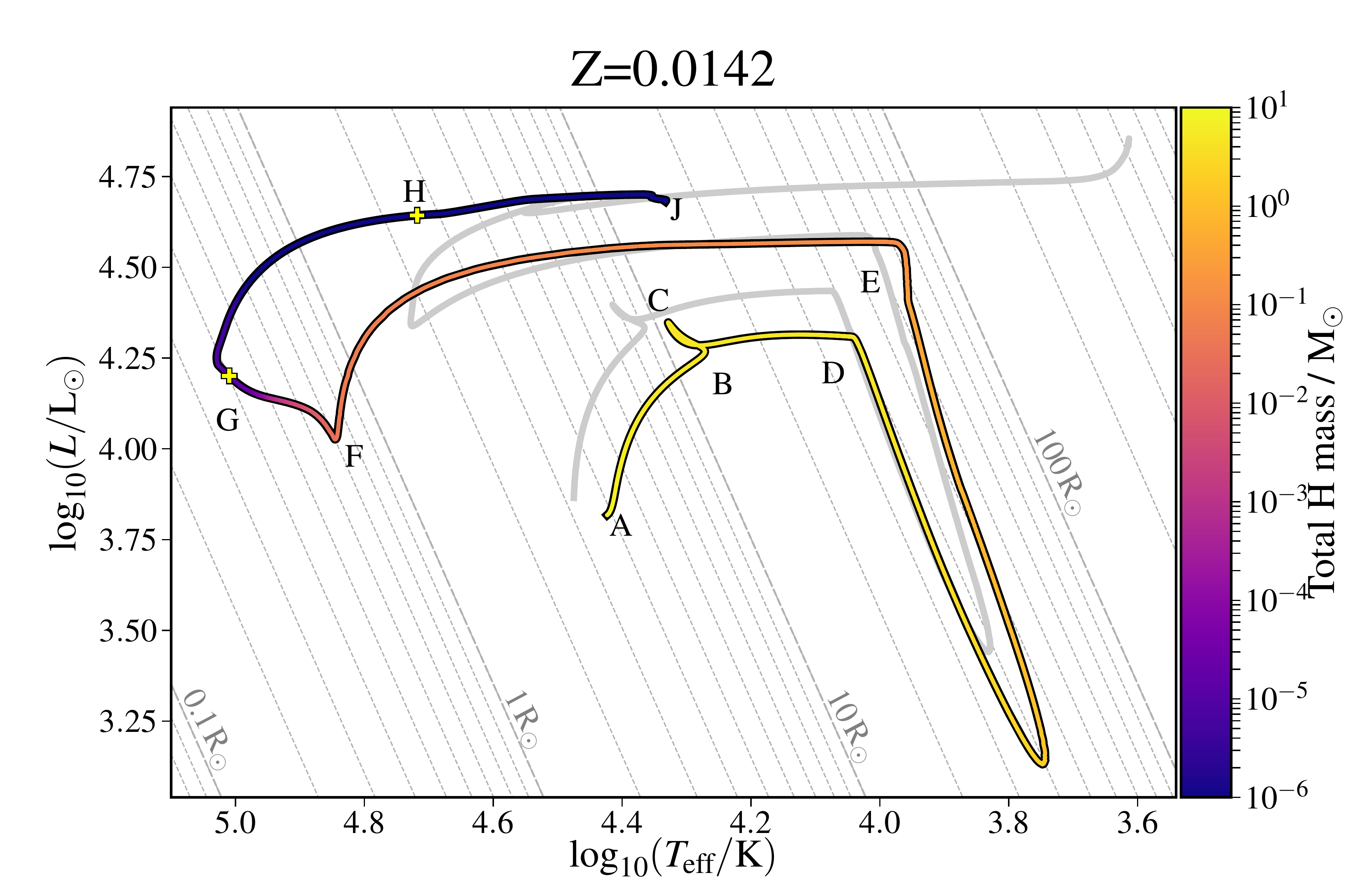}
		\includegraphics[width=\textwidth]{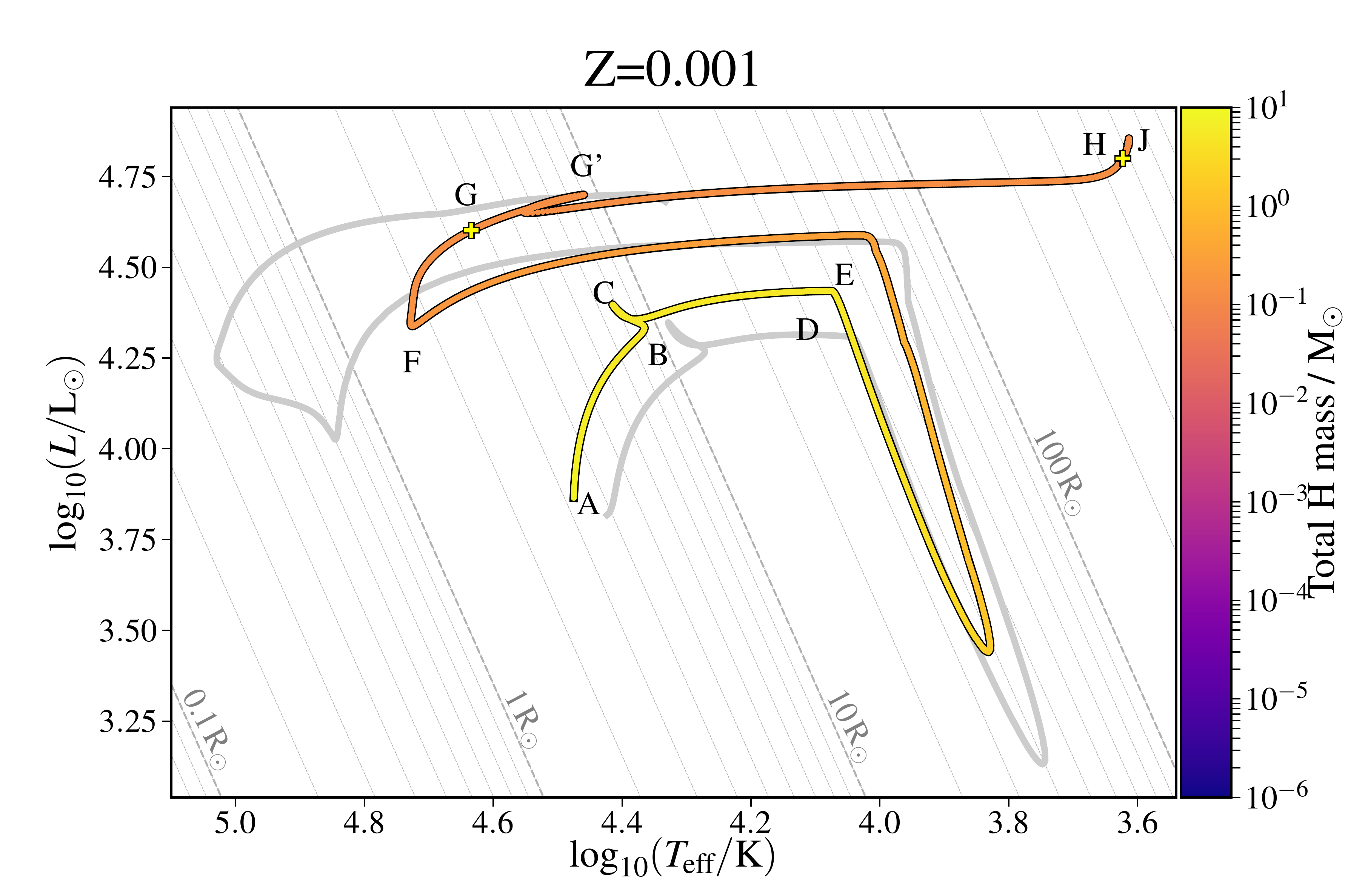}
	}
	\resizebox{\hsize}{!}{
		\includegraphics[width=0.8\textwidth]{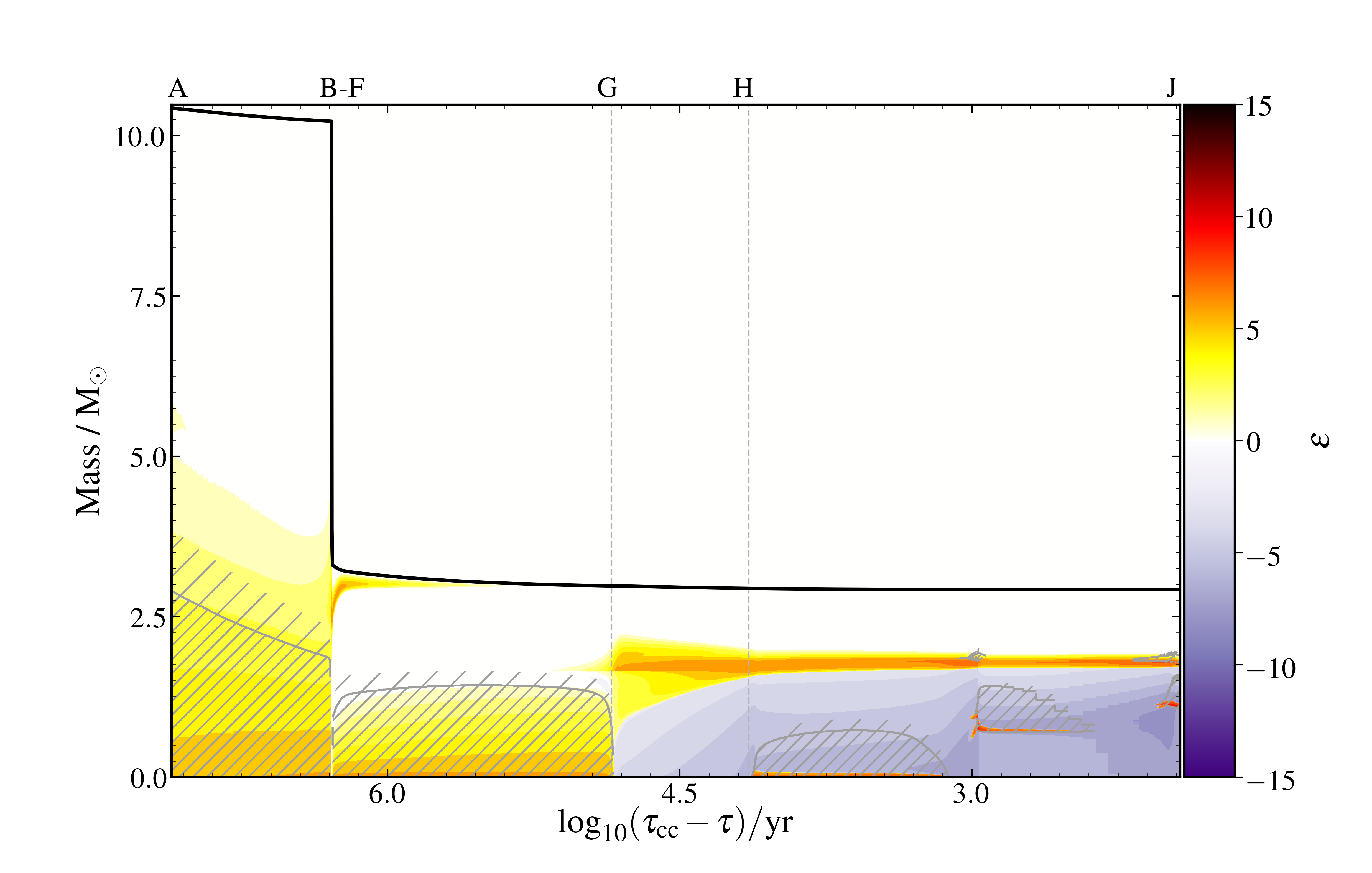}
		\includegraphics[width=0.8\textwidth]{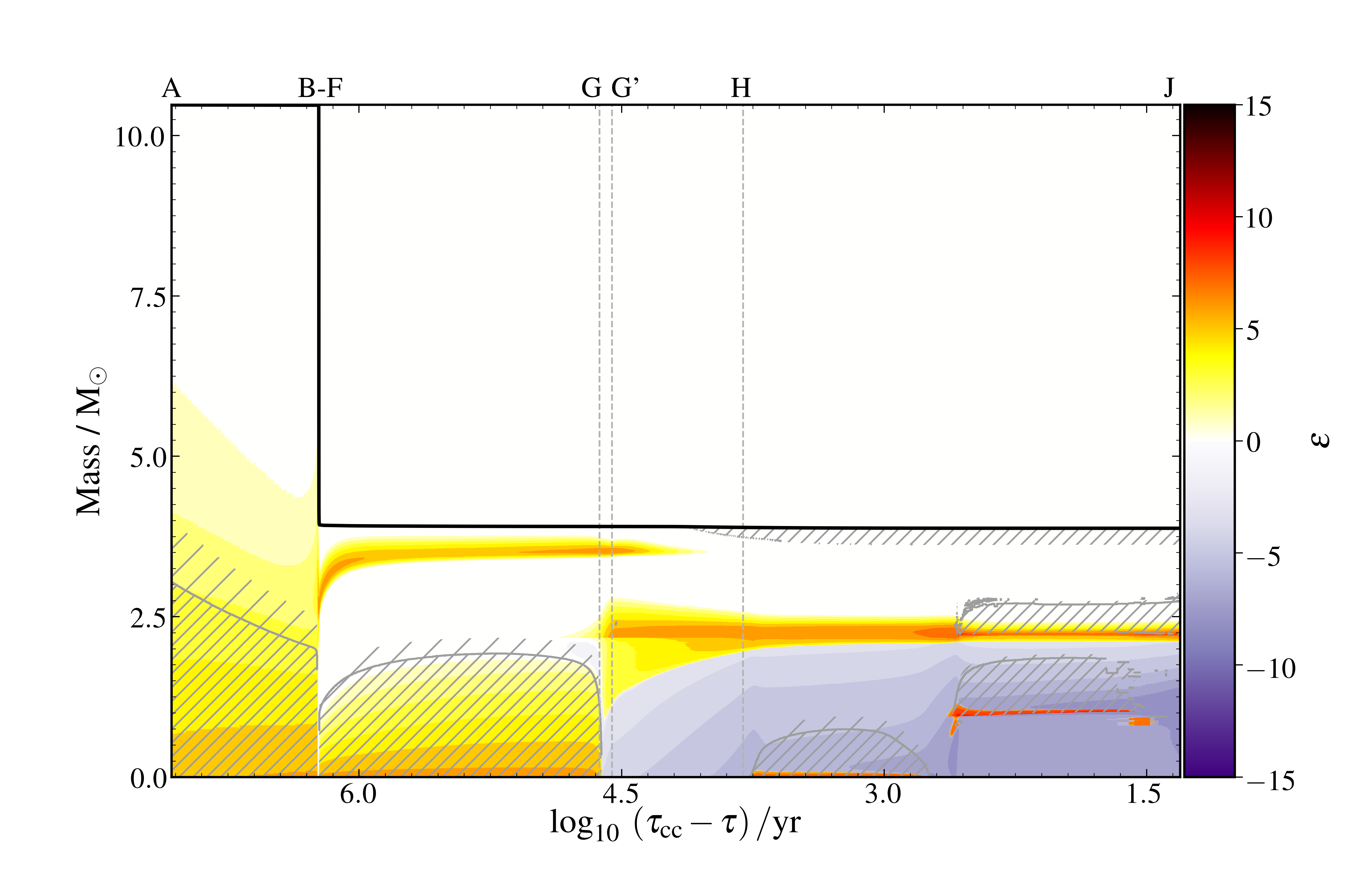}
	}
	\resizebox{\hsize}{!}{
	\includegraphics[width=0.8\textwidth]{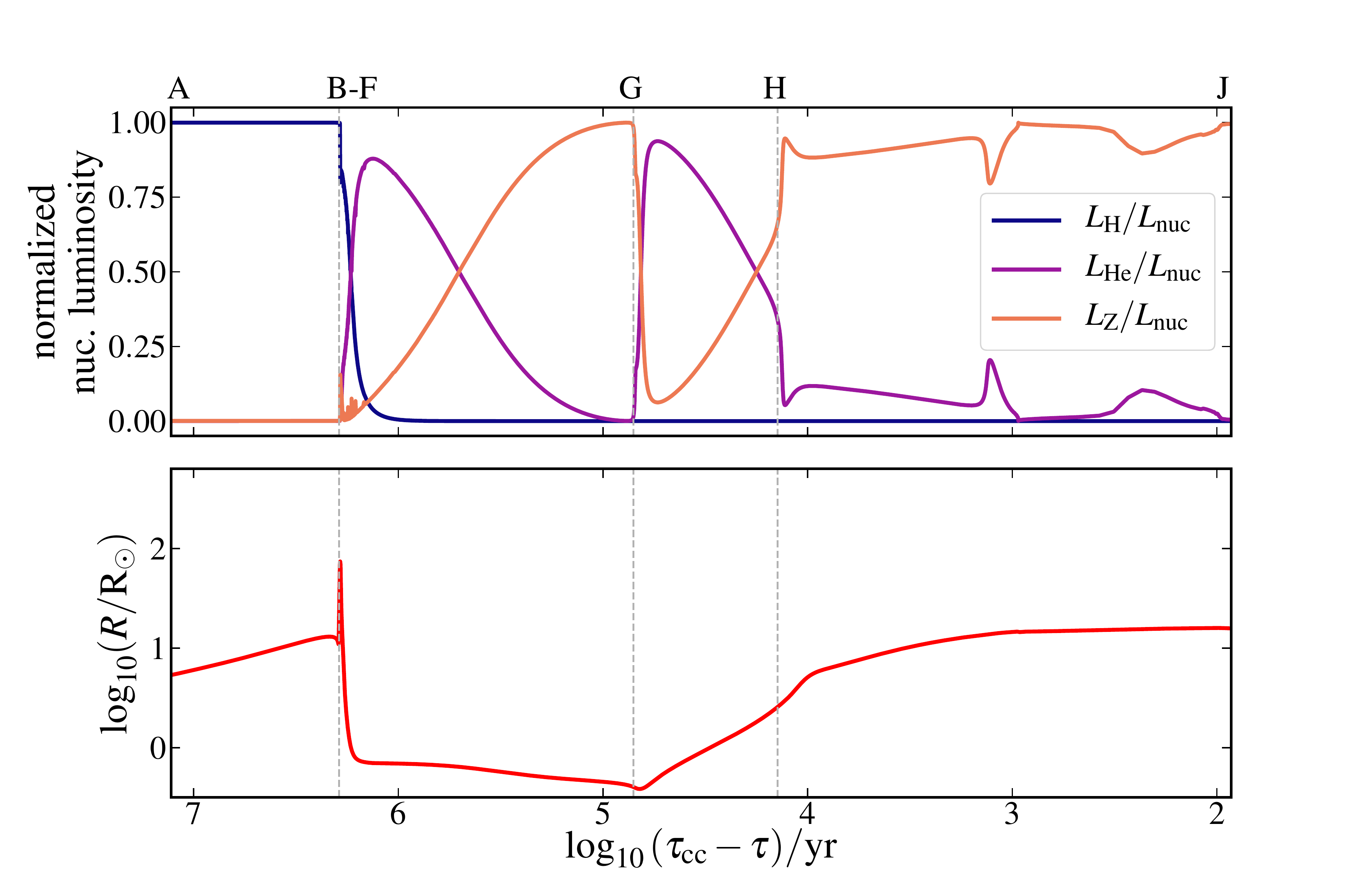}
	\includegraphics[width=0.8\textwidth]{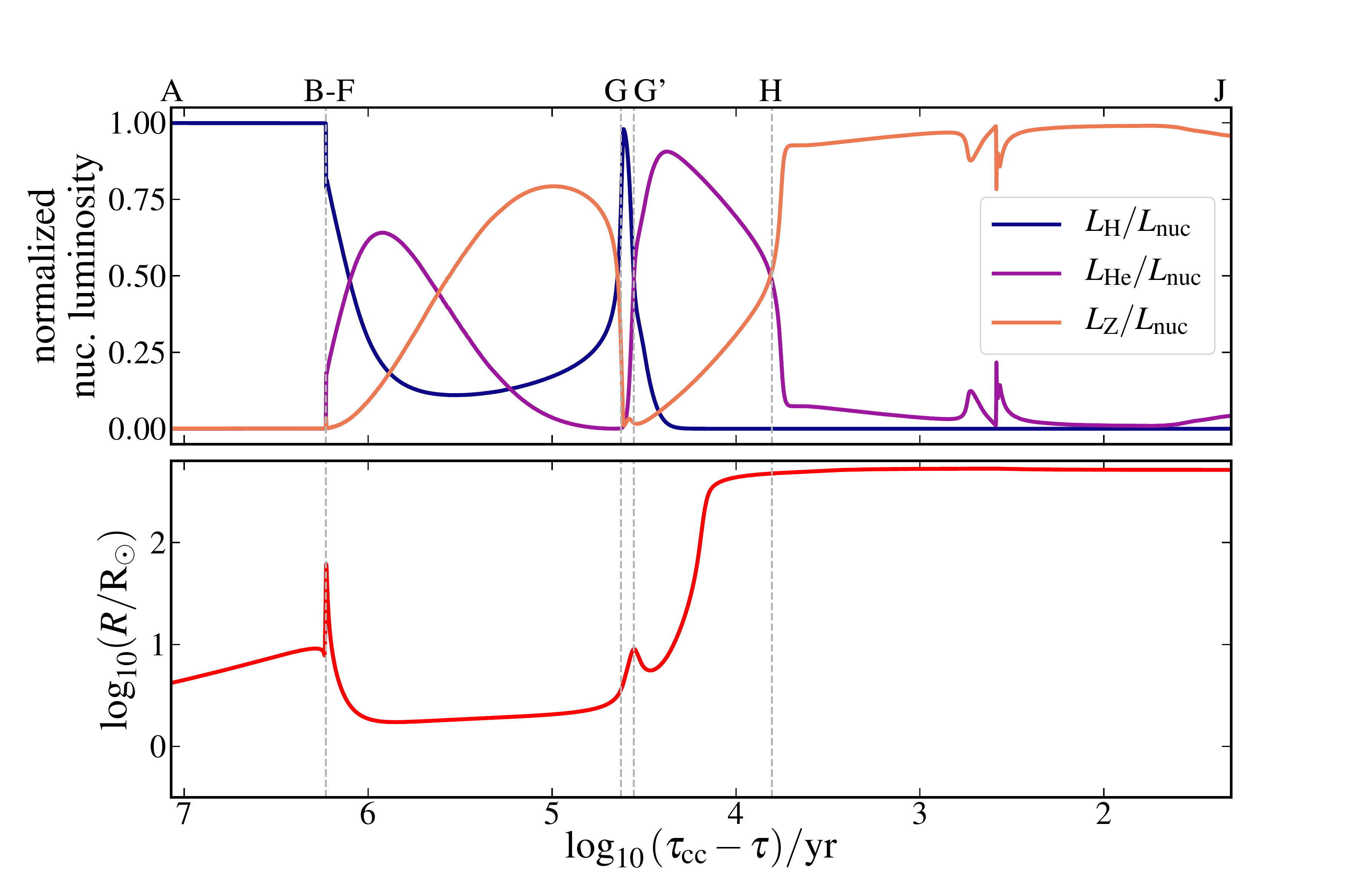}
	}
	\caption{
		The evolution of a high 
		(left) and low 
		(right) metallicity star of initially $10.5\, \rm{M_{\odot}}$ that is stripped due to binary interaction. Letters A to J mark evolutionary points discussed in section \ref{sec:ev}. 
		\textbf{Top:} Evolutionary tracks on the Hertzsprung-Russell diagram. The color indicates the total mass of hydrogen present at a given moment. For comparison we show lines of constant radii as well as the evolution of the alternate-metallicity model in gray.   
		\textbf{Middle:} Evolution of the stellar structure of the stars in mass coordinate (Kippenhahn diagrams), given as a function of time until core collapse ($\tau_{\rm cc}$). Convective and overshooting regions are marked with double- and single-hatched regions, respectively. Colors indicate zones dominated by nuclear burning (yellow) or neutrino cooling (purple) where  $\epsilon=\rm{sign}\left(\epsilon_{\rm{nuc}}-\epsilon_{\nu}\right)\log_{10}\left(\rm{max}\left(1.0,|\epsilon_{\rm{nuc}}-\epsilon_{\nu}|\right) / [\rm{erg\,g^{-1}\,s^{-1}}]\right)$. Here $\epsilon_{\rm{nuc}}$ is the nuclear energy generation rate and $\epsilon_{\nu}$ the neutrino energy. The black line indicates the location of the stellar surface. 
		\textbf{Bottom:} Luminosity produced by hydrogen-, helium- and metal-burning as a fraction of the total luminosity produced by nuclear reactions (upper) and stellar radius (lower) as a function of time until core-collapse.
}
\label{fig:example_models}
\end{figure*}
Before presenting our full grid of models, we first describe the evolution and properties of two representative stellar models, with identical initial component masses and orbital periods, at solar ($Z=0.0142$) and low metallicity ($Z=0.001$). The initial mass of the hydrogen-rich primary star in these models is $10.5$\Msun, corresponding to the mean initial mass in our grid. Following the method described in the previous section, we place this star in an orbit with an initial period of $25 \,\rm{d}$, with an $8.4$\Msun companion. For the purposes of our investigation, the companion is removed after the mass-transfer phase, when the primary star that has become a stripped star has reached core helium depletion. 

Figure \ref{fig:example_models} compares the evolution of these primary-star models (left panels: solar metallicity, right panels: low metallicity). The top pair of panels presents their evolution in Hertzsprung-Russell diagrams (HRDs). The middle panels present their Lagrangian internal structures as a function of time, known as Kippenhahn diagrams. The lower sets of plots give the time evolution of the stellar radii, along with the fractional contributions of different forms of nuclear luminosity. Key points are labelled with letters in each panel, and is discussed in the following subsections. 

\subsection{Evolution until core helium burning (A--F)}
\label{sec:ev_a_f}
The early evolution of the primary star in the binary is effectively the same as that of a single star. While we include the effects of rotation and tides, they have a negligible impact on the evolution. During their main-sequence evolution (labeled as A--B in Fig. \ref{fig:example_models}) the stars burn hydrogen into helium in a convective core. As they do so, they increase in luminosity. The higher-metallicity star is cooler and larger due to a higher opacity in the outer layers \citep[due to increased bound-free and bound-bound absorption, e.g., ][]{maeder_tables_1990, schaller_new_1992}. After leaving the main sequence (B), the stars contract (see also the middle and lower panel of Fig. \ref{fig:example_models}) until hydrogen is ignited in a shell (C). They then expand on their thermal time-scale until they fill their Roche lobe, which leads to transfer of matter to their companion.\\
During mass transfer (D--E), the stars lose the majority of their hydrogen-rich envelopes on a thermal time-scale \citep[see, e.g.,][for more details]{kippenhahn_entwicklung_1967}. The mass-transfer phase occurs at similar effective temperatures in both models presented here, since the size of the Roche lobe is the same, although the lower metallicity model has expanded slightly more relatively to the higher metallicity model since the end of the main sequence. More hydrogen is retained after the end of the Roche lobe overflow phase (B--F) in the lower-metallicity model (0.27\Msun, compared to 0.12\Msun at solar metallicity, see also the upper panels of Fig.~\ref{fig:example_models}). This is because the stripping process becomes less effective at low metallicity due to the lower opacity in the outermost layers (see especially \citealt{gotberg_ionizing_2017} and references therein; also \citealt{klencki_high_2018}). With a lower opacity, more of the hydrogen-rich envelope remains within the Roche lobe once the stars detach, leaving a thicker layer of hydrogen on top of the metal-poor stripped star. This is clearest in the Kippenhahn diagrams in Fig. \ref{fig:example_models}, in which the lower-metallicity model has more mass outside the hydrogen-burning shell.\\
After the end of the mass-transfer phase the stripped star shrinks towards a new gravothermal equilibrium structure. Meanwhile, convective helium burning has already started in the core. From the luminosity minimum (F), the dominant driver of structural change is once more nuclear burning.

\subsection{Core helium burning phase (F--G)}
\label{sec:ev_f_g}
Point F in Fig. \ref{fig:example_models} marks the longest-lived phase of these stripped stars (about 10\% of the stellar lifetime), with helium burning in a convective core. 
The metal-rich model is hotter and more compact than the metal-poor model because the metal-poor model retains more hydrogen in its envelope, which allows it to be larger \citep{cox_equilibrium_1961}.
From here onward the evolution varies considerably between the two models. The metal-rich star shrinks monotonically during the core helium burning phase, while the metal-poor model first shrinks, then stays approximately constant, before starting to expand again \citep[e.g., ][]{gotberg_ionizing_2017}.
 
\subsection{Expansion phase after core helium depletion (G--J)}
\label{sec:ev_g_j}
\begin{figure*}[ht!]
	\resizebox{\hsize}{!}{
		\includegraphics[width=\textwidth]{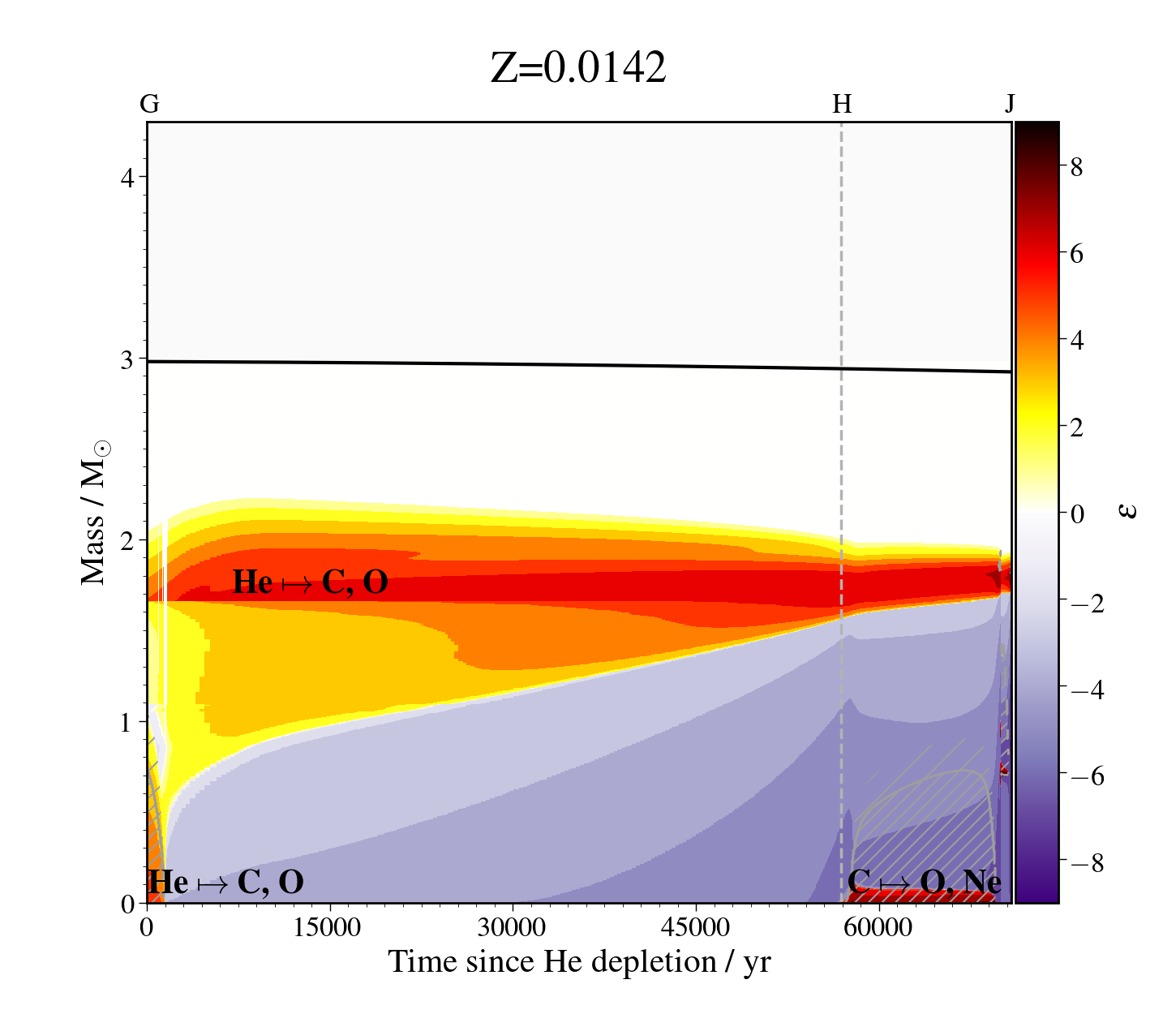}
		\includegraphics[width=\textwidth]{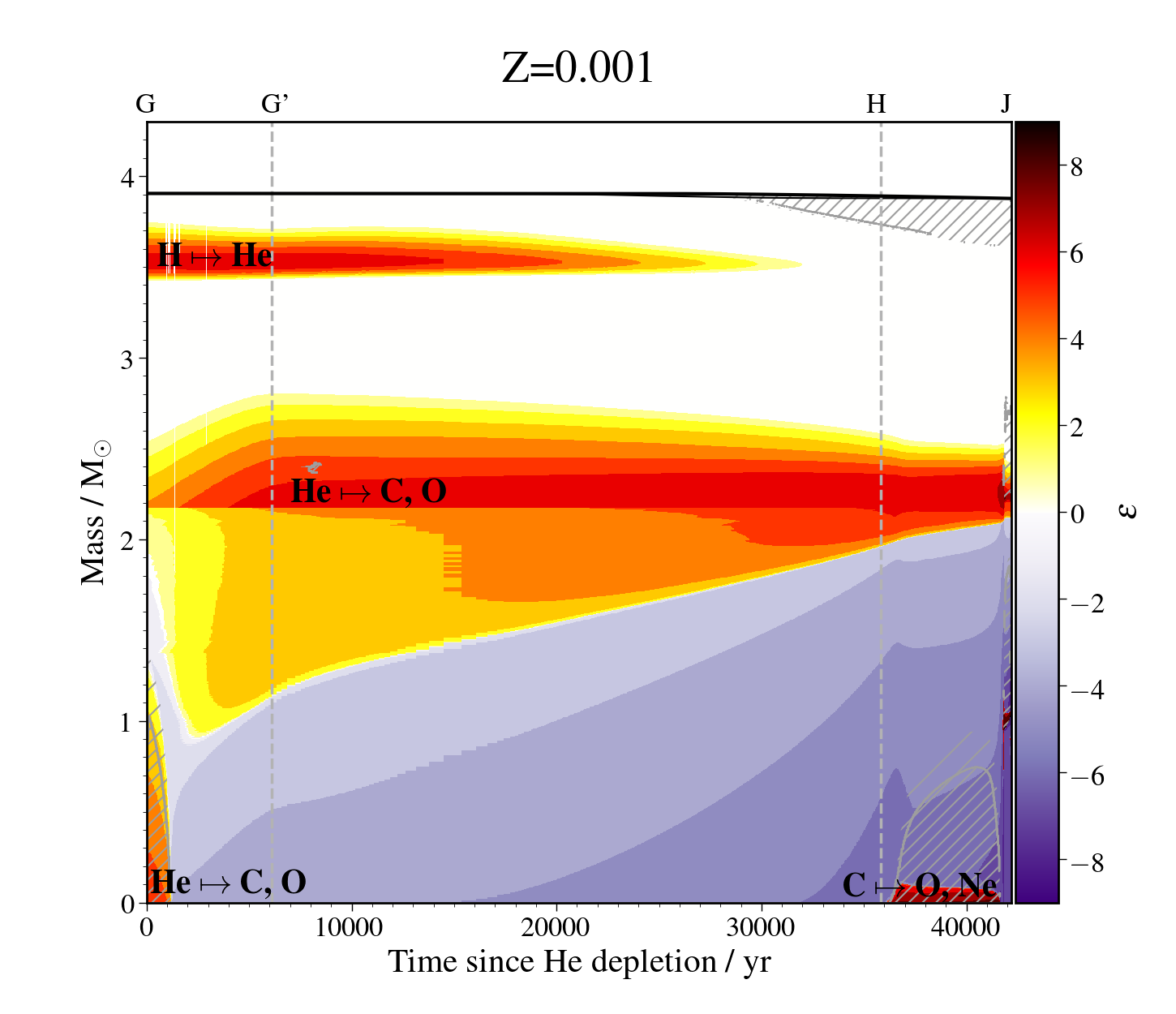}
	}
	\resizebox{\hsize}{!}{
	\includegraphics[width=\textwidth]{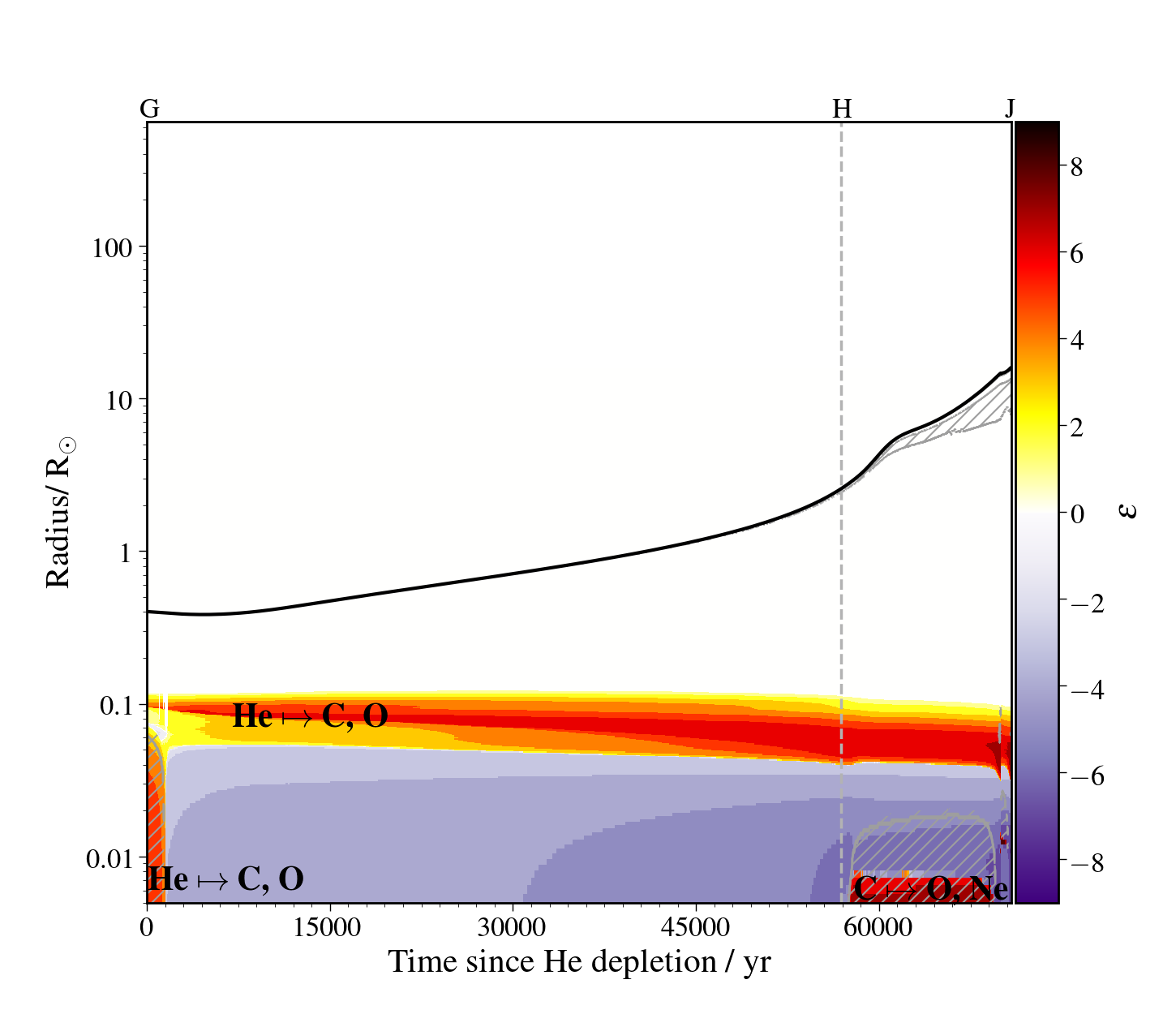}
	\includegraphics[width=\textwidth]{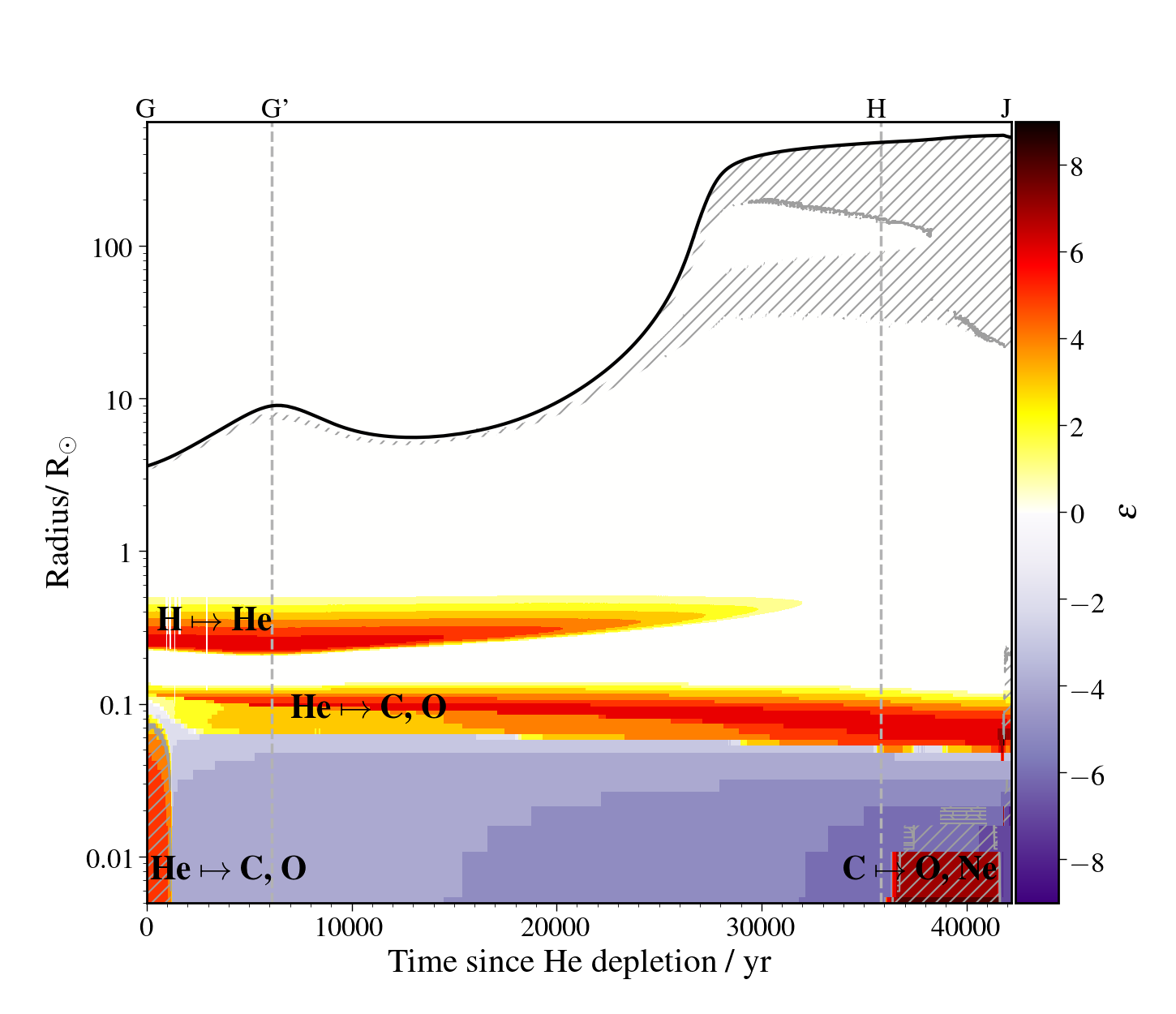}
}
	\caption{
		Kippenhahn diagrams showing the evolution of the stellar structure for our two example models at solar (left panels) and low (right panels) metallicity. They are shown using mass coordinate as the vertical axis (top panels) and radius coordinate (bottom panels). The horizontal axis indicates the time since helium depletion up to core carbon depletion. The black line represents the surface of the star. Regions of mixing by convection and overshooting are marked with double- and single-hatches, respectively. Colors indicate zones dominated by nuclear burning or neutrino cooling. See Fig. \ref{fig:example_models} and Section~\ref{sec:ev} for details and discussion.}
	\label{fig:kipp_example}
\end{figure*}
After core helium depletion, the stars begin the short last phase of their lives, which lasts for less than 1\% of their total stellar lifetime. Points labelled G in Fig.~\ref{fig:example_models} are at the moment when the central helium mass fraction drops below $10^{-4}$. At this point, the metal-rich model has a total mass of 2.98\Msun and lost all its remaining hydrogen due to stellar winds, whereas the metal-poor model has a total mass of 3.90\Msun and a total mass of hydrogen of 0.14\Msun. The structures during the post-core-helium-burning phase are presented in more detail in Fig.~\ref{fig:kipp_example}, using both mass and radius coordinate systems (the top and bottom panels, respectively).\\
We see large differences between the high- and low-metallicity models in this last evolutionary phase. The radius evolution becomes very distinct, with a monotonic increase of the radius for the metal-rich model until a maximum of 15\Rsun is reached, and a non-monotonic increase of the radius for the metal-poor model until a maximum of 530\Rsun (see the upper and lower panels of Fig.~\ref{fig:example_models}). We further find that while the burning processes in the inner core are fairly similar, the metal-poor model develops prominent convection zones in the outer envelope and around the helium burning shell, and still experiences hydrogen shell-burning, while the metal-rich model has lost all its remaining hydrogen layer by this point due to winds (see Fig.~\ref{fig:kipp_example} and the middle panels of Fig.~\ref{fig:example_models}).\\

Some general features are common to the evolution at both metallicities, e.g., the narrowing of the helium-burning shell (in mass extent, see upper panels of Fig.~\ref{fig:kipp_example}) as the size of the carbon-oxygen (CO) core decreases (in radial extent, see lower panels of Fig.~\ref{fig:kipp_example}), and the increasing rate of neutrino cooling from the CO core as it contracts. During this phase of core contraction, both models overall expand. Since the mass of the CO core is somewhat higher in the less-stripped low-metallicity model, that evolution occurs more rapidly. 

Fig.~\ref{fig:kipp_example} shows that the low-metallicity model also retains a significant hydrogen layer which can sustain a burning shell, unlike the high-metallicity model that loses its hydrogen-rich envelope due to stellar winds. For the low-metallicity model hydrogen shell burning dominates the nuclear luminosity around the time of core helium exhaustion, and there is a local maximum in the radius evolution (G'), as shown in the lower-right panels of Figs.~\ref{fig:example_models} and \ref{fig:kipp_example}. In contrast, by point G hydrogen burning is not relevant for the higher-metallicity model, and there is soon a shallow minimum in the radius evolution.  

The radius expansion during the shell-burning phases, including the non-monotonic expansion of the low-metallicity model, might be interpreted in terms of the "mirror principle" (see, e.g., \citealt{kippenhahn_stellar_2012}). Examining the middle and lower panels of Fig. \ref{fig:example_models} and Fig.~\ref{fig:kipp_example}, respectively, the phases of radius expansion occur when one shell-burning source dominates the nuclear luminosity. The lower-right panels of Fig. \ref{fig:example_models} show that the first peak in the radius expansion of the low-metallicity model occurs when two shell sources are releasing roughly equivalent luminosities, analogous to the blue loop observed in some models for intermediate mass stars \citep[see e.g., ][]{kippenhahn_stellar_2012}. The layer above the helium-burning shell expands, as seen in Fig.~\ref{fig:kipp_example}, which leads to the cooling of the hydrogen burning shell. Eventually the temperature and density of the hydrogen-rich material are too low to sustain hydrogen burning and the shell is extinguished. 

Carbon is later ignited in the center, leading to the development of a new convective core. (Point H indicates this, showing when the core carbon abundance drops by 2\% from the post-core-helium-burning value). This is associated with a sharp rise in the burning luminosity of elements heavier than helium, as seen in the lower panels of Fig. \ref{fig:example_models}. By this time both stars, of 2.98\Msun and 3.90\Msun for the metal-rich and the metal-poor model, respectively, have only one shell source, helium. The hydrogen-rich envelope of the low-metallicity model still means that the star is more than an order of magnitude larger than the solar-metallicity model. 

At the end of the evolution, more than 90\% of the envelope has become convective and extends over 400\Rsun, while only containing a few tenths of a solar mass of material. In reality we do not expect the stripped star to be able to expand this much. It will likely interact again with its companion, which is still expected to be present. This would lead to another phase of mass transfer, with potentially important consequences for the type of supernova \citep[see, e.g.][]{yoon_type_2010,yoon_type_2017,dessart_supernovae_2018}. After core carbon depletion is reached (as marked with letter J), less than 100 yr remains before these models reach core-collapse. Since the thermal timescale of the stars are much longer than 100 yrs, their radii, masses and surface composition is not significantly changed after central carbon depletion. We can therefore regard the properties the stripped stars have at point J as the properties the stripped stars will have at explosion.

\section{Comparing the metal-poor and metal-rich grids}
\label{sec:grid}
\begin{figure*}[h!]
	\centering
\includegraphics[width=0.65\textwidth]{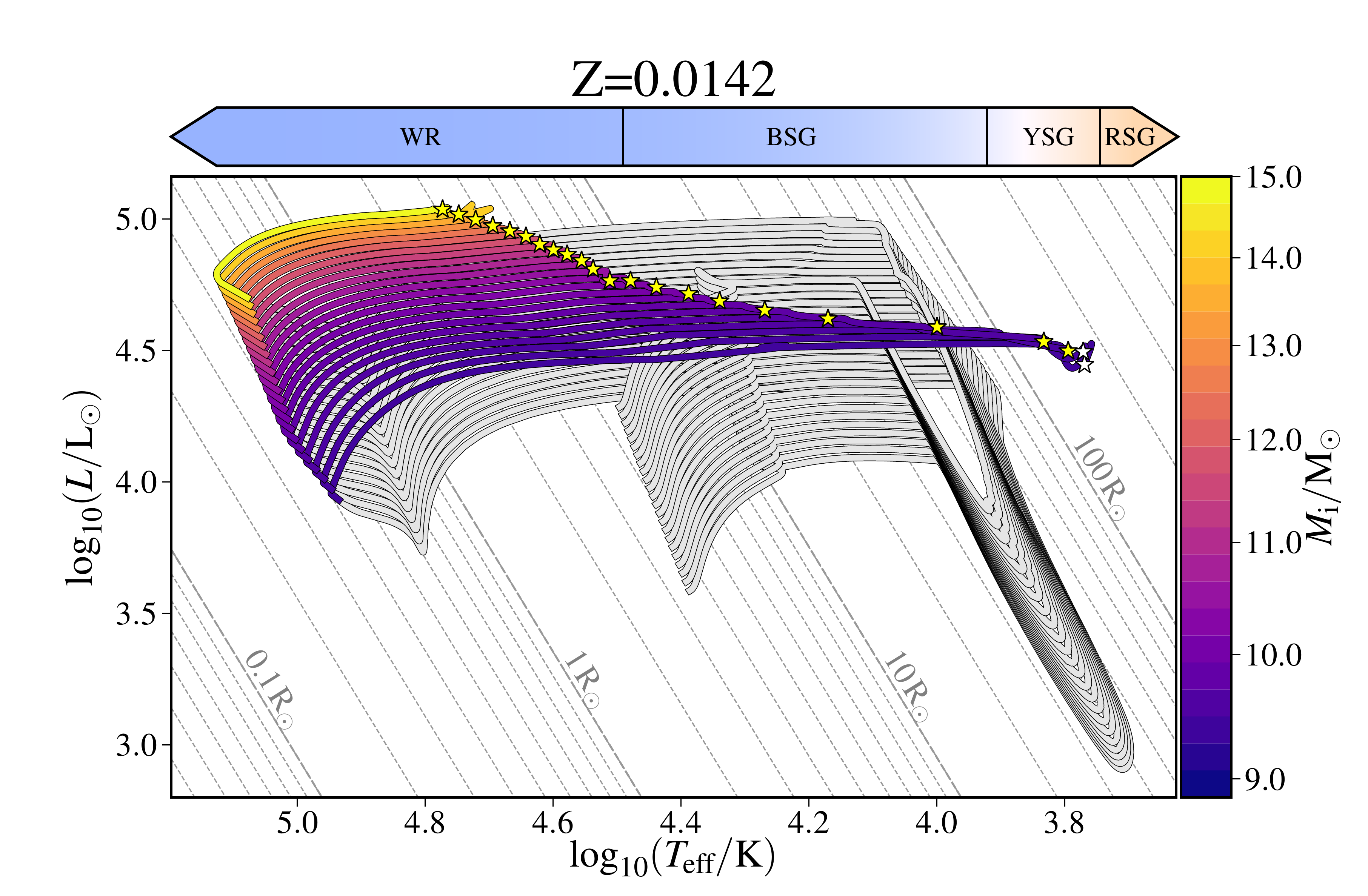}
\includegraphics[width=0.65\textwidth]{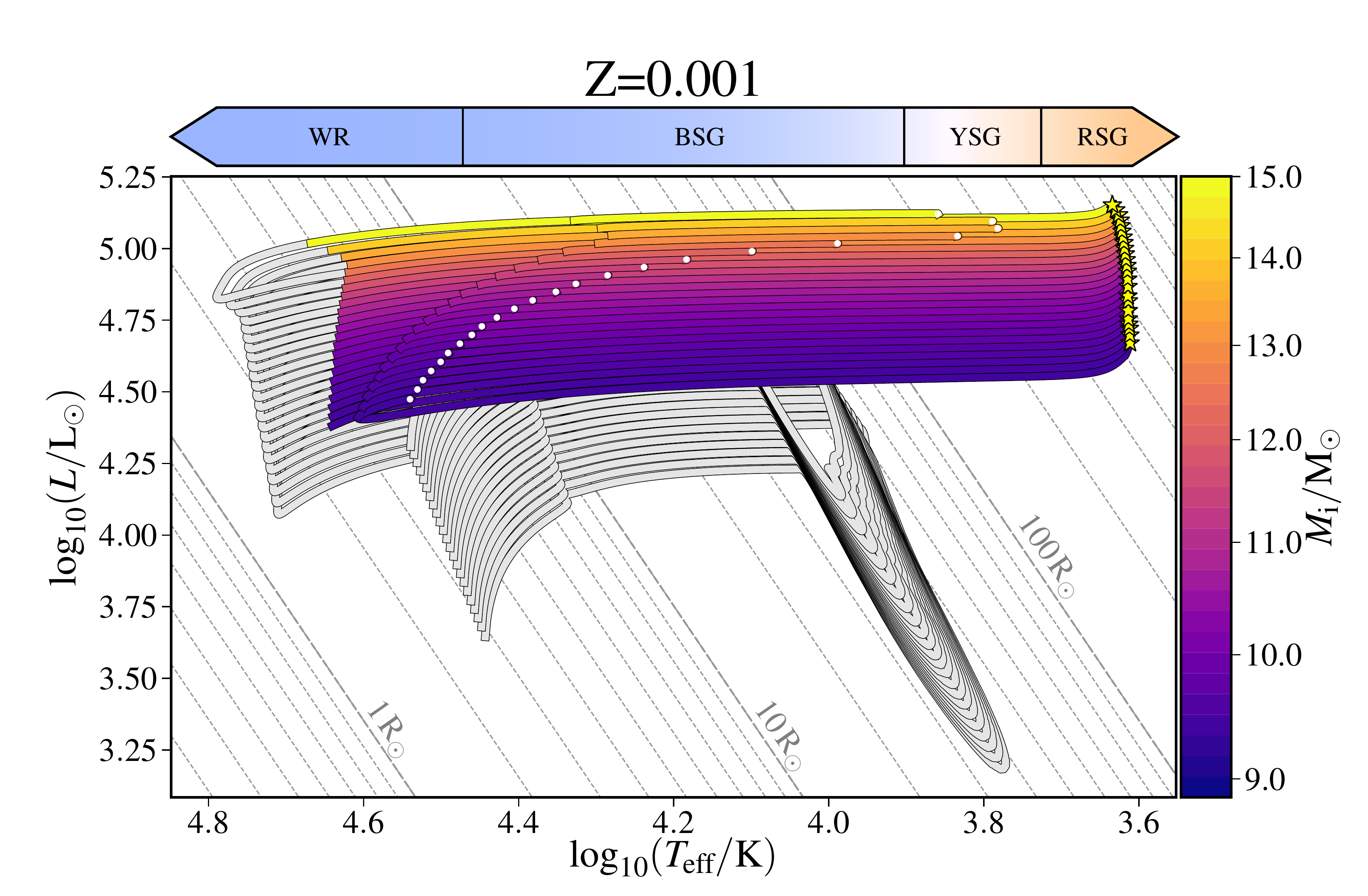}
	\caption{Evolutionary tracks of stars at solar (top) and sub-solar metallicity (bottom) showing their luminosity, $L$ as a function of effective temperature, $T_{\rm eff}$. Lines of constant radius are provided for reference. We highlight the evolution after core helium depletion in color, where the color refers to the initial mass. 
	Yellow star symbols indicate core carbon exhaustion, and white star symbols denote the final time step calculated for models that did not reach core carbon exhaustion.
	The white circles in the lower panel indicate when the radius reaches a first maximum after core helium depletion.
	Above each panel we indicate approximate ranges of effective temperatures typical for Wolf Rayet stars (WR), blue, yellow and red supergiants (BSG, YSG and RSG respectively). The horizontal color scale above each panel 
	indicates, for each effective temperature, the color as perceived by the human eye.
}
	\label{fig:HRD_grid}
\end{figure*}

Here we present results from two grids of evolutionary models of stars that are stripped through binary mass transfer. The grids are computed at two different metallicities (solar and $Z=0.001$) and each consists of twenty-three models with different initial masses for the primary star, ranging from $8.8$ to $15$\Msun. Tables~\ref{tab:zsun} and \ref{tab:zlow} provide an overview of the key parameters.

\subsection{Evolutionary tracks in the Hertzsprung-Russell diagram}
\label{sec:grid_HRD}
The evolutionary tracks are presented in Fig. \ref{fig:HRD_grid}, with the top panel showing the results for solar metallicity and the bottom panel for low metallicity. The tracks show the evolution from the onset of hydrogen burning up to the completion of helium burning in light grey, for a detailed description of which see Section~\ref{sec:ev_a_f} and \ref{sec:ev_f_g}. Here we highlight in color the last phases of the evolution, from the completion of central helium burning until central carbon depletion, where the color indicates the initial mass of the progenitor. 

The most striking feature in these plots is how far the evolutionary tracks extend to the right during the late stages of their evolution, i.e., how much their effective temperature decreases and their radius expands before carbon is depleted in the core (which is marked with a yellow star symbol). 
In our solar metallicity models, shown in the top panel, there is a wide range of final effective temperatures and thus final radii, in general agreement with earlier studies \citep[cf.][]{habets_evolution_1986-1,dewi_late_2003, yoon_type_2010,yoon_type_2017}. The lowest mass models in our grid ($M_{\rm{ZAMS}} < 9.5$\Msun) reach the lowest final effective temperatures and largest final radii of $\log_{10} T_{\rm{eff}}/\rm{K} \approx 3.8$ and $R\approx150\Rsun$, respectively. These temperatures are typical for yellow supergiants \citep[YSG, e.g., ][]{drout_yellow_2009}.
For the intermediate mass models in our grid we find final temperatures and radii that are more typical for blue supergiants \citep[BSG, e.g., ][]{fitzpatrick_properties_1988}. 
The highest mass models in the grid reach final effective temperatures and radii of up to $\log_{10} T_{\rm{eff}}/\rm{K} \approx 4.7$ and as low as $R\approx3.5\Rsun$, respectively. Their wind mass-loss rate is of the order of $10^{-6} \Msun / \rm{yr}$ or more. These properties are characteristic for classic Wolf-Rayet stars \cite[WR, e.g., ][]{crowther_physical_2007}. 
All models in our low-metallicity grid end their lives as cool ($\log_{10} T_{\rm{eff}}/\rm{K} \approxeq 3.55 $) and large ($R > 400-700 \,R_{\odot}$) stars typical for red giants or red supergiants \citep[RSG, e.g.,][]{groh_fundamental_2013}. 

\subsection{Evolution of the radius and its connection to the nuclear burning shells} \label{sec:grid_radius}

\begin{figure*}[htp!]
\centering
		\includegraphics[width=0.45\textwidth]{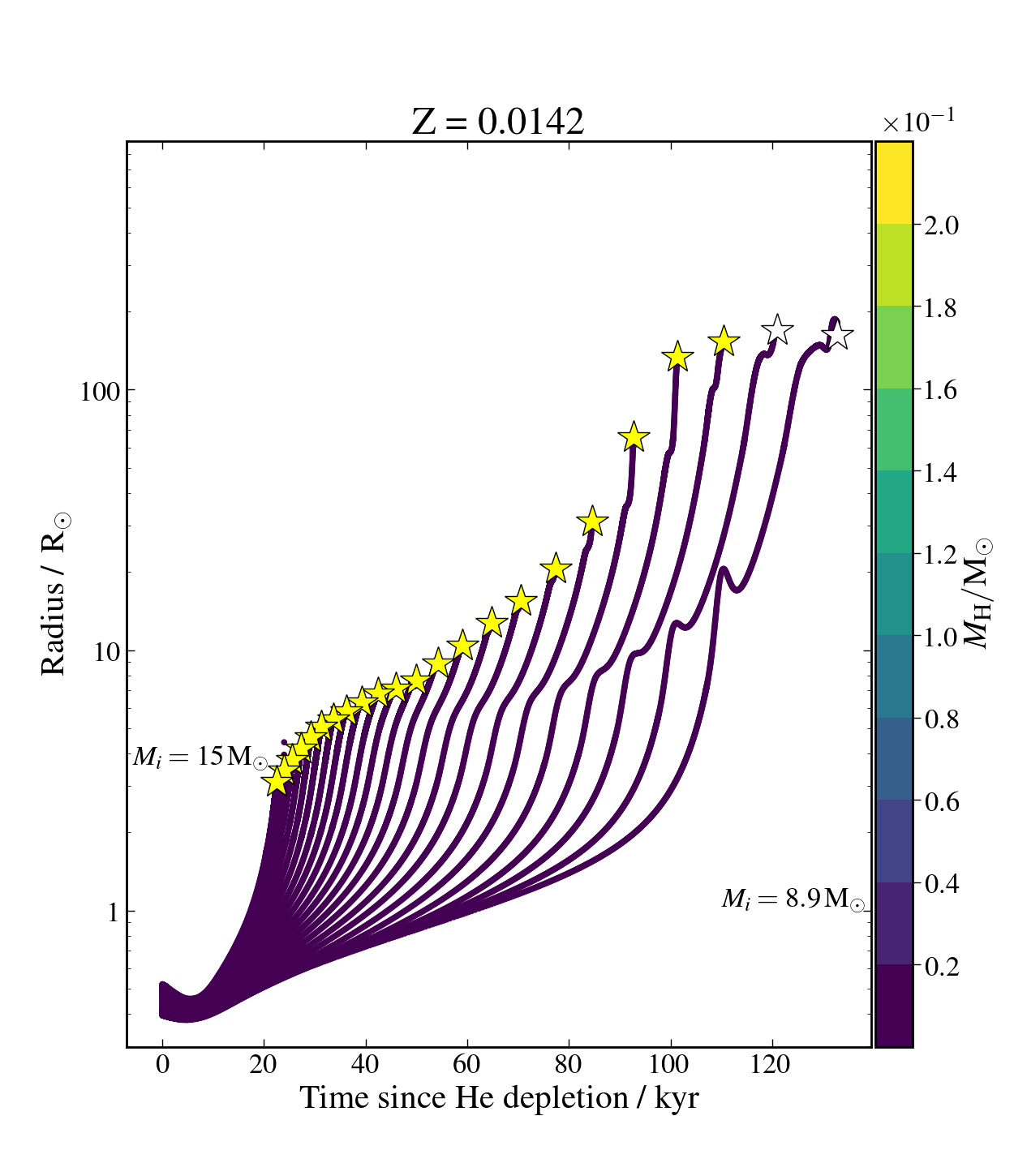}%
		\includegraphics[width=0.45\textwidth]{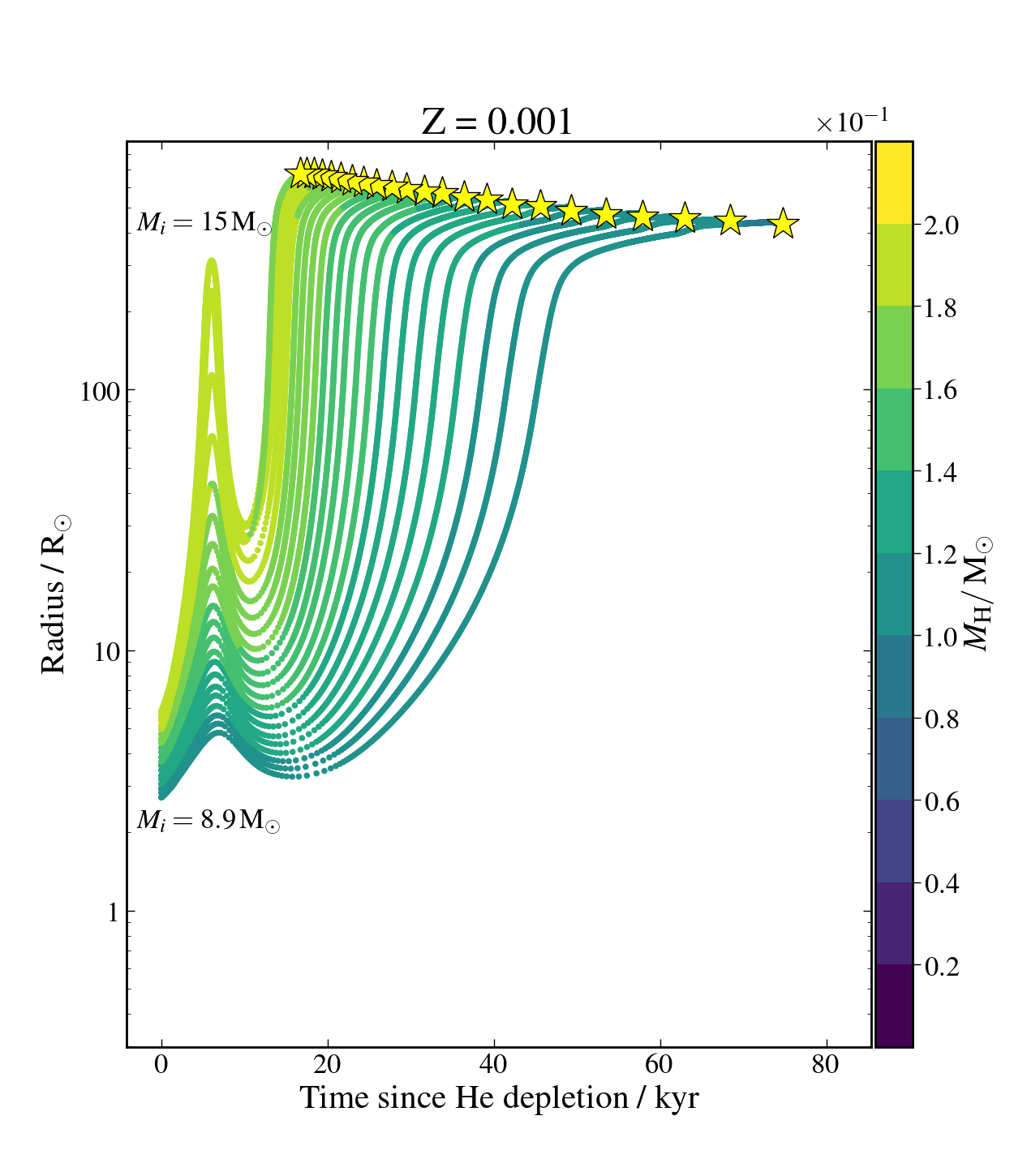}
		\includegraphics[width=0.45\textwidth]{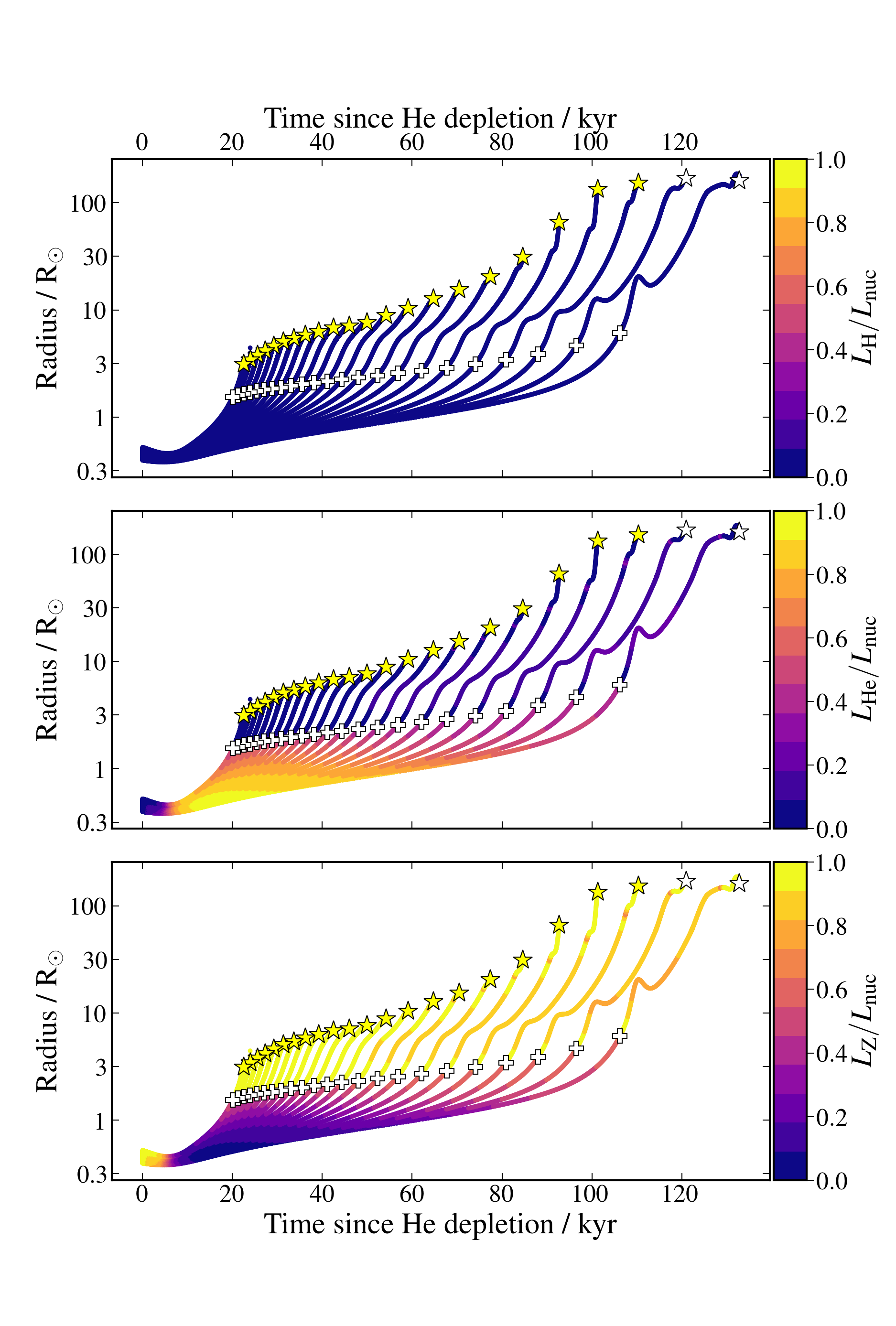}%
		\includegraphics[width=0.45\textwidth]{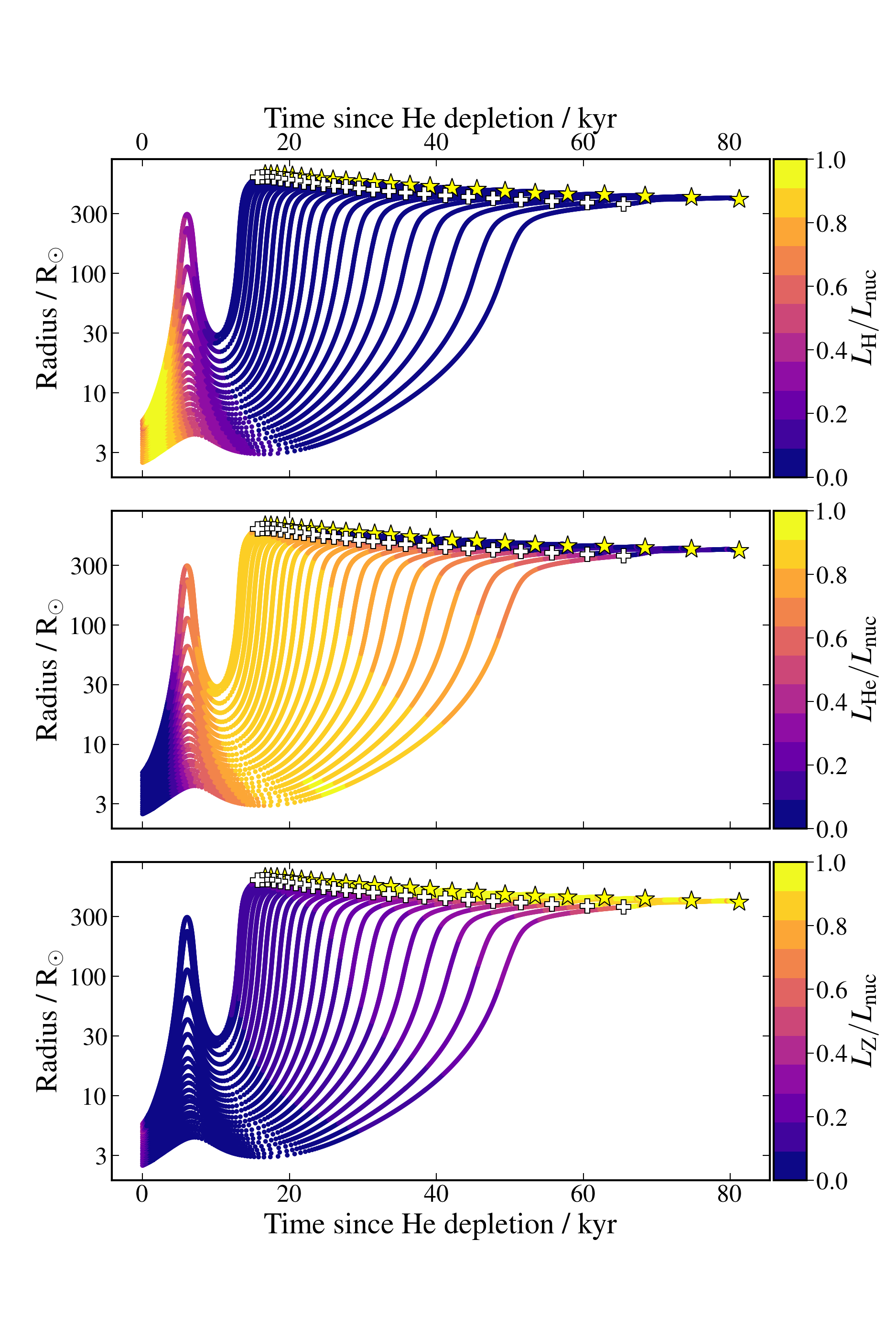}
	\caption{Evolution of the radius as a function of time after core helium depletion for the stars in our model grids at solar (left) and lower (right) metallicity. From top to bottom, colors indicate the remaining mass of hydrogen and the normalized luminosity from nuclear burning of hydrogen, helium, and heavier elements, see text for details. More massive stars evolve faster and are thus located towards the left of the figures. White crosses indicate the moment of core carbon ignition. Yellow star symbols indicate core carbon exhaustion, and white star symbols denote the final time step calculated for models that did not reach core carbon exhaustion.}
	\label{fig:Rad_ev}
\end{figure*}

The evolution of the radii is shown in the panels in Fig.~\ref{fig:Rad_ev} as a function of time since helium depletion. The solar metallicity models are shown on the left and low metallicity models on the right. The color bars indicate relevant physical quantities, different for each row, which we discuss further below. Each panel show a sequence of twenty-three tracks which correspond to our models for different initial masses. The tracks for the massive models can be readily identified as they evolve faster and complete their final phases of evolution in about 20\,kyr. The lower-mass models in our grid take 125\,kyr at solar metallicity and about 80\,kyr at low metallicity. 

We see that the solar-metallicity models expand from about 0.5\Rsun up to final radii of 3\Rsun for the more massive models, while the lower mass models expand to radii of 180\Rsun. In contrast, the low-metallicity models start at radii of 3 to 5\Rsun and reach final radii of 400-700\Rsun. For an overview see also Tables~\ref{tab:zsun} and \ref{tab:zlow}.  

For the low-metallicity models, the post-core-helium-burning radius expansion is clearly not monotonic with time, as already discussed in Section \ref{sec:ev}. The stars expand to a first maximum around 8\,kyr after helium depletion for all models, followed by a contraction and re-expansion phase. The first expansion phase is most significant for the most massive models, which reach radii of up to 300\Rsun. The total radius expansion for the low-metallicity models is very significant (two orders of magnitude). The implication of this is that these stars are expected to exceed the size of their Roche lobe and, at low metallicity, they may already do so during the first expansion phase. This would trigger at least one additional phase of mass transfer shortly after core helium depletion; those are not modeled here. Moreover, they may well fill their Roche lobe shortly before or at the moment of explosion.

To understand the origin of the difference in behavior between solar and low metallicity models it is helpful to inspect further physical parameters. The main difference between the solar and low metallicity models is the remaining hydrogen mass, which is indicated in color in the panels in the top row of Fig.~\ref{fig:Rad_ev}. The solar metallicity models contain almost no hydrogen, i.e. less than about 0.05\Msun after the mass transfer ceases and even less at the end their evolution (see $M_{\rm H}^{\rm postMT}$ and $M_{\rm H}^{\rm f}$ in Table~\ref{tab:zsun}). In contrast, the low-metallicity models contain about 0.2\Msun of hydrogen after the mass transfer ceases and the stellar winds are too weak to substantially reduce this afterwards. 

We can gain further insight when considering the nuclear burning processes that contribute to the total luminosity by nuclear burning ($L_{\rm nuc}$). Colors in the lower panels on row 2, 3 and 4 of Fig. \ref{fig:Rad_ev} show the relative contribution to the total nuclear luminosity resulting from burning of hydrogen ($L_{\rm H}$), burning of helium through the triple-alpha reaction ($L_{\rm He}$), and the collective burning of heavier elements ($L_{\rm Z}$), including helium burning by alpha captures onto carbon. 

During the first 10,000\,years after helium depletion the radii of the solar-metallicity models change only slowly, while the low-metallicity models expand rapidly. At solar metallicity, the thin hydrogen-rich layer that is left after Roche-lobe stripping, if any, is not sufficient to support hydrogen shell burning. By contrast, in the low metallicity models enough hydrogen is retained to sustain hydrogen shell burning. Hydrogen burning dominates the nuclear luminosity during the first expansion phase, as can be seen on the second row of Fig.~\ref{fig:Rad_ev}. At the peak of the first stellar expansion hydrogen burning contributes about half of the total nuclear luminosity. At that time the contribution from the helium-burning shell is increasing as can be seen in the panel on the third row of Fig.~\ref{fig:Rad_ev}. The turning point in radial expansion occurs when the stars have roughly equivalent luminosity contributions from two shell sources. The first expansion in the low-metallicity models is thus associated with hydrogen shell-burning, and the subsequent contraction is consistent with the "double mirror effect" (see also Sect.~\ref{sec:ev_g_j}).

Thereafter, both high- and low-metallicity models show radius expansion when helium-shell burning dominates the nuclear luminosity. The low-metallicity models reach much larger radii, and most of their expansion is during this burning phase. The plus symbols in Fig. \ref{fig:Rad_ev} indicate the start of core carbon burning, which we define as the moment when the core carbon abundance drops by $2\%$ below its post-core-helium-burning value. The solar-metallicity models show significant further expansion after this time, for the lowest-mass models by more than an order of magnitude in radius, as a result of to the left-over hydrogen layer. The low-metallicity models have completed most of their expansion by the onset of core carbon burning.

\subsection{Binding energy 
\label{sec:grid_E_B}}

\begin{figure}[tbh]
\centering
 \includegraphics[width=0.5\textwidth]{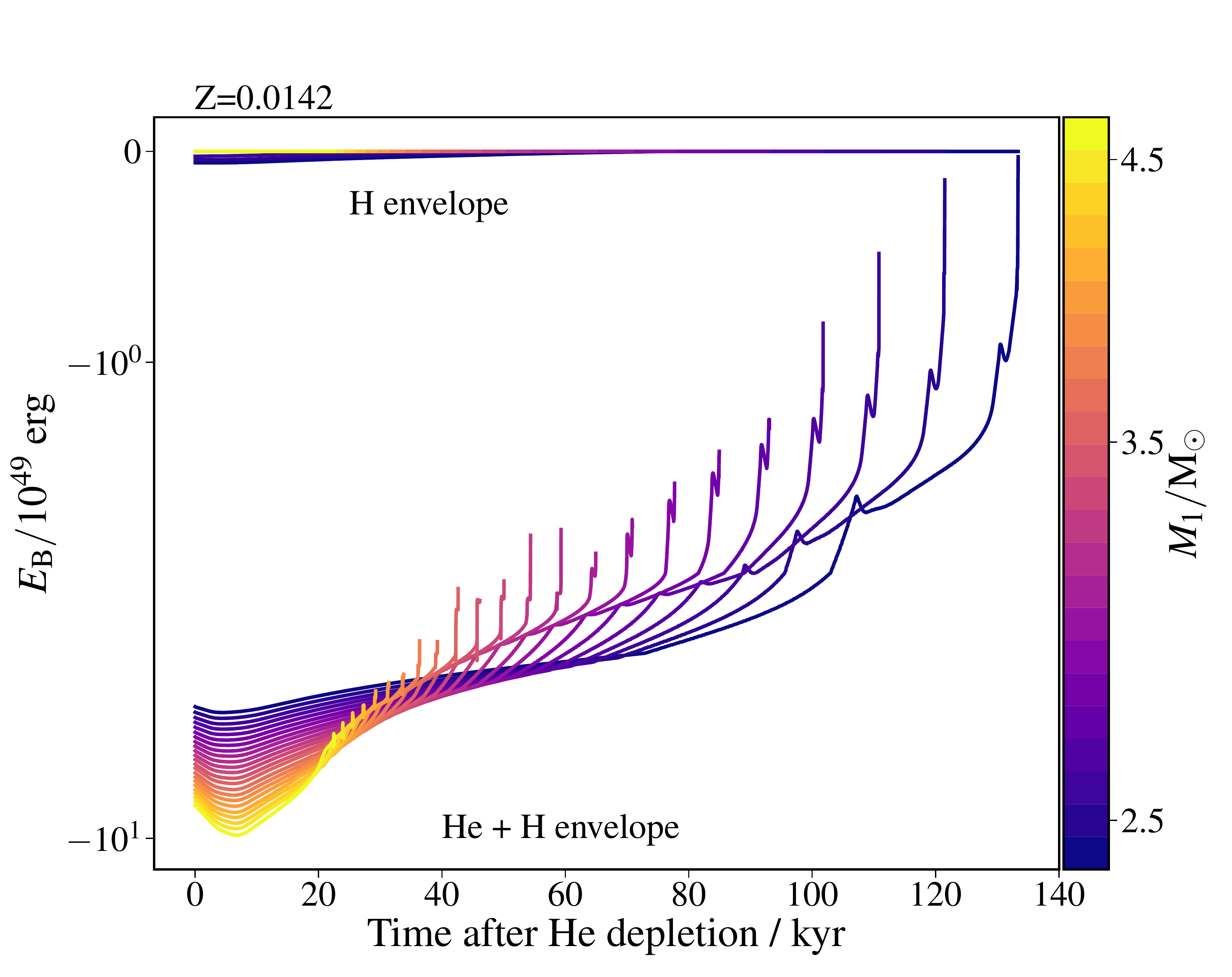}
\includegraphics[width=0.5\textwidth]{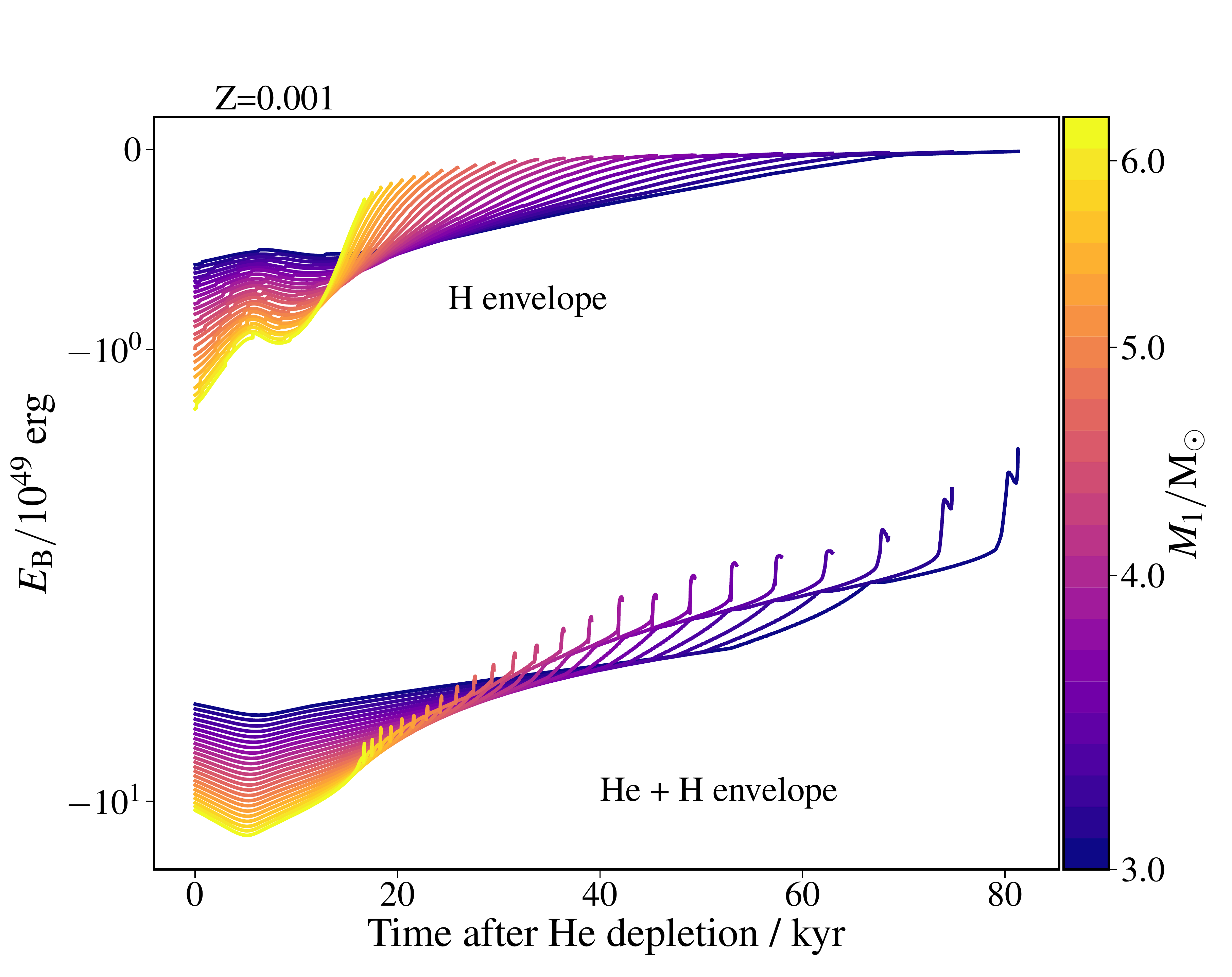}
\caption{ Binding energy of the envelope as a function of radius after core helium depletion at solar (top panel) and low (bottom panel) metallicity. We show values for the hydrogen-rich envelope alone, as well as the hydrogen and helium envelope combined, as labelled. Colors indicate the mass of the stripped star. \label{fig:Eb_time}}
\end{figure}

In the previous sections we discussed the expansion of stripped stars in their final evolutionary phases. As a result of this expansion stripped stars can fill their Roche lobe again, initiating a subsequent phase of mass transfer. In the case of unstable mass transfer, the system is expected to enter a common-envelope phase. This would shrink the orbit and thus have important potential consequences for the final fate of the binary, possibly as source of gravitational waves. We briefly discuss the most important stellar property affecting the outcome of a common envelope phase that we can compute, namely, the binding energy of the envelope \citep[e.g.,][]{Webbink1984}.

We define the binding energy $E_{\rm{B}}$ of the (hydrogen or helium) envelope as:
\begin{equation}
\label{eq:E_b}
\centering E_{\rm{B}} = -\int_{\rm{core}}^{\rm{surface}}\left(-\frac{G m}{r} + \epsilon(m) \right) dm \textrm{,}
\end{equation}
where G is the gravitational constant, $m$ the mass coordinate, and $r$ the radius coordinate of the star, and $\epsilon$ the specific internal energy. This internal energy includes not only the thermal energy terms, but also the potential energy stored in ionised species and dissociated molecules (see, e.g., \citealt{han_possible_1994,ivanova_common_2013}). We define the relevant core-envelope boundary as the point at which the abundance of the dominant element in the envelope drops below 10\%. The chosen definition is somewhat arbitrary, but we consider it reasonable for our current purposes (see also Appendix \ref{sec:appendix_eb}).

The results for our solar metallicity models are shown in the top panel of Fig.~\ref{fig:Eb_time}. The binding energy of the H envelope is negligible, as expected, since most hydrogen has been removed efficiently during Roche-lobe overflow and subsequently by the stellar wind. Shortly after helium depletion, the He+H envelope has a binding energy of the order of about $-5$ to $-10 \times 10^{49}$\,erg, where the more massive stars in our grid are more tightly bound (i.e., their binding energies are more negative). As time proceeds, the stars evolve and expand. Their envelopes loosen to reach values of the binding energy of about $-2$ to $-6 \times 10^{49}$ erg at the end of their evolution, when the envelope starts to become convective.

For our low metallicity models, shown in the bottom panel, we find the same qualitative trend with mass and time. However, the binding energies are about $-1 \times 10^{49}$\,erg for the H envelope and about ten times more than that for the H+He envelope. All these models develop a convective envelope when their radius reaches 200--300\Rsun. We discuss the consequences for the stability of mass-transfer and gravitational-wave progenitors in Section \ref{sec:dis:unstableNS}.

\section{Comparison with the radii for helium stars
adopted in population synthesis simulations}

\label{sec:comparison}

\begin{figure*}[ht]
\centering
\includegraphics[width=0.65\textwidth]{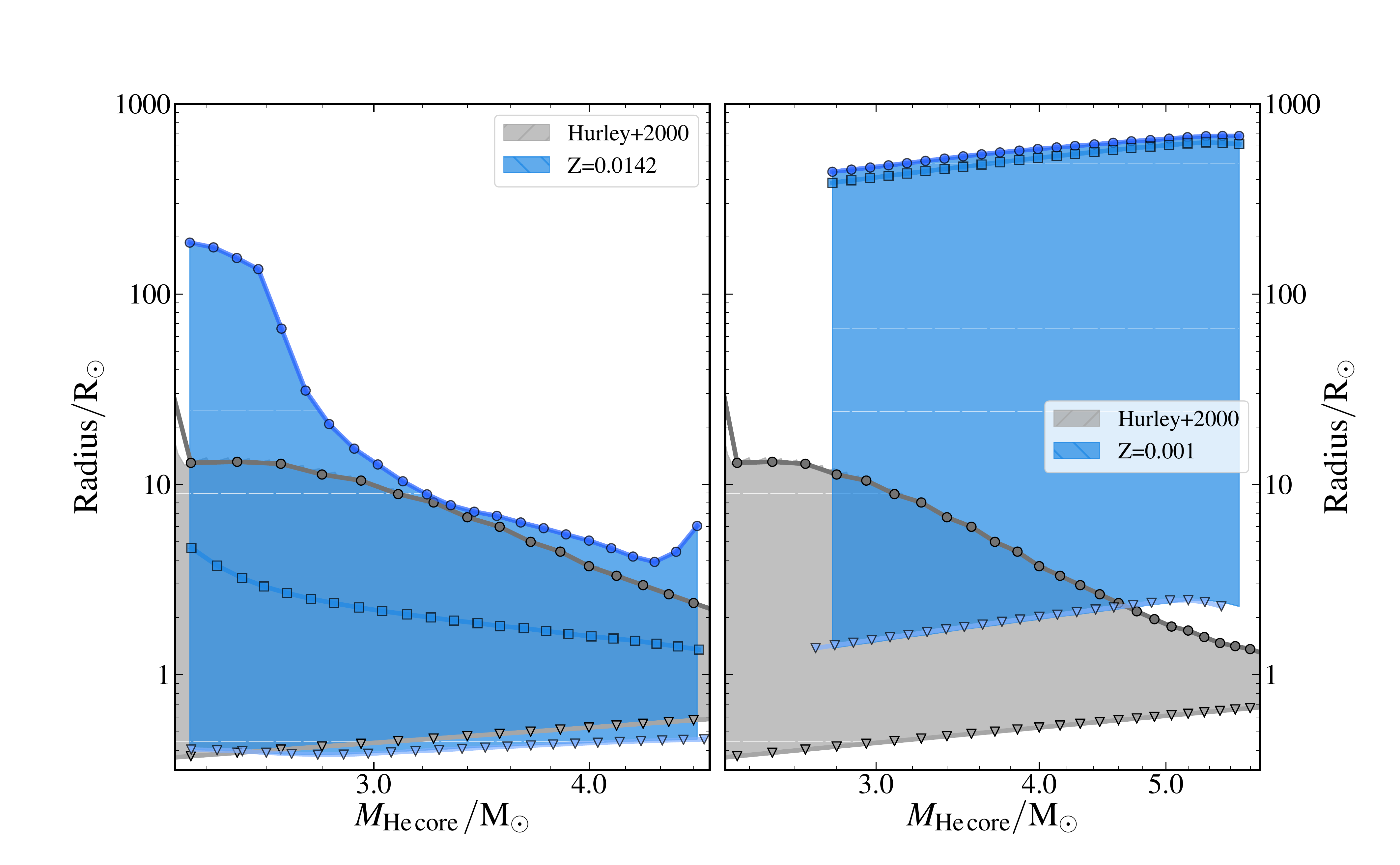}
\includegraphics[width=0.65\textwidth]{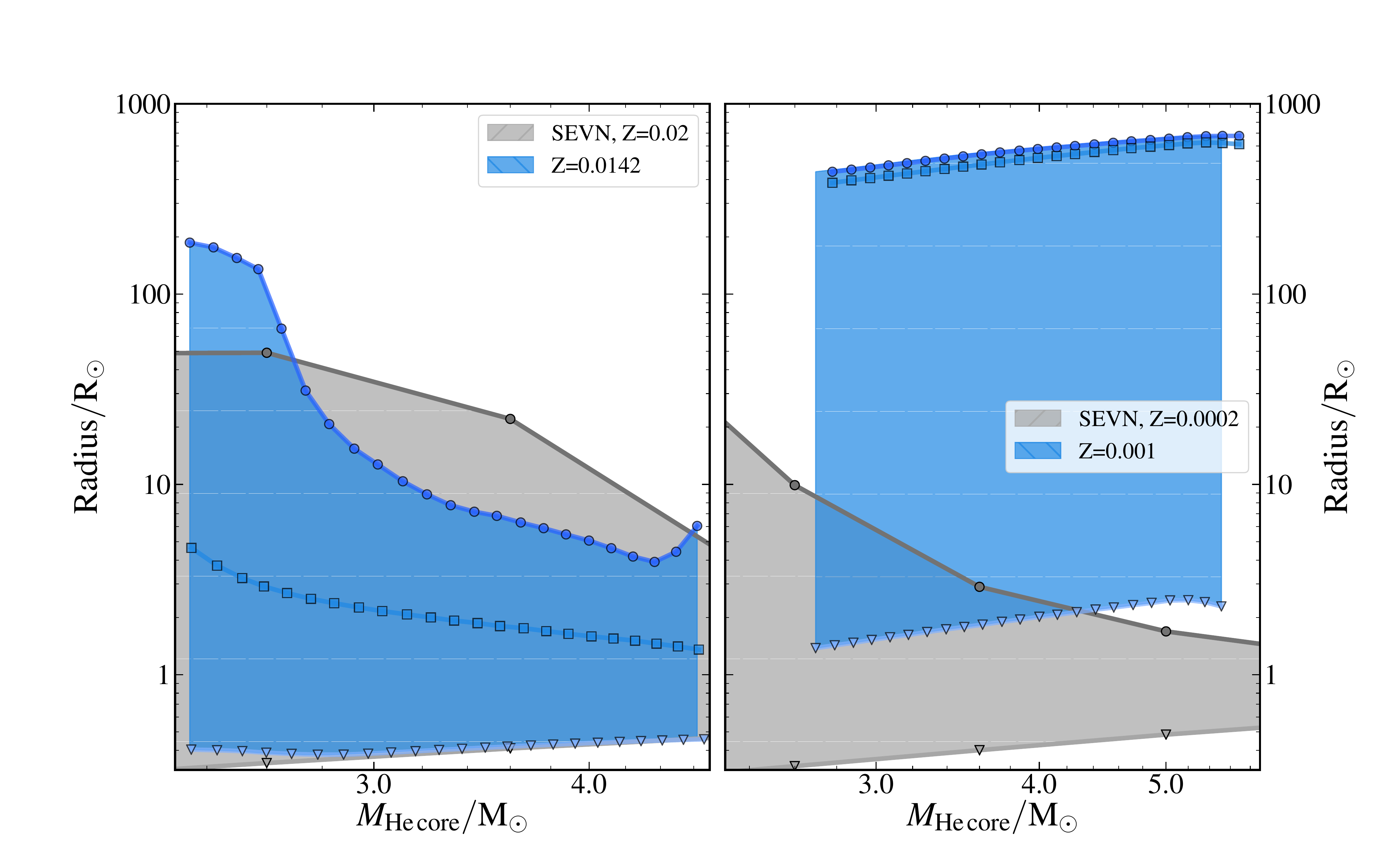}
	\caption{ Comparison of our results (shown in blue in each panel) with (1) the widely-used prescriptions by \citet{hurley_comprehensive_2000} (shown in grey in the panels on the top rown) and (2) the PARSEC models \citep{bressan_parsec:_2012} used in the SEVN code \citep{spera_merging_2019} (shown in grey in the panels on the bottom row). Panels on the left show a comparison at solar metallicity (or its closest available equivalent, i.e. Z= 0.02 for SEVN) and panels on the right show a comparison for low metallicity (Z= 0.0002 for SEVN). We show the maximum radius (circles) and the minimum radius (triangles). For our models we also mark the radius at core carbon ignition (squares).}
	\label{fig:MaxRad}
\end{figure*}

Having accurate estimates for the minimum and maximum radii
of stars is crucial in the context of binary systems. For an individual
binary system, the radial expansion determines whether
a star can swell to fill its Roche lobe and start a (new) phase of
mass transfer. For a full population of binary systems, the radial
expansion is an essential factor in determining what fraction of this
population interacts.

To simulate large populations of binary systems and their exotic
end products as X-ray binaries and gravitational wave progenitors,
prescriptions are needed that can be evaluated quickly.
Generally, two approaches are commonly adopted. Many population
synthesis codes make use of the analytic formulae
by \citet{hurley_comprehensive_2000}. We compare our results with these prescriptions in Section
\ref{sec:comparison_hurley}. Alternatively, one can make use of interpolation in grids
of pre-computed stellar evolution models. One example of this is the SEVN code,
with which we compare in Section \ref{sec:comparison_SEVN}.

\subsection{Analytic prescriptions by \citet{hurley_comprehensive_2000} \label{sec:comparison_hurley}}

The Hurley fitting formula are adopted in many population synthesis codes that are in active use. These include, but are not limited to, \texttt{BSE} \citep{hurley_evolution_2002}, \texttt{StarTrack} \citep{Belczynski+2008}, \texttt{Binary\_c} \citep{Izzard+2004, Izzard+2006}, \texttt{COMPAS} \citep{Stevenson+2017, vigna-gomez_formation_2018}, and the latest arrival \texttt{COSMIC} \citep{Breivik+2019}. The discussion below thus applies to these studies and others that make use of these prescriptions. 


The Hurley fitting formulae are based on evolutionary models by \citet{pols_stellar_1998} which solve for the full set of stellar structure equations using the Eggleton code \citep{Eggleton1971,Eggleton1972, Eggleton+1973} with updates to the equation of state by \citet{Pols+1995}. Specifically, the fitting formulae for helium stars \citep[provided in Section 6.1 of][]{hurley_evolution_2002} are based on detailed models for single stars computed with the same input
physics as the grid presented in \citet{pols_stellar_1998}. These models
assume a homogeneous initial composition with a helium
mass fraction of $Y = 0.98$, a mass fraction of heavier elements
of $Z = 0.02$ and no hydrogen, $X = 0$, \citep[see][for a discussion]{dewi_evolution_2002,dewi_late_2003}.

The main difference from our models is that we self-consistently account for the stripping process due to a binary companion, assuming a representative initial orbital period. In our models, the resulting stripped star can still contain a remaining layer of hydrogen at the surface. The effect of such a remaining layer of hydrogen is thus not accounted for in the Hurley prescriptions. A further difference is that the original stellar models behind the Hurley prescriptions were provided for only one fixed value of the metallicity. In our calculations we find large differences between stripped stars at solar and low metallicity, which are not accounted for in the Hurley prescriptions.

The top row of Fig.~\ref{fig:MaxRad} compares our results (in blue) with the Hurley prescriptions (in grey). For solar metallicity, shown in the top left panel, we find that minimum radii given by the Hurley prescription are comparable to the minimum radii we find in our detailed models. 
The maximum radii from \citet{hurley_comprehensive_2000} show a similar trend with mass as we find in our models, but we also observe important and significant differences, in particular at the low-mass end (helium core masses of 2--2.5\Msun). Our models reach maximum radii of about 200\Rsun, which is an order of magnitude larger than the maximum radii for the Hurley prescriptions, which reach only about 20\Rsun. 

At low metallicity, shown in the top right panel of Fig.~\ref{fig:MaxRad}, we find that the Hurley prescription under-estimates the minimum radii by about a factor of three. The differences are much larger for the maximum radii. The maximum radii in our models increase with increasing mass and reach radii of 400--600\Rsun, which is 2--3 orders of magnitude larger than the maximum radii from the Hurley prescriptions. We find the largest differences in radii at the high mass end, where our models have radii that are 500 times larger than the Hurley maximum radii. These differences are due to the presence of the remaining hydrogen layer in our stripped models, which is not accounted for in the Hurley prescriptions.

\subsection{Example of grid based interpolations: SEVN code}
\label{sec:comparison_SEVN}
An example of a code that uses interpolation within a pre-computed stellar model grid is SEVN, a grid-based population synthesis code \citep{spera_merging_2019} which has also been used to make predictions for gravitational wave progenitors. The stellar models behind this code are single stellar models computed with the PARSEC stellar evolution code \citep{bressan_parsec:_2012}.

The bottom panels of Fig.~\ref{fig:MaxRad} compare our results to those of SEVN that are closest in metallicity. In the lower left panel we compare our solar metallicity models (Z=0.0142) to their models which assume a slightly higher value for the solar metallicity, $Z=0.02$, which is the old canonical value for solar metallicity. In the lower right panel we show the PARSEC models for $Z=0.0002$ which is closest to the metallicity of Z=0.001 that we adopted for our low metallicity grid. 

We find similar general differences as described above when comparing to the Hurley prescriptions. For solar metallicity we find again that the minimum radii are in fairly good agreement. For the maximum radii we again note significant differences depending on the mass. We find significantly larger maximum radii of about 200\Rsun, compared to about 60\Rsun for the PARSEC models at the lower mass end (for masses below about 2.7\Msun). For larger masses, the models used in SEVN reach maximum radii of 10 to 50~\Rsun, which is substantially larger than the maximum radii we find of 5 to 30\Rsun in our models. Understanding the origin of these differences would require further investigation. It may be due, in part, to the difference in metallicity, but differences in the micro-physics or treatment of convection may also play a role. At low metallicity we again find very large differences, similar to but even more pronounced than the differences we find with the Hurley prescription, as shown in the lower right panel of Fig.~\ref{fig:MaxRad}.

We provide new analytic fits to our models for use in population synthesis calculations in Appendix \ref{sec:appendix_fits}.

\section{Discussion \label{sec:discussion}}

As shown in Section~\ref{sec:comparison}, we find systematically larger radii for stripped stars than those commonly used in population synthesis calculations. The large radii can trigger additional phases of mass and angular momentum transfer \citep[traditionally referred to as Case BB or Case BC mass transfer,][and references therein]{dewi_evolution_2002}.

Such additional interaction can impact the final masses and orbital separation and are thus important for modelling the populations of binaries. Specifically, these later phases of interaction are thought to be very important in the formation of peculiar supernovae and gravitational wave progenitors \citep{Ivanova_The_Role_2003, dewi_late_2003, Tauris2013}. Moreover, \citet{zevin_can_2019} argue that these additional mass-transfer phases are necessary to explain enrichment in globular clusters, assuming that r-process enrichment primarily originates from double neutron star systems.


A full assessment of the implications would require extended grids of models that follow these additional phases of mass transfer self-consistently. We provide a first estimate of the additional number of systems affected compared to the widely used prescriptions in Section~\ref{sec:impactGW}. In Section~\ref{sec:dis:unstableNS} we discuss the question whether the late phase of mass transfer would be stable or lead to a common envelope phase involving a neutron star. We examine the implications for the observability of these stars in Section~\ref{sec:dis:observability} and for supernova progenitors in Section~\ref{sec:dis:supernovae}. In Section~\ref{sec:dis:uncertainties} we discuss the main uncertainties that affect the results presented in
this work.

\subsection{Expected increase in the number of binary systems that interact with a
helium donor \label{sec:impactGW}}

Whether or not the stripped star will fill its Roche lobe anew depends on the size of the Roche lobe, which scales linearly with the separation $a$ for a given mass ratio \citep[e.g.][]{eggleton_aproximations_1983}. The separation of a particular binary system depends on its initial separation, the amount of mass that is transferred, and the amount of angular momentum that is lost from the system during the first mass transfer. Given the large uncertainties in the mass-transfer process, the distribution of separations is not well known. 

In order to make a simple estimate we make the agnostic assumption that the separations (or more precisely the Roche-lobe radii) are distributed according to the standard \citet{opik_notitle_1924} law, i.e., a distribution that is flat in the logarithm. We further assume that the separations span the full ``range of interest'', i.e. are such that the Roche radii span from $R_{\min}$ to at least $R_{\max}$, where 
\[
 R_{\min} \equiv \min(R_{\min}^{\rm this\,study}, R_{\min}^{\rm Hurley}) \text{ and }
 R_{\max} \equiv \max(R_{\max}^{\rm this\,study}, R_{\max}^{\rm Hurley}). 
\]
where we use the superscripts to indicate the origin of the minimum and maximum radii. 

Assuming \"Opik's law leads to the following expression for the relative number of systems that interact with our new estimates for the radial expansion, compared to what would have been obtained with the Hurley prescriptions: 
\begin{equation}
f \approx \frac{\log R_{\max}^{\rm this\,study} - \log R_{\min}}{\log R_{\max}^{\rm Hurley} - \log R_{\min}}.
\end{equation}

With this, we find that a stripped star is about twice as likely to interact relative to \cite{hurley_comprehensive_2000} if we consider solar metallicity progenitors in the stripped-star mass range $2-2.5$\Msun. 
At low metallicity, this fraction rises even more due to the larger increases in radius. With this simple estimate, we find that stripped stars between 2 and 6\Msun are 2-30 times more likely to interact relative to the \citet{hurley_comprehensive_2000} prescriptions.

The numbers quoted here should be taken with a grain of salt. We have little reason to expect a logarithmically flat separation distribution to be realistic for systems that have already gone through a phase of interaction. Moreover, this simple estimate does not take into account the dependence of the low metallicity models on the initial orbital parameters. For short enough orbits, stars at lower metallicity would lose their hydrogen-rich envelopes after the first binary interaction, leading to a smaller increase in radius \citep[see, e.g., low metallicity models with short orbital periods of][]{yoon_type_2017}. 
However, it is noteworthy that at low metallicity, $100\%$ of all stripped stars computed with these initial orbital parameters will fill their Roche lobe anew (see Fig.~\ref{fig:HRD_grid}). This is in stark contrast with what follows from the Hurley prescription that predicts no stripped stars in this mass range to fill their Roche lobe again.

\subsection{Unstable mass transfer in systems with neutron star companions \label{sec:dis:unstableNS}}

The large radii of stripped stars may allow them to fill their Roche lobe anew and start to transfer mass to their companion. The case where the companion is already a neutron star is of particular interest, since such a system is a possible immediate progenitor of a double neutron star system \citep[e.g., ][]{fragos_the_complete_2019}.

When the stripped star fills its Roche lobe, its mass is still expected to exceed that of a typical neutron star of 1.4\Msun. If the mass transfer is stable, the orbit is expected to shrink because mass is transferred from a more massive star to a less massive companion \citep{paczynski_evolutionary_1971}. Secondly, mass lost from the system is likely to be emitted primarily from the vicinity of the less massive neutron star. The mass lost thus likely carries a specific angular momentum that is similar to or larger than the specific orbital angular momentum of the neutron star. This is larger than the average specific orbital angular momentum and we thus expect the orbital separation to shrink \citep{van-den-Heuvel+2017}.

A more dramatic shrinking of the orbit is expected when mass transfer is unstable. In this case the neutron star becomes engulfed in the envelope of the donor \citep[known as common envelope (CE) evolution, for a review see][]{ivanova_common_2013}, and the orbital separation can be shortened drastically, depending on the binding energy of the envelope and the efficiency with which it is ejected.

To know whether or not mass transfer is unstable would require further detailed calculations. For a first estimate, we assume that unstable mass transfer occurs in these systems only if the donor star has a convective envelope. The stripped stars we consider have masses in the range 2-6\Msun hence, if the companion is a 1.4\Msun neutron star, only the highest-mass of these stripped stars would canonically be expected to undergo unstable mass transfer when they have radiative envelopes.

\begin{figure*}[tbh]
\includegraphics[width=0.5\textwidth]{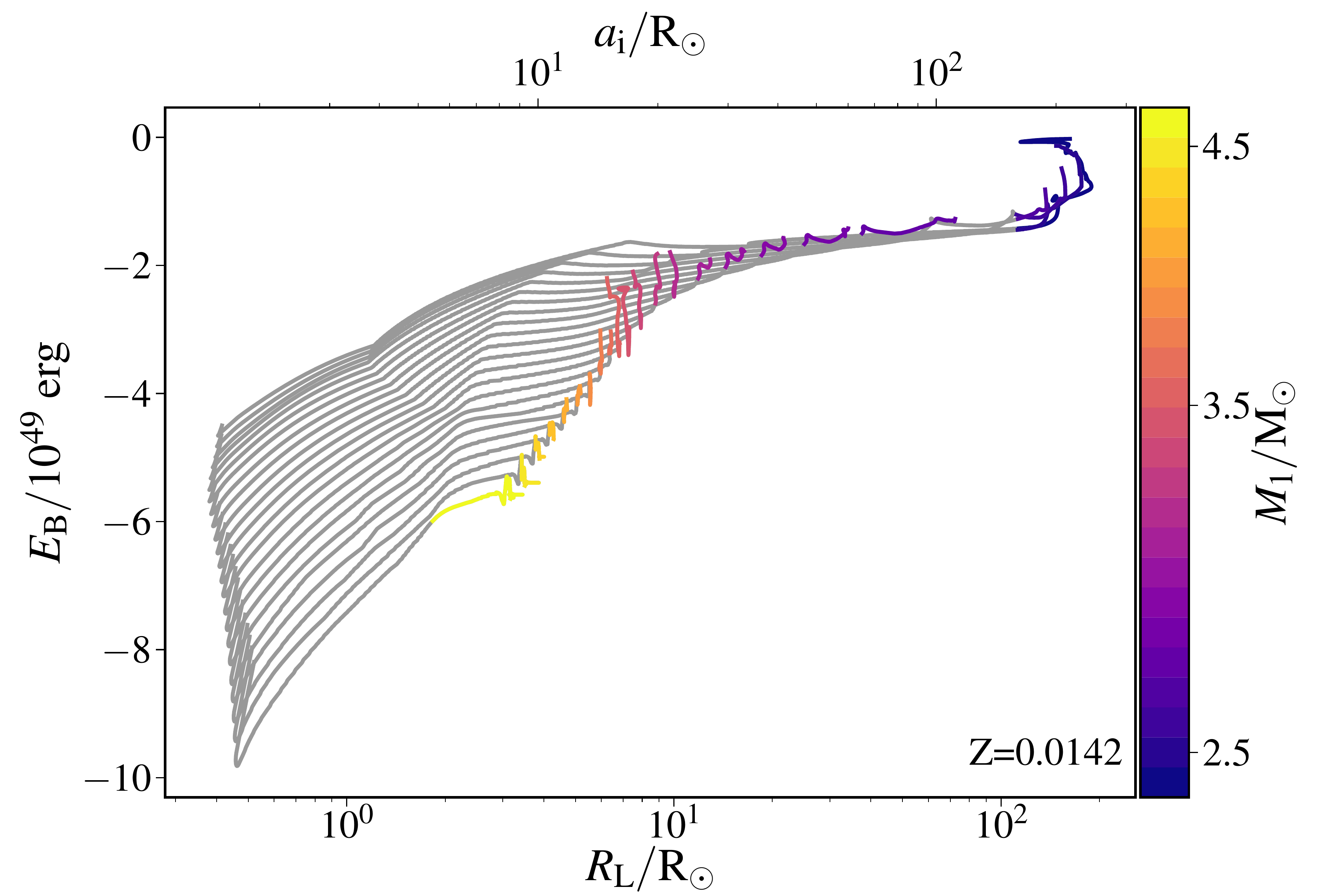}%
\includegraphics[width=0.5\textwidth]{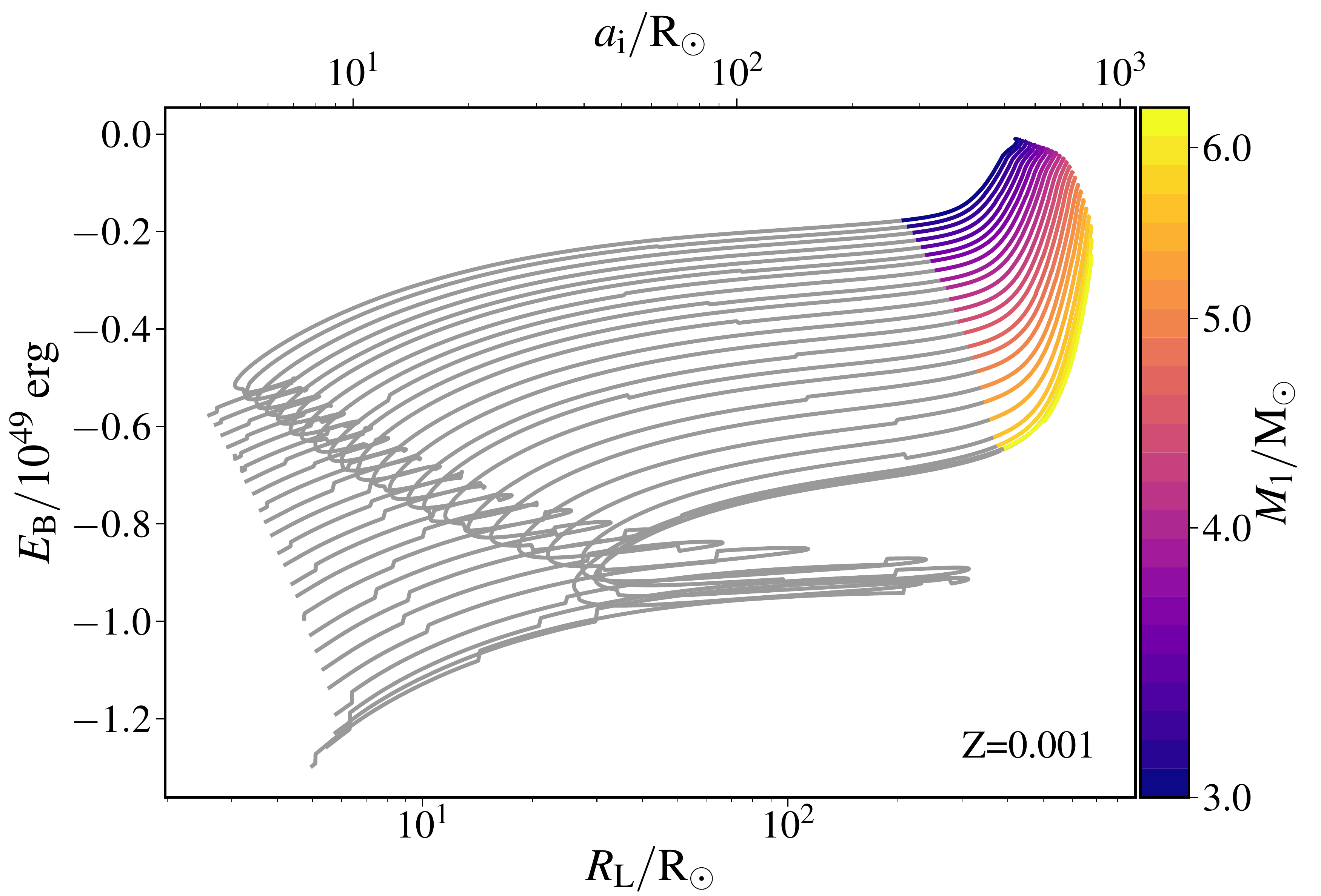}
\includegraphics[width=0.5\textwidth]{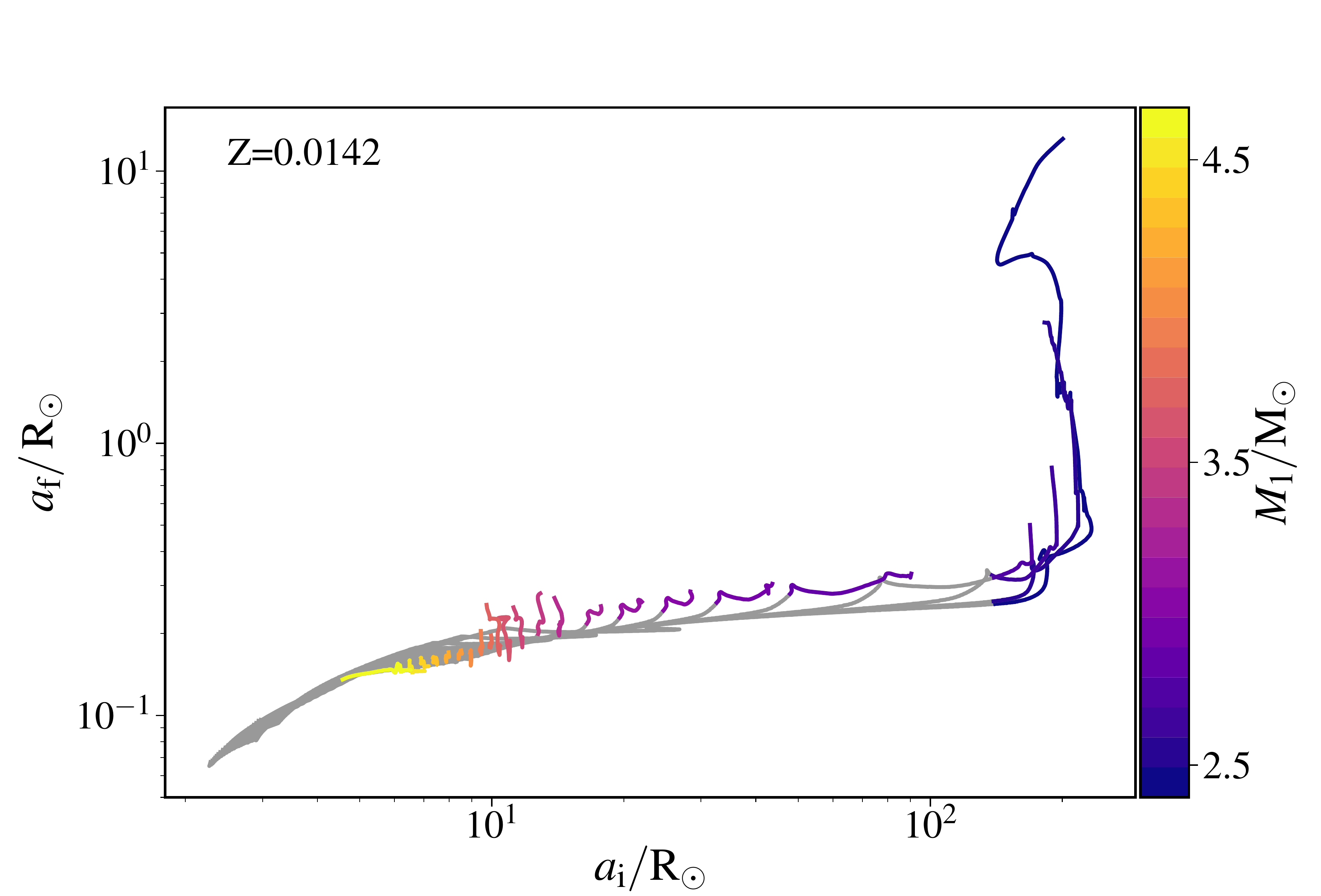}%
\includegraphics[width=0.5\textwidth]{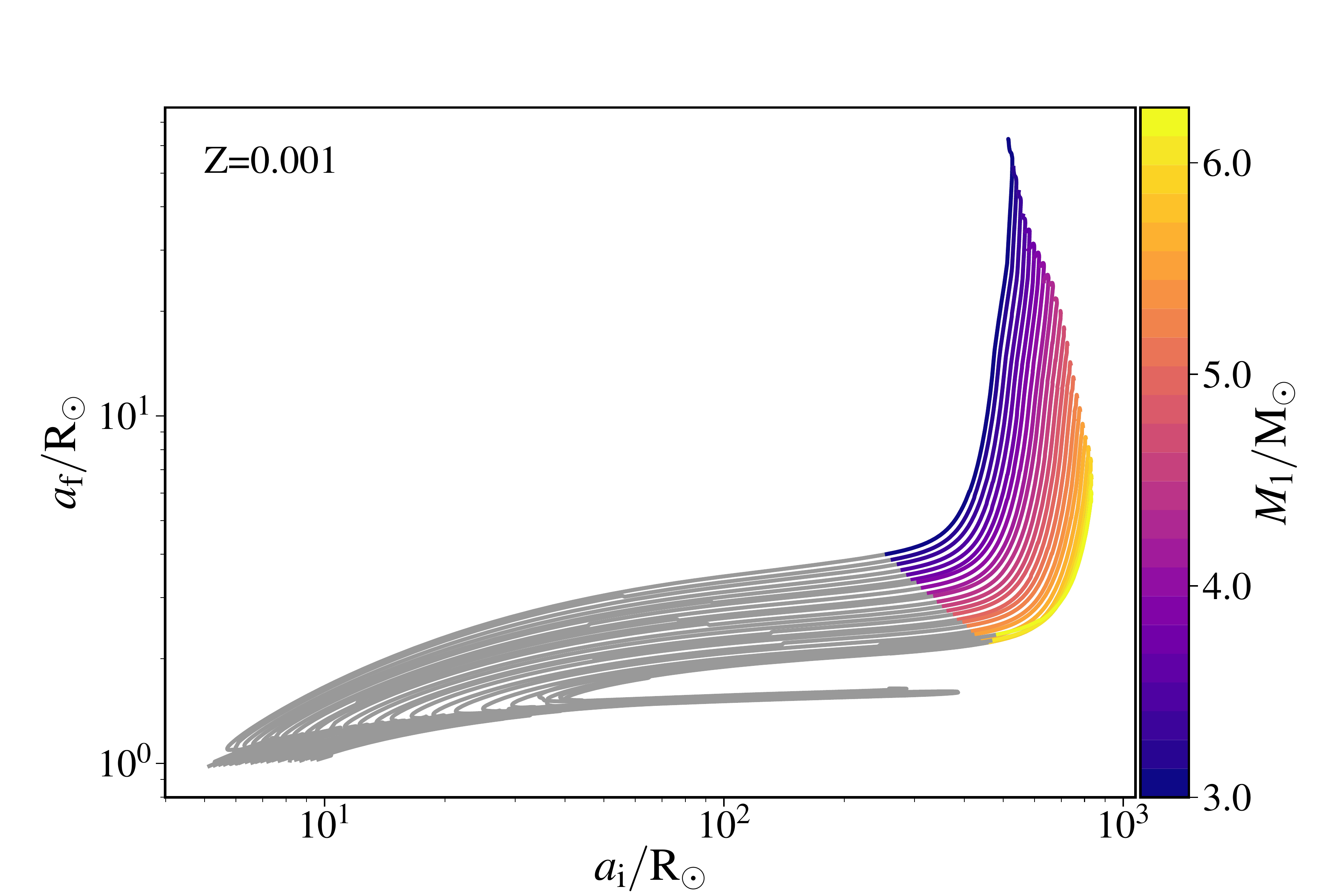}
	\caption{Top: Binding energy of the envelope plotted as a function of radius. On the top x-axis we have indicated the orbital separation assuming a neutron star companion with a canonical mass of 1.4\Msun, for reference. Our solar models are shown in the top left panel where we show the binding energy of the helium envelope. Our low metallicity models are show in the top right panel where we show the binding energy of the hydrogen envelope (cf. Section~\ref{sec:ev_g_j} and Fig.~\ref{fig:Eb_time}). Bottom: final separation after the common envelope evolution of stripped stars with neutron star companions as a function of the initial separation at solar (left) and low (right) metallicity. Stars with a pre-CE convective envelope are marked with colors, which indicate the mass of the stripped star, assuming a common envelope efficiency parameter $\alpha_{\rm{CE}} = 1$.}
	\label{fig:E_B_x}
\end{figure*}

In Fig.~\ref{fig:E_B_x} we show various tracks for our stripped stars where we highlight systems with convective envelopes at the onset of mass-transfer with colors. Stripped stars with radii larger than about 200\Rsun have convective envelopes. Assuming again that these systems are distributed flat in $\log a$ implies that about a fifth of the systems would begin mass transfer with a convective envelope. However, as a caveat we note that these stars do not develop very massive convective envelopes (see for example Fig.~\ref{fig:kipp_example}). We also note that if the remaining nuclear-burning lifetime is very short; \citet{tauris_ultra-stripped_2015} argue that there may not be sufficient time to complete the common-envelope inspiral before core-collapse.

The top panels of Fig. \ref{fig:E_B_x} show the binding energy as a function of radius and, for reference, the corresponding separation if the companion were a neutron star. In the bottom panels of Fig. \ref{fig:E_B_x}, we present the predicted post-CE separation as a function of the pre-CE separation, if mass transfer were unstable. For this calculation, we take the standard assumption that the orbital energy is completely converted into the binding energy of the envelope \citep[common envelope efficiency parameter $\alpha_{\rm{CE}} = 1$,][]{Webbink1984} and that the envelope is ejected with exactly the necessary escape velocity. We compute the envelope structure parameter $\lambda$ calculations in the appendix \ref{sec:appendix_eb}.

 At low metallicity, we find that the final separation is between 2 and 3\Rsun for initial separations smaller than 300\Rsun through stable mass-transfer. This is of potential interest for gravitational wave progenitors. Above these initial separations, the envelope becomes convective and the final separations reach values between 3 and 100\Rsun for the lowest mass models. These systems could become gravitational-wave sources if the final explosion marking the formation of the second neutron star results in a tighter orbit.

Our detailed simulations show strong metallicity effects which might affect the formation of double neutron stars. These effects are currently not included in the vast majority of binary population synthesis codes and may alter the rates and distributions of compact object mergers. Population synthesis simulations typically predict that double-neutron star merger rates are only very weakly metallicity dependent \citep[e.g.,][]{Neijssel+2019}.

\subsection{Observability}
\label{sec:dis:observability}
Stars stripped in binaries are notoriously difficult to detect during their longest-lived phase of core helium burning with current instruments. Not only are they compact, with typical sizes of about 1\Rsun, but most of their radiation is emitted in the extreme ultraviolet. Their companions typically outshine them in optical wavelengths \citep{gotberg_ionizing_2017,gotberg_spectral_2018}. However, this is no longer the case for the later evolutionary phases on which this work focuses. Their radius and luminosity increases until they reach giant sizes, in agreement with previous studies \citep[e.g.][]{habets_evolution_1986, yoon_type_2010, Yoon+_detectability_2012}. \citet{schootemeijer_clues_2018} discuss several systems which probably contain a helium-shell burning stripped star. Of those, the closest to the models we present is $\upsilon$ Sgr, with an inferred stripped-star mass of 2.5\Msun \citep{Dudley+Jeffery_Mass_1990}.

All our model stars reach effective temperatures of $4,000$ to $10,000 \,\rm{K}$, which spans typical ranges for WR stars, YSGs, and reaches the high temperature end of RSGs (see Fig. \ref{fig:HRD_grid}). We can estimate their observational characteristics based on their compositions, luminosities, and effective temperatures.\\
At high metallicity, typical spectra should be helium and nitrogen-rich, and hydrogen-poor. We expect surface gravity values ranging from $0.3 < \log_{10}{\left(g / [\rm{cm}\, \rm{s}^{-2}]\right)} < 4$ (see Table \ref{tab:zsun}). As helium lines can be challenging to measure, these spectra could be confused with those from nitrogen-enhanced B-stars.

At low metallicity, typical spectra should be similar, but the amount of hydrogen varies depending on the mass-transfer history and the orbital separation. We find extremely low surface gravity values of $\log_{10}{\left(g / [\rm{cm}\, \rm{s}^{-2}]\right)}\approxeq -0.4$ (see Table \ref{tab:zlow}) and expect narrow absorption features in their spectra, provided they do not experience strong mass-loss episodes. Given the lack of resolved stellar populations at such low metallicity, it may well be challenging to measure spectra of systems containing such giant stripped stars with present facilities. The Magellanic Clouds may provide examples sufficiently different in composition from the Galaxy to test the metallicity trends we describe.

Stripped stars with the lowest effective temperatures and highest luminosities are expected to be easiest to detect at optical wavelengths, and may well dominate the total emission of their binary system. These may be observed as helium red giants \citep{trimble_helium_1973, Yoon+_detectability_2012}. The most promising way to detect these systems is by detecting sources that are overluminous for their (Keplerian) masses, just like the system $\phi$ Persei discussed by \citet{schootemeijer_clues_2018}. These may be identified by searching for discrepancies between spectroscopic, evolutionary, and Keplerian masses, if available.

\subsection{Supernova progenitors}
\label{sec:dis:supernovae}
Our findings are relevant for understanding the properties of core-collapse supernovae, in particular stripped-envelope supernovae of type Ibc and type IIb (for pioneering work, see \citealt{podsiadlowski_presupernova_1992}; for interpretations of the class-defining Type IIb SN 1993J see \citealt{Podsiadlowski+_The_Progenitor_1993, Nomoto+_A_type_IIb_1993}; for recent studies see, e.g., \citealt{Bersten+_The_Type_IIb_2012, Eldridge2015, dessart_supernovae_2018, sravan_progenitors_2018}.). 

At the moment of explosion, the supernovae from our progenitor models may be classified as type Ibc or type IIb supernovae, depending on the amount of hydrogen and helium retained \citep[see also][]{yoon_type_2017}. The large radii of many of our models can be identified through shock cooling signatures in early light-curves \citep[e.g.,][]{Schawinski+_Supernova_2008, yoon_type_2010, piro_what_2013}. We next discuss two potential consequences of the evolution of these stars that, at least to our knowledge, have not been pointed out before.
 
\subsubsection{Circumstellar material prior to explosion}
At low metallicity, we find that the stars experience two phases of expansion. The first occurs shortly after helium depletion and is associated with hydrogen shell burning. The expansion is most significant in our models for higher mass stars, which expand by more than an order of magnitude. After this phase the stars contract until they expand again at the end of their life. The first phase of expansion is so severe for the more massive progenitors that we expect them to briefly fill their Roche lobe. If the resulting mass-transfer event is non-conservative we would expect ejection of mass prior to the explosion. For the most massive progenitors this occurs about 10,000\,years before their terminal explosion. This is short enough that the ejected material may still be close enough to the star at the moment of explosion to interact with the supernova shock. If so, these systems might be progenitors of at least some type Ibn supernovae \citep[see, e.g.,][]{Foley_SN_2006jc_2007, Pastorello_A_Giant_2007, Pastorello_Massive_2008, Hosseinzadeh_Type_Ibn_2019}.

\subsubsection{Asymmetric supernova progenitors}
Due to the large radial expansion we expect many stripped stars to fill their Roche lobe again shortly before their death or even at the time they explode. This implies that the supernova progenitor is not spherical, but instead has a non-axisymmetric "pear-like" shape as imposed by the shape of the Roche lobe. We expect this to have implications for the explosion, in particular when the supernova shock reaches the outer layers where the deformation is strongest.

\citet{afsariardchi_aspherical_2018} study explosions in non-spherical progenitors. They find that aspherical progenitors may have different shock-breakout signatures that are viewing-angle dependent. They further find that asphericity leads to collisions that would otherwise not have occurred and affect the observables. The case of a non-axisymmetric progenitor has not been modelled in their study. Although they argue the effects of asphericity in extended progenitors would be weak, they do not model Roche-lobe filling stars at explosion. Given that these configurations may be common, we encourage further detailed studies of these effects.

\subsection{Uncertainties}
\label{sec:dis:uncertainties}
The evolutionary models presented in this work are affected by several uncertainties. We have considered a fixed typical initial orbital period and mass ratio. For our high-metallicity models we have verified that the exact choice of the initial orbital period and companion mass has very little effect on the maximum radius of stripped stars if varied within reasonable limits \citep[cf.][]{gotberg_ionizing_2017,gotberg_spectral_2018}. At low metallicity, \cite{yoon_type_2010,yoon_type_2017}, \cite{claeys_binary_2011}, and \cite{ouchi_radii_2017} show that the amount of hydrogen left at the surface of the donor star is a function of the initial orbital separation. In appendix \ref{sec:appendix_porb}, we demonstrate that, for case B mass transfer, our general findings are robust against variations of the orbital period at low metallicity.

We further emphasize that we have modelled the late evolution of stripped stars by allowing them to fully expand, since our objective was to determine their maximum radii. In reality, we expect these stars to still be in orbit around a companion star. This would truncate their expansion and initiate a new phase of Roche-lobe overflow.

We consider the effects of internal mixing by convection and overshooting. Internal mixing is one of the main uncertainties in stellar evolution and also affects our results. \citet{yoon_type_2017} has pointed out the importance of mixing in the region above the retreating convective core of the donor star while it is on the main sequence. The choice of overshooting, efficiency of semi-convection and other potential mixing processes affects the details of the chemical profile. This in turn modifies the response of the donor star to stripping process and determines the mass of the hydrogen-rich layer that is left after the stripping process. Observations of stripped stars, for example as proposed by \citet{gotberg_spectral_2018}, may help to constrain these uncertainties in the future. The presence or absence of the expansion described in Section~\ref{sec:dis:supernovae}, which is equivalent to a blue loop, probably depends on the treatment of internal mixing (although similar additional expansion phases can be also seen in models from \citealt{yoon_type_2017}).

When estimating the number of systems that interact anew we assume a uniform distribution of orbital separations in log space. The actual distribution of separations is uncertain and depends on both the distribution of initial separations and the amount of orbital shrinking and widening during the first phase of interaction. A more advanced assumption would be to use the output of population synthesis simulations \citep{dewi_late_2003, Ivanova_The_Role_2003}.

We presented estimates for the binding energy of the envelope. These depend on the chosen definition of the core-envelope boundary \citep[e.g.][]{tauris_research_2001}, and on which energy terms are included \citep[e.g.][]{ivanova_common_2013}. This is discussed further in the appendix \ref{sec:appendix_eb}.

Stellar wind mass loss is also an important uncertainty for the evolution of massive stars \citep{smith_mass_2014, renzo_systematic_2017}. In the cases studied here, mass is primarily removed by Roche-lobe overflow and not by the stellar winds, so we do not expect a large impact on our results. However, we stress that the mass-loss rates for stripped stars are uncertain \citep{gotberg_ionizing_2017, gotberg_spectral_2018, vink_winds_2017,gilkis_effects_2019}. This should not have a large effect on our solar metallicity models, except perhaps for the highest-mass models in our grid. The effect of wind mass loss at low metallicity is negligible, as we demonstrate in appendix \ref{sec:appendix_winds}.

The spatial and temporal resolutions, and the nuclear network that we have adopted, should be sufficient for our purposes. We note that these models have not been optimized as input for supernova simulations, as this requires higher resolution, a more extended nuclear network, and calculations of the final evolutionary steps \citep[e.g.][]{Farmer2016}. However, we will present and discuss such models in a subsequent paper (Laplace et al. in prep).

\section{Summary and conclusion}
\label{sec:conclusion}
In this paper, we study the radius evolution of stars that have lost most of their hydrogen-rich envelope due to interaction with a companion. We consider stars with initial masses of 8--15\Msun at solar and low metallicity. We investigate how the internal composition profile, and changes in the nuclear burning phases, are linked with the radial evolution of the star.  Our results are in general agreement with previous studies.
Our findings can be summarized as follows:

\begin{enumerate}

\item Stars stripped in binaries can swell up to giant sizes, despite having lost most of their envelope, as has also been shown in earlier studies. This implies they can fill their Roche lobe again (sometimes referred to as Case BB or BC mass transfer).\\

\item The maximum radius achieved strongly depends on whether or not these stars retain a hydrogen layer. At solar metallicity, mass stripping by Roche-lobe overflow is effective in removing most of the hydrogen envelope. Winds play a minor role, but are strong enough to remove the remaining hydrogen layer. At low metallicity, all our models retain a significant hydrogen-rich envelope until the end of their evolution, in agreement with earlier findings. \\

\item At high metallicity, we find that the maximum radius (10--100\Rsun) is inversely proportional to the stellar mass. Only for the low mass end (progenitors with initial masses between 8 and 9\Msun) do we expect stripped stars to swell enough to interact anew with their companion, in agreement with earlier studies. \\

\item At low metallicity, for case B mass transfer, stripped stars can reach sizes of up to 400-700\Rsun, unless they fill their Roche-lobe anew. This maximum expansion is robust against variations of the wind mass loss and of the orbital period.\\

\item Population synthesis studies that rely on the \citet{hurley_comprehensive_2000} predictions or on interpolation of grids of detailed single stellar models do not properly account for the structure of stripped stars, in particular the effect of a remaining hydrogen layer. We find large discrepancies at solar metallicity in the mass range important for neutron star progenitors. At low metallicity we find discrepancies for the full mass range and we expect progenitors of both neutron stars and black holes to be affected.\\

\item We estimate, with simple assumptions, that population synthesis studies under-predict the number of systems that interact by a factor of 2 at solar metallicity (for stripped stars of about 2--3\Msun). For low metallicity the discrepancy is much worse. The fraction of systems that re-interact is underestimated by a factor of 10--30 (for stripped stars of about 2--6\Msun).\\

\item We draw attention to an additional expansion phase that occurs exclusively in low metallicity models shortly after central helium exhaustion. This phase is associated with hydrogen shell burning and only lasts a few thousand years. The star shrinks again briefly once helium shell-burning dominates, followed by the final expansion phase. \\

\item The first radius expansion we find at low metallicity may also
have important consequences. The low binding energy of the
hydrogen envelope suggests that mass may be ejected only a few tens of thousands of years prior to the final explosion, giving rise to a hydrogen-rich shell around the progenitor. This could impact the observable properties of the resulting supernova.\\

\item Many stripped stars are expected to be filling their Roche lobe at the moment their core collapses. This means that they would not be spherical at the moment they explode, but have a ``pear shape'' enforced by their Roche lobe. This may have interesting consequences for the observable characteristics of the final explosion. \\

\item Our results pose important concerns about the validity of rapid simulations of gravitational-wave sources. Our detailed simulations show metallicity effects that are not accounted for in rapid population synthesis simulations. Specifically, we expect the rates and channels for the formation of double neutron stars to be dependent on metallicity, in contrast to recent claims.

\end{enumerate}
We anticipate progress as new observational constraints become available for core-collapse supernovae from transient surveys and for double compact object mergers from gravitational-wave detectors. Robust model predictions will be needed to interpret these observations and learn about the physics of their binary progenitors.
Our findings call for detailed investigations to better understand the outcome of additional mass-transfer phases in binary systems that have already experienced previous interactions.


\begin{acknowledgements}
We thank the referee Christopher Tout for detailed comments. This work has benefited from helpful input by M. Renzo and E. Zapartas. The authors further acknowledge N. Afsariardchi, J. J. Eldridge, A. Gilkis, J. Klencki, C. Matzner, T. Moriya, A. Piro, O. R. Pols, T. Tauris, J. Yongje, and S.-Ch. Yoon for useful discussions.
The authors acknowledge funding by the European Union's Horizon 2020 research and innovation program from the European Research Council (ERC, grant agreement No.\ 715063), and by the Netherlands Organisation for Scientific Research (NWO) as part of the Vidi research program BinWaves with project number 639.042.728. YG acknowledges the funding from the Alvin E.\ Nashman fellowship for Theoretical Astrophysics. RF is supported by the Netherlands Organisation for Scientific Research (NWO) through a top module 2 grant with project number 614.001.501 (PI de Mink). The simulations were computed on the Dutch national e-infrastructure (Cartesius, project number 17234) with the support of the SURF Cooperative.\\

\emph{Software:} \texttt{mesaPlot} \citep{Farmer_mesaPlot_2019}, \texttt{matplotlib} \citep{Hunter_matplotlib_2007}, \texttt{numpy} \citep{vanderWalt_numpy_2011}, \texttt{ipython/jupyter} \citep{perez_ipython_2007}, \href{http://markkness.net/colorpy/ColorPy.html}{\texttt{ColorPy}} $\textcopyright$ by Mark Kness and \texttt{MESA} \citep{paxton_modules_2011,paxton_modules_2013,paxton_modules_2015,paxton_modules_2018, paxton_modules_2019}.

\end{acknowledgements}

\bibliographystyle{bibtex/aa} 
\bibliography{bibtex/strippedStars.bib} 

\begin{thebibliography}{122}
\expandafter\ifx\csname natexlab\endcsname\relax\def\natexlab#1{#1}\fi

\bibitem[{Abbott {et~al.}(2016)Abbott, Abbott, Abbott, Abernathy, Acernese,
  Ackley, Adams, Adams, Addesso, Adhikari, Adya, Affeldt, Agathos, Agatsuma,
  Aggarwal, Aguiar, Aiello, Ain, Ajith, Allen, Allocca, Altin, Anderson,
  Anderson, Arai, Arain, Araya, Arceneaux, Areeda, Arnaud, Arun, Ascenzi,
  Ashton, Ast, Aston, Astone, Aufmuth, Aulbert, Babak, Bacon, Bader, Baker,
  Baldaccini, Ballardin, Ballmer, Barayoga, Barclay, Barish, Barker, Barone,
  Barr, Barsotti, Barsuglia, Barta, Bartlett, Barton, Bartos, Bassiri, Basti,
  Batch, Baune, Bavigadda, Bazzan, Behnke, Bejger, Belczynski, Bell, Bell,
  Berger, Bergman, Bergmann, Berry, Bersanetti, Bertolini, Betzwieser, Bhagwat,
  Bhandare, Bilenko, Billingsley, Birch, Birney, Birnholtz, Biscans, Bisht,
  Bitossi, Biwer, Bizouard, Blackburn, Blair, Blair, Blair, Bloemen, Bock,
  Bodiya, Boer, Bogaert, Bogan, Bohe, Bojtos, Bond, Bondu, Bonnand, Boom, Bork,
  Boschi, Bose, Bouffanais, Bozzi, Bradaschia, Brady, Braginsky, Branchesi,
  Brau, Briant, Brillet, Brinkmann, Brisson, Brockill, Brooks, Brown, Brown,
  Brown, Buchanan, Buikema, Bulik, Bulten, Buonanno, Buskulic, Buy, Byer,
  Cabero, Cadonati, Cagnoli, Cahillane, Bustillo, Callister, Calloni, Camp,
  Cannon, Cao, Capano, Capocasa, Carbognani, Caride, Casanueva~Diaz, Casentini,
  Caudill, Cavaglià, Cavalier, Cavalieri, Cella, Cepeda, Baiardi, Cerretani,
  Cesarini, Chakraborty, Chalermsongsak, Chamberlin, Chan, Chao, Charlton,
  Chassande-Mottin, Chen, Chen, Cheng, Chincarini, Chiummo, Cho, Cho, Chow,
  Christensen, Chu, Chua, Chung, Ciani, Clara, Clark, Cleva, Coccia, Cohadon,
  Colla, Collette, Cominsky, Constancio, Conte, Conti, Cook, Corbitt, Cornish,
  Corsi, Cortese, Costa, Coughlin, Coughlin, Coulon, Countryman, Couvares,
  Cowan, Coward, Cowart, Coyne, Coyne, Craig, Creighton, Creighton, Cripe,
  Crowder, Cruise, Cumming, Cunningham, Cuoco, Dal~Canton, Danilishin,
  D'Antonio, Danzmann, Darman, Da~Silva~Costa, Dattilo, Dave, Daveloza, Davier,
  Davies, Daw, Day, De, DeBra, Debreczeni, Degallaix, De~Laurentis, Deléglise,
  Del~Pozzo, Denker, Dent, Dereli, Dergachev, DeRosa, De~Rosa, DeSalvo,
  Dhurandhar, Díaz, Di~Fiore, Di~Giovanni, Di~Lieto, Di~Pace, Di~Palma,
  Di~Virgilio, Dojcinoski, Dolique, Donovan, Dooley, Doravari, Douglas, Downes,
  Drago, Drever, Driggers, Du, Ducrot, Dwyer, Edo, Edwards, Effler, Eggenstein,
  Ehrens, Eichholz, Eikenberry, Engels, Essick, Etzel, Evans, Evans, Everett,
  Factourovich, Fafone, Fair, Fairhurst, Fan, Fang, Farinon, Farr, Farr,
  Favata, Fays, Fehrmann, Fejer, Feldbaum, Ferrante, Ferreira, Ferrini,
  Fidecaro, Finn, Fiori, Fiorucci, Fisher, Flaminio, Fletcher, Fong, Fournier,
  Franco, Frasca, Frasconi, Frede, Frei, Freise, Frey, Frey, Fricke, Fritschel,
  Frolov, Fulda, Fyffe, Gabbard, Gair, Gammaitoni, Gaonkar, Garufi, Gatto,
  Gaur, Gehrels, Gemme, Gendre, Genin, Gennai, George, Gergely, Germain, Ghosh,
  Ghosh, Ghosh, Giaime, Giardina, Giazotto, Gill, Glaefke, Gleason, Goetz,
  Goetz, Gondan, González, Castro, Gopakumar, Gordon, Gorodetsky, Gossan,
  Gosselin, Gouaty, Graef, Graff, Granata, Grant, Gras, Gray, Greco, Green,
  Greenhalgh, Groot, Grote, Grunewald, Guidi, Guo, Gupta, Gupta, Gushwa,
  Gustafson, Gustafson, Hacker, Hall, Hall, Hammond, Haney, Hanke, Hanks,
  Hanna, Hannam, Hanson, Hardwick, Harms, Harry, Harry, Hart, Hartman, Haster,
  Haughian, Healy, Heefner, Heidmann, Heintze, Heinzel, Heitmann, Hello,
  Hemming, Hendry, Heng, Hennig, Heptonstall, Heurs, Hild, Hoak, Hodge, Hofman,
  Hollitt, Holt, Holz, Hopkins, Hosken, Hough, Houston, Howell, Hu, Huang,
  Huerta, Huet, Hughey, Husa, Huttner, Huynh-Dinh, Idrisy, Indik, Ingram, Inta,
  Isa, Isac, Isi, Islas, Isogai, Iyer, Izumi, Jacobson, Jacqmin, Jang, Jani,
  Jaranowski, Jawahar, Jiménez-Forteza, Johnson, Johnson-McDaniel, Jones,
  Jones, Jonker, Ju, Haris, Kalaghatgi, Kalogera, Kandhasamy, Kang, Kanner,
  Karki, Kasprzack, Katsavounidis, Katzman, Kaufer, Kaur, Kawabe, Kawazoe,
  Kéfélian, Kehl, Keitel, Kelley, Kells, Kennedy, Keppel, Key, Khalaidovski,
  Khalili, Khan, Khan, Khan, Khazanov, Kijbunchoo, Kim, Kim, Kim, Kim, Kim,
  Kim, King, King, Kinzel, Kissel, Kleybolte, Klimenko, Koehlenbeck, Kokeyama,
  Koley, Kondrashov, Kontos, Koranda, Korobko, Korth, Kowalska, Kozak, Kringel,
  Krishnan, Królak, Krueger, Kuehn, Kumar, Kumar, Kuo, Kutynia, Kwee, Lackey,
  Landry, Lange, Lantz, Lasky, Lazzarini, Lazzaro, Leaci, Leavey, Lebigot, Lee,
  Lee, Lee, Lee, Lenon, Leonardi, Leong, Leroy, Letendre, Levin, Levine, Li,
  Libson, Littenberg, Lockerbie, Logue, Lombardi, London, Lord, Lorenzini,
  Loriette, Lormand, Losurdo, Lough, Lousto, Lovelace, Lück, Lundgren, Luo,
  Lynch, Ma, MacDonald, Machenschalk, MacInnis, Macleod, Magaña-Sandoval,
  Magee, Mageswaran, Majorana, Maksimovic, Malvezzi, Man, Mandel, Mandic,
  Mangano, Mansell, Manske, Mantovani, Marchesoni, Marion, Márka, Márka,
  Markosyan, Maros, Martelli, Martellini, Martin, Martin, Martynov, Marx,
  Mason, Masserot, Massinger, Masso-Reid, Matichard, Matone, Mavalvala,
  Mazumder, Mazzolo, McCarthy, McClelland, McCormick, McGuire, McIntyre,
  McIver, McManus, McWilliams, Meacher, Meadors, Meidam, Melatos, Mendell,
  Mendoza-Gandara, Mercer, Merilh, Merzougui, Meshkov, Messenger, Messick,
  Meyers, Mezzani, Miao, Michel, Middleton, Mikhailov, Milano, Miller,
  Millhouse, Minenkov, Ming, Mirshekari, Mishra, Mitra, Mitrofanov,
  Mitselmakher, Mittleman, Moggi, Mohan, Mohapatra, Montani, Moore, Moore,
  Moraru, Moreno, Morriss, Mossavi, Mours, Mow-Lowry, Mueller, Mueller, Muir,
  Mukherjee, Mukherjee, Mukherjee, Mukund, Mullavey, Munch, Murphy, Murray,
  Mytidis, Nardecchia, Naticchioni, Nayak, Necula, Nedkova, Nelemans, Neri,
  Neunzert, Newton, Nguyen, Nielsen, Nissanke, Nitz, Nocera, Nolting,
  Normandin, Nuttall, Oberling, Ochsner, O'Dell, Oelker, Ogin, Oh, Oh, Ohme,
  Oliver, Oppermann, Oram, O'Reilly, O'Shaughnessy, Ott, Ottaway, Ottens,
  Overmier, Owen, Pai, Pai, Palamos, Palashov, Palomba, Pal-Singh, Pan, Pan,
  Pankow, Pannarale, Pant, Paoletti, Paoli, Papa, Paris, Parker, Pascucci,
  Pasqualetti, Passaquieti, Passuello, Patricelli, Patrick, Pearlstone,
  Pedraza, Pedurand, Pekowsky, Pele, Penn, Perreca, Pfeiffer, Phelps, Piccinni,
  Pichot, Pickenpack, Piergiovanni, Pierro, Pillant, Pinard, Pinto, Pitkin,
  Poeld, Poggiani, Popolizio, Post, Powell, Prasad, Predoi, Premachandra,
  Prestegard, Price, Prijatelj, Principe, Privitera, Prix, Prodi, Prokhorov,
  Puncken, Punturo, Puppo, Pürrer, Qi, Qin, Quetschke, Quintero,
  Quitzow-James, Raab, Rabeling, Radkins, Raffai, Raja, Rakhmanov, Ramet,
  Rapagnani, Raymond, Razzano, Re, Read, Reed, Regimbau, Rei, Reid, Reitze,
  Rew, Reyes, Ricci, Riles, Robertson, Robie, Robinet, Rocchi, Rolland,
  Rollins, Roma, Romano, Romano, Romanov, Romie, Rosińska, Rowan, Rüdiger,
  Ruggi, Ryan, Sachdev, Sadecki, Sadeghian, Salconi, Saleem, Salemi, Samajdar,
  Sammut, Sampson, Sanchez, Sandberg, Sandeen, Sanders, Sanders, Sassolas,
  Sathyaprakash, Saulson, Sauter, Savage, Sawadsky, Schale, Schilling, Schmidt,
  Schmidt, Schnabel, Schofield, Schönbeck, Schreiber, Schuette, Schutz, Scott,
  Scott, Sellers, Sengupta, Sentenac, Sequino, Sergeev, Serna, Setyawati,
  Sevigny, Shaddock, Shaffer, Shah, Shahriar, Shaltev, Shao, Shapiro, Shawhan,
  Sheperd, Shoemaker, Shoemaker, Siellez, Siemens, Sigg, Silva, Simakov,
  Singer, Singer, Singh, Singh, Singhal, Sintes, Slagmolen, Smith, Smith,
  Smith, Smith, Son, Sorazu, Sorrentino, Souradeep, Srivastava, Staley,
  Steinke, Steinlechner, Steinlechner, Steinmeyer, Stephens, Stevenson, Stone,
  Strain, Straniero, Stratta, Strauss, Strigin, Sturani, Stuver, Summerscales,
  Sun, Sutton, Swinkels, Szczepańczyk, Tacca, Talukder, Tanner, Tápai,
  Tarabrin, Taracchini, Taylor, Theeg, Thirugnanasambandam, Thomas, Thomas,
  Thomas, Thorne, Thorne, Thrane, Tiwari, Tiwari, Tokmakov, Tomlinson, Tonelli,
  Torres, Torrie, Töyrä, Travasso, Traylor, Trifirò, Tringali, Trozzo, Tse,
  Turconi, Tuyenbayev, Ugolini, Unnikrishnan, Urban, Usman, Vahlbruch, Vajente,
  Valdes, Vallisneri, van Bakel, van Beuzekom, van~den Brand, Van Den~Broeck,
  Vander-Hyde, van~der Schaaf, van Heijningen, van Veggel, Vardaro, Vass,
  Vasúth, Vaulin, Vecchio, Vedovato, Veitch, Veitch, Venkateswara, Verkindt,
  Vetrano, Viceré, Vinciguerra, Vine, Vinet, Vitale, Vo, Vocca, Vorvick, Voss,
  Vousden, Vyatchanin, Wade, Wade, Wade, Waldman, Walker, Wallace, Walsh, Wang,
  Wang, Wang, Wang, Wang, Ward, Ward, Warner, Was, Weaver, Wei, Weinert,
  Weinstein, Weiss, Welborn, Wen, Weßels, Westphal, Wette, Whelan, Whitcomb,
  White, Whiting, Wiesner, Wilkinson, Willems, Williams, Williams, Williamson,
  Willis, Willke, Wimmer, Winkelmann, Winkler, Wipf, Wiseman, Wittel, Woan,
  Worden, Wright, Wu, Yablon, Yakushin, Yam, Yamamoto, Yancey, Yap, Yu, Yvert,
  ZadroŻny, Zangrando, Zanolin, Zendri, Zevin, Zhang, Zhang, Zhang, Zhang,
  Zhao, Zhou, Zhou, Zhu, Zucker, Zuraw, Zweizig, Collaboration, \&
  Collaboration}]{abbott_observation_2016}
Abbott, B.~P., Abbott, R., Abbott, T.~D., {et~al.} 2016, Physical Review
  Letters, 116, 061102

\bibitem[{{Abbott} {et~al.}(2019){Abbott}, {Abbott}, {Abbott}, {Abraham},
  {Acernese}, {Ackley}, {Adams}, {Adhikari}, {Adya}, {Affeldt}, {Agathos},
  {Agatsuma}, {Aggarwal}, {Aguiar}, {Aiello}, {Ain}, {Ajith}, {Allen},
  {Allocca}, {Aloy}, {Altin}, {Amato}, {Ananyeva}, {Anderson}, {Anderson},
  {Angelova}, {Antier}, {Appert}, {Arai}, {Araya}, {Areeda}, {Ar{\`e}ne},
  {Arnaud}, {Arun}, {Ascenzi}, {Ashton}, {Aston}, {Astone}, {Aubin}, {Aufmuth},
  {AultONeal}, {Austin}, {Avendano}, {Avila-Alvarez}, {LIGO Scientific
  Collaboration}, \& {Virgo
  Collaboration}}]{ligo_&_virgo_collaboration_gwtc-1:_2019}
{Abbott}, B.~P., {Abbott}, R., {Abbott}, T.~D., {et~al.} 2019, Physical Review
  X, 9, 031040

\bibitem[{Abbott {et~al.}(2017)Abbott, Abbott, Abbott, Acernese, Ackley, Adams,
  Adams, Addesso, Adhikari, Adya, Affeldt, Afrough, Agarwal, Agathos, Agatsuma,
  Aggarwal, Aguiar, Aiello, Ain, Ajith, Allen, Allen, Allocca, Altin, Amato,
  Ananyeva, Anderson, Anderson, Angelova, Antier, Appert, Arai, Araya, Areeda,
  Arnaud, Arun, Ascenzi, Ashton, Ast, Aston, Astone, Atallah, Aufmuth, Aulbert,
  AultONeal, Austin, Avila-Alvarez, Babak, Bacon, Bader, Bae, Bailes, Baker,
  Baldaccini, Ballardin, Ballmer, Banagiri, Barayoga, Barclay, Barish, Barker,
  Barkett, Barone, Barr, Barsotti, Barsuglia, Barta, Barthelmy, Bartlett,
  Bartos, Bassiri, Basti, Batch, Bawaj, Bayley, Bazzan, Bécsy, Beer, Bejger,
  Belahcene, Bell, Berger, Bergmann, Bernuzzi, Bero, Berry, Bersanetti,
  Bertolini, Betzwieser, Bhagwat, Bhandare, Bilenko, Billingsley, Billman,
  Birch, Birney, Birnholtz, Biscans, Biscoveanu, Bisht, Bitossi, Biwer,
  Bizouard, Blackburn, Blackman, Blair, Blair, Blair, Bloemen, Bock, Bode,
  Boer, Bogaert, Bohe, Bondu, Bonilla, Bonnand, Boom, Bork, Boschi, Bose,
  Bossie, Bouffanais, Bozzi, Bradaschia, Brady, Branchesi, Brau, Briant,
  Brillet, Brinkmann, Brisson, Brockill, Broida, Brooks, Brown, Brown, Brunett,
  Buchanan, Buikema, Bulik, Bulten, Buonanno, Buskulic, Buy, Byer, Cabero,
  Cadonati, Cagnoli, Cahillane, Calderón~Bustillo, Callister, Calloni, Camp,
  Canepa, Canizares, Cannon, Cao, Cao, Capano, Capocasa, Carbognani, Caride,
  Carney, Carullo, Casanueva~Diaz, Casentini, Caudill, Cavaglià, Cavalier,
  Cavalieri, Cella, Cepeda, Cerdá-Durán, Cerretani, Cesarini, Chamberlin,
  Chan, Chao, Charlton, Chase, Chassande-Mottin, Chatterjee, Chatziioannou,
  Cheeseboro, Chen, Chen, Chen, Cheng, Chia, Chincarini, Chiummo, Chmiel, Cho,
  Cho, Chow, Christensen, Chu, Chua, Chua, Chung, Chung, Ciani, Ciolfi,
  Cirelli, Cirone, Clara, Clark, Clearwater, Cleva, Cocchieri, Coccia, Cohadon,
  Cohen, Colla, Collette, Cominsky, Constancio, Conti, Cooper, Corban, Corbitt,
  Cordero-Carrión, Corley, Cornish, Corsi, Cortese, Costa, Coughlin, Coughlin,
  Coulon, Countryman, Couvares, Covas, Cowan, Coward, Cowart, Coyne, Coyne,
  Creighton, Creighton, Cripe, Crowder, Cullen, Cumming, Cunningham, Cuoco,
  Dal~Canton, Dálya, Danilishin, D'Antonio, Danzmann, Dasgupta,
  Da~Silva~Costa, Dattilo, Dave, Davier, Davis, Daw, Day, De, DeBra, Degallaix,
  De~Laurentis, Deléglise, Del~Pozzo, Demos, Denker, Dent, De~Pietri,
  Dergachev, De~Rosa, DeRosa, De~Rossi, DeSalvo, de~Varona, Devenson,
  Dhurandhar, Díaz, Dietrich, Di~Fiore, Di~Giovanni, Di~Girolamo, Di~Lieto,
  Di~Pace, Di~Palma, Di~Renzo, Doctor, Dolique, Donovan, Dooley, Doravari,
  Dorrington, Douglas, Dovale~Álvarez, Downes, Drago, Dreissigacker, Driggers,
  Du, Ducrot, Dudi, Dupej, Dwyer, Edo, Edwards, Effler, Eggenstein, Ehrens,
  Eichholz, Eikenberry, Eisenstein, Essick, Estevez, Etienne, Etzel, Evans,
  Evans, Factourovich, Fafone, Fair, Fairhurst, Fan, Farinon, Farr, Farr,
  Fauchon-Jones, Favata, Fays, Fee, Fehrmann, Feicht, Fejer, Fernandez-Galiana,
  Ferrante, Ferreira, Ferrini, Fidecaro, Finstad, Fiori, Fiorucci, Fishbach,
  Fisher, Fitz-Axen, Flaminio, Fletcher, Fong, Font, Forsyth, Forsyth,
  Fournier, Frasca, Frasconi, Frei, Freise, Frey, Frey, Fries, Fritschel,
  Frolov, Fulda, Fyffe, Gabbard, Gadre, Gaebel, Gair, Gammaitoni, Ganija,
  Gaonkar, Garcia-Quiros, Garufi, Gateley, Gaudio, Gaur, Gayathri, Gehrels,
  Gemme, Genin, Gennai, George, George, Gergely, Germain, Ghonge, Ghosh, Ghosh,
  Ghosh, Giaime, Giardina, Giazotto, Gill, Glover, Goetz, Goetz, Gomes,
  Goncharov, González, Gonzalez~Castro, Gopakumar, Gorodetsky, Gossan,
  Gosselin, Gouaty, Grado, Graef, Granata, Grant, Gras, Gray, Greco, Green,
  Gretarsson, Groot, Grote, Grunewald, Gruning, Guidi, Guo, Gupta, Gupta,
  Gushwa, Gustafson, Gustafson, Halim, Hall, Hall, Hamilton, Hammond, Haney,
  Hanke, Hanks, Hanna, Hannam, Hannuksela, Hanson, Hardwick, Harms, Harry,
  Harry, Hart, Haster, Haughian, Healy, Heidmann, Heintze, Heitmann, Hello,
  Hemming, Hendry, Heng, Hennig, Heptonstall, Heurs, Hild, Hinderer, Ho, Hoak,
  Hofman, Holt, Holz, Hopkins, Horst, Hough, Houston, Howell, Hreibi, Hu,
  Huerta, Huet, Hughey, Husa, Huttner, Huynh-Dinh, Indik, Inta, Intini, Isa,
  Isac, Isi, Iyer, Izumi, Jacqmin, Jani, Jaranowski, Jawahar, Jiménez-Forteza,
  Johnson, Johnson-McDaniel, Jones, Jones, Jonker, Ju, Junker, Kalaghatgi,
  Kalogera, Kamai, Kandhasamy, Kang, Kanner, Kapadia, Karki, Karvinen,
  Kasprzack, Kastaun, Katolik, Katsavounidis, Katzman, Kaufer, Kawabe,
  Kéfélian, Keitel, Kemball, Kennedy, Kent, Key, Khalili, Khan, Khan, Khan,
  Khazanov, Kijbunchoo, Kim, Kim, Kim, Kim, Kim, Kim, Kimbrell, King, King,
  Kinley-Hanlon, Kirchhoff, Kissel, Kleybolte, Klimenko, Knowles, Koch,
  Koehlenbeck, Koley, Kondrashov, Kontos, Korobko, Korth, Kowalska, Kozak,
  Krämer, Kringel, Krishnan, Królak, Kuehn, Kumar, Kumar, Kumar, Kuo,
  Kutynia, Kwang, Lackey, Lai, Landry, Lang, Lange, Lantz, Lanza, Larson,
  Lartaux-Vollard, Lasky, Laxen, Lazzarini, Lazzaro, Leaci, Leavey, Lee, Lee,
  Lee, Lee, Lee, Lehmann, Lenon, Leon, Leonardi, Leroy, Letendre, Levin, Li,
  Linker, Littenberg, Liu, Liu, Lo, Lockerbie, London, Lord, Lorenzini,
  Loriette, Lormand, Losurdo, Lough, Lousto, Lovelace, Lück, Lumaca, Lundgren,
  Lynch, Ma, Macas, Macfoy, Machenschalk, MacInnis, Macleod, Magaña~Hernandez,
  Magaña-Sandoval, Magaña~Zertuche, Magee, Majorana, Maksimovic, Man, Mandic,
  Mangano, Mansell, Manske, Mantovani, Marchesoni, Marion, Márka, Márka,
  Markakis, Markosyan, Markowitz, Maros, Marquina, Marsh, Martelli, Martellini,
  Martin, Martin, Martynov, Marx, Mason, Massera, Masserot, Massinger,
  Masso-Reid, Mastrogiovanni, Matas, Matichard, Matone, Mavalvala, Mazumder,
  McCarthy, McClelland, McCormick, McCuller, McGuire, McIntyre, McIver,
  McManus, McNeill, McRae, McWilliams, Meacher, Meadors, Mehmet, Meidam,
  Mejuto-Villa, Melatos, Mendell, Mercer, Merilh, Merzougui, Meshkov,
  Messenger, Messick, Metzdorff, Meyers, Miao, Michel, Middleton, Mikhailov,
  Milano, Miller, Miller, Miller, Millhouse, Milovich-Goff, Minazzoli,
  Minenkov, Ming, Mishra, Mitra, Mitrofanov, Mitselmakher, Mittleman, Moffa,
  Moggi, Mogushi, Mohan, Mohapatra, Molina, Montani, Moore, Moraru, Moreno,
  Morisaki, Morriss, Mours, Mow-Lowry, Mueller, Muir, Mukherjee, Mukherjee,
  Mukherjee, Mukund, Mullavey, Munch, Muñiz, Muratore, Murray, Nagar, Napier,
  Nardecchia, Naticchioni, Nayak, Neilson, Nelemans, Nelson, Nery, Neunzert,
  Nevin, Newport, Newton, Ng, Nguyen, Nguyen, Nichols, Nielsen, Nissanke, Nitz,
  Noack, Nocera, Nolting, North, Nuttall, Oberling, O'Dea, Ogin, Oh, Oh, Ohme,
  Okada, Oliver, Oppermann, Oram, O'Reilly, Ormiston, Ortega, O'Shaughnessy,
  Ossokine, Ottaway, Overmier, Owen, Pace, Page, Page, Pai, Pai, Palamos,
  Palashov, Palomba, Pal-Singh, Pan, Pan, Pang, Pang, Pankow, Pannarale, Pant,
  Paoletti, Paoli, Papa, Parida, Parker, Pascucci, Pasqualetti, Passaquieti,
  Passuello, Patil, Patricelli, Pearlstone, Pedraza, Pedurand, Pekowsky, Pele,
  Penn, Perez, Perreca, Perri, Pfeiffer, Phelps, Piccinni, Pichot,
  Piergiovanni, Pierro, Pillant, Pinard, Pinto, Pirello, Pitkin, Poe, Poggiani,
  Popolizio, Porter, Post, Powell, Prasad, Pratt, Pratten, Predoi, Prestegard,
  Prijatelj, Principe, Privitera, Prix, Prodi, Prokhorov, Puncken, Punturo,
  Puppo, Pürrer, Qi, Quetschke, Quintero, Quitzow-James, Raab, Rabeling,
  Radkins, Raffai, Raja, Rajan, Rajbhandari, Rakhmanov, Ramirez, Ramos-Buades,
  Rapagnani, Raymond, Razzano, Read, Regimbau, Rei, Reid, Reitze, Ren, Reyes,
  Ricci, Ricker, Rieger, Riles, Rizzo, Robertson, Robie, Robinet, Rocchi,
  Rolland, Rollins, Roma, Romano, Romano, Romel, Romie, Rosińska, Ross, Rowan,
  Rüdiger, Ruggi, Rutins, Ryan, Sachdev, Sadecki, Sadeghian, Sakellariadou,
  Salconi, Saleem, Salemi, Samajdar, Sammut, Sampson, Sanchez, Sanchez,
  Sanchis-Gual, Sandberg, Sanders, Sassolas, Sathyaprakash, Saulson, Sauter,
  Savage, Sawadsky, Schale, Scheel, Scheuer, Schmidt, Schmidt, Schnabel,
  Schofield, Schönbeck, Schreiber, Schuette, Schulte, Schutz, Schwalbe, Scott,
  Scott, Seidel, Sellers, Sengupta, Sentenac, Sequino, Sergeev, Shaddock,
  Shaffer, Shah, Shahriar, Shaner, Shao, Shapiro, Shawhan, Sheperd, Shoemaker,
  Shoemaker, Siellez, Siemens, Sieniawska, Sigg, Silva, Singer, Singh, Singhal,
  Sintes, Slagmolen, Smith, Smith, Smith, Somala, Son, Sonnenberg, Sorazu,
  Sorrentino, Souradeep, Spencer, Srivastava, Staats, Staley, Steinke,
  Steinlechner, Steinlechner, Steinmeyer, Stevenson, Stone, Stops, Strain,
  Stratta, Strigin, Strunk, Sturani, Stuver, Summerscales, Sun, Sunil, Suresh,
  Sutton, Swinkels, Szczepańczyk, Tacca, Tait, Talbot, Talukder, Tanner,
  Tápai, Taracchini, Tasson, Taylor, Taylor, Tewari, Theeg, Thies, Thomas,
  Thomas, Thomas, Thorne, Thorne, Thrane, Tiwari, Tiwari, Tokmakov, Toland,
  Tonelli, Tornasi, Torres-Forné, Torrie, Töyrä, Travasso, Traylor,
  Trinastic, Tringali, Trozzo, Tsang, Tse, Tso, Tsukada, Tsuna, Tuyenbayev,
  Ueno, Ugolini, Unnikrishnan, Urban, Usman, Vahlbruch, Vajente, Valdes,
  Vallisneri, van Bakel, van Beuzekom, van~den Brand, Van Den~Broeck,
  Vander-Hyde, van~der Schaaf, van Heijningen, van Veggel, Vardaro, Varma,
  Vass, Vasúth, Vecchio, Vedovato, Veitch, Veitch, Venkateswara, Venugopalan,
  Verkindt, Vetrano, Viceré, Viets, Vinciguerra, Vine, Vinet, Vitale, Vo,
  Vocca, Vorvick, Vyatchanin, Wade, Wade, Wade, Walet, Walker, Wallace, Walsh,
  Wang, Wang, Wang, Wang, Wang, Ward, Warner, Was, Watchi, Weaver, Wei,
  Weinert, Weinstein, Weiss, Wen, Wessel, Weßels, Westerweck, Westphal, Wette,
  Whelan, Whitcomb, Whiting, Whittle, Wilken, Williams, Williams, Williamson,
  Willis, Willke, Wimmer, Winkler, Wipf, Wittel, Woan, Woehler, Wofford, Wong,
  Worden, Wright, Wu, Wysocki, Xiao, Yamamoto, Yancey, Yang, Yap, Yazback, Yu,
  Yu, Yvert, ZadroŻny, Zanolin, Zelenova, Zendri, Zevin, Zhang, Zhang, Zhang,
  Zhang, Zhao, Zhou, Zhou, Zhu, Zhu, Zimmerman, Zucker, Zweizig, Collaboration,
  \& Collaboration}]{abbott_gw170817:_2017}
Abbott, B.~P., Abbott, R., Abbott, T.~D., {et~al.} 2017, Physical Review
  Letters, 119, 161101

\bibitem[{Afsariardchi \& Matzner(2018)}]{afsariardchi_aspherical_2018}
Afsariardchi, N. \& Matzner, C.~D. 2018, The Astrophysical Journal, 856, 146

\bibitem[{Asplund {et~al.}(2009)Asplund, Grevesse, Sauval, \&
  Scott}]{asplund_chemical_2009}
Asplund, M., Grevesse, N., Sauval, A.~J., \& Scott, P. 2009, Annual Review of
  Astronomy and Astrophysics, 47, 481

\bibitem[{{Belczynski} {et~al.}(2008){Belczynski}, {Kalogera}, {Rasio}, {Taam},
  {Zezas}, {Bulik}, {Maccarone}, \& {Ivanova}}]{Belczynski+2008}
{Belczynski}, K., {Kalogera}, V., {Rasio}, F.~A., {et~al.} 2008, \apjs, 174,
  223

\bibitem[{{Bersten} {et~al.}(2014){Bersten}, {Benvenuto}, {Folatelli},
  {Nomoto}, {Kuncarayakti}, {Srivastav}, {Anupama}, {Quimby}, \&
  {Sahu}}]{Bersten+2014}
{Bersten}, M.~C., {Benvenuto}, O.~G., {Folatelli}, G., {et~al.} 2014, \aj, 148,
  68

\bibitem[{{Bersten} {et~al.}(2012){Bersten}, {Benvenuto}, {Nomoto}, {Ergon},
  {Folatelli}, {Sollerman}, {Benetti}, {Botticella}, {Fraser}, {Kotak},
  {Maeda}, {Ochner}, \& {Tomasella}}]{Bersten+_The_Type_IIb_2012}
{Bersten}, M.~C., {Benvenuto}, O.~G., {Nomoto}, K., {et~al.} 2012, \apj, 757,
  31

\bibitem[{Bestenlehner {et~al.}(2014)Bestenlehner, Gräfener, Vink, Najarro,
  de~Koter, Sana, Evans, Crowther, Hénault-Brunet, Herrero, Langer, Schneider,
  Simón-Díaz, Taylor, \& Walborn}]{bestenlehner_vlt-flames_2014}
Bestenlehner, J.~M., Gräfener, G., Vink, J.~S., {et~al.} 2014, Astronomy and
  Astrophysics, 570, A38

\bibitem[{{Breivik} {et~al.}(2019){Breivik}, {Coughlin}, {Zevin}, {Rodriguez},
  {Kremer}, {Ye}, {Andrews}, {Kurkowski}, {Digman}, {Larson}, \&
  {Rasio}}]{Breivik+2019}
{Breivik}, K., {Coughlin}, S.~C., {Zevin}, M., {et~al.} 2019, arXiv e-prints,
  arXiv:1911.00903

\bibitem[{Bressan {et~al.}(2012)Bressan, Marigo, Girardi, Salasnich, Dal~Cero,
  Rubele, \& Nanni}]{bressan_parsec:_2012}
Bressan, A., Marigo, P., Girardi, L., {et~al.} 2012, Monthly Notices of the
  Royal Astronomical Society, 427, 127

\bibitem[{Brott {et~al.}(2011)Brott, de~Mink, Cantiello, Langer, de~Koter,
  Evans, Hunter, Trundle, \& Vink}]{brott_rotating_2011}
Brott, I., de~Mink, S.~E., Cantiello, M., {et~al.} 2011, Astronomy and
  Astrophysics, 530, A115

\bibitem[{{Buchler} \& {Yueh}(1976)}]{buchler_compton_1976}
{Buchler}, J.~R. \& {Yueh}, W.~R. 1976, \apj, 210, 440

\bibitem[{Böhm-Vitense(1958)}]{bohm-vitense_uber_1958}
Böhm-Vitense, E. 1958, Zeitschrift fur Astrophysik, 46, 108

\bibitem[{{Cao} {et~al.}(2013){Cao}, {Kasliwal}, {Arcavi}, {Horesh}, {Hancock},
  {Valenti}, {Cenko}, {Kulkarni}, {Gal-Yam}, {Gorbikov}, {Ofek}, {Sand},
  {Yaron}, {Graham}, {Silverman}, {Wheeler}, {Marion}, {Walker}, {Mazzali},
  {Howell}, {Li}, {Kong}, {Bloom}, {Nugent}, {Surace}, {Masci}, {Carpenter},
  {Degenaar}, \& {Gelino}}]{Cao+2013}
{Cao}, Y., {Kasliwal}, M.~M., {Arcavi}, I., {et~al.} 2013, \apjl, 775, L7

\bibitem[{{Cassisi} {et~al.}(2007){Cassisi}, {Potekhin}, {Pietrinferni},
  {Catelan}, \& {Salaris}}]{cassisi_updated_2007}
{Cassisi}, S., {Potekhin}, A.~Y., {Pietrinferni}, A., {Catelan}, M., \&
  {Salaris}, M. 2007, \apj, 661, 1094

\bibitem[{Claeys {et~al.}(2011)Claeys, de~Mink, Pols, Eldridge, \&
  Baes}]{claeys_binary_2011}
Claeys, J. S.~W., de~Mink, S.~E., Pols, O.~R., Eldridge, J.~J., \& Baes, M.
  2011, Astronomy and Astrophysics, 528, A131

\bibitem[{Cox \& Salpeter(1961)}]{cox_equilibrium_1961}
Cox, J.~P. \& Salpeter, E.~E. 1961, The Astrophysical Journal, 133, 764

\bibitem[{Crowther(2007)}]{crowther_physical_2007}
Crowther, P.~A. 2007, Annual Review of Astronomy and Astrophysics, 45, 177

\bibitem[{de~Jager {et~al.}(1988)de~Jager, Nieuwenhuijzen, \& van~der
  Hucht}]{de_jager_mass_1988}
de~Jager, C., Nieuwenhuijzen, H., \& van~der Hucht, K.~A. 1988, Astronomy and
  Astrophysics Supplement Series, 72, 259

\bibitem[{de~Kool(1990)}]{de_kool_common_1990}
de~Kool, M. 1990, The Astrophysical Journal, 358, 189

\bibitem[{de~Mink {et~al.}(2007)de~Mink, Pols, \&
  Hilditch}]{de_mink_efficiency_2007}
de~Mink, S.~E., Pols, O.~R., \& Hilditch, R.~W. 2007, Astronomy and
  Astrophysics, 467, 1181

\bibitem[{Dessart {et~al.}(2018)Dessart, Yoon, Livne, \&
  Waldman}]{dessart_supernovae_2018}
Dessart, L., Yoon, S.-C., Livne, E., \& Waldman, R. 2018, Astronomy and
  Astrophysics, 612, A61

\bibitem[{Dewi \& Pols(2003)}]{dewi_late_2003}
Dewi, J. D.~M. \& Pols, O.~R. 2003, Monthly Notices of the Royal Astronomical
  Society, 344, 629

\bibitem[{Dewi {et~al.}(2002)Dewi, Pols, Savonije, \& van~den
  Heuvel}]{dewi_evolution_2002}
Dewi, J. D.~M., Pols, O.~R., Savonije, G.~J., \& van~den Heuvel, E. P.~J. 2002,
  Monthly Notices of the Royal Astronomical Society, 331, 1027

\bibitem[{Divine(1965)}]{divine_structure_1965}
Divine, N. 1965, The Astrophysical Journal, 142, 824

\bibitem[{Drout {et~al.}(2009)Drout, Massey, Meynet, Tokarz, \&
  Caldwell}]{drout_yellow_2009}
Drout, M.~R., Massey, P., Meynet, G., Tokarz, S., \& Caldwell, N. 2009, The
  Astrophysical Journal, 703, 441

\bibitem[{{Dudley} \& {Jeffery}(1990)}]{Dudley+Jeffery_Mass_1990}
{Dudley}, R.~E. \& {Jeffery}, C.~S. 1990, \mnras, 247, 400

\bibitem[{{Eggleton}(1971)}]{Eggleton1971}
{Eggleton}, P.~P. 1971, \mnras, 151, 351

\bibitem[{{Eggleton}(1972)}]{Eggleton1972}
{Eggleton}, P.~P. 1972, \mnras, 156, 361

\bibitem[{Eggleton(1983)}]{eggleton_aproximations_1983}
Eggleton, P.~P. 1983, The Astrophysical Journal, 268, 368

\bibitem[{{Eggleton} {et~al.}(1973){Eggleton}, {Faulkner}, \&
  {Flannery}}]{Eggleton+1973}
{Eggleton}, P.~P., {Faulkner}, J., \& {Flannery}, B.~P. 1973, \aap, 23, 325

\bibitem[{{Eldridge} {et~al.}(2015){Eldridge}, {Fraser}, {Maund}, \&
  {Smartt}}]{Eldridge2015}
{Eldridge}, J.~J., {Fraser}, M., {Maund}, J.~R., \& {Smartt}, S.~J. 2015,
  \mnras, 446, 2689

\bibitem[{Eldridge {et~al.}(2013)Eldridge, Fraser, Smartt, Maund, \&
  Crockett}]{eldridge_death_2013}
Eldridge, J.~J., Fraser, M., Smartt, S.~J., Maund, J.~R., \& Crockett, R.~M.
  2013, Monthly Notices of the Royal Astronomical Society, 436, 774

\bibitem[{{Eldridge} \& {Maund}(2016)}]{Eldridge+2016}
{Eldridge}, J.~J. \& {Maund}, J.~R. 2016, \mnras, 461, L117

\bibitem[{{Eldridge} \& {Stanway}(2016)}]{eldridge_BPASS_2016}
{Eldridge}, J.~J. \& {Stanway}, E.~R. 2016, \mnras, 462, 3302

\bibitem[{Eldridge {et~al.}(2017)Eldridge, Stanway, Xiao, McClelland, Taylor,
  Ng, Greis, \& Bray}]{eldridge_binary_2017}
Eldridge, J.~J., Stanway, E.~R., Xiao, L., {et~al.} 2017, Publications of the
  Astronomical Society of Australia, 34, e058

\bibitem[{{Farmer}(2019)}]{Farmer_mesaPlot_2019}
{Farmer}, R. 2019, {rjfarmer/mesaplot: Bug fixes}

\bibitem[{{Farmer} {et~al.}(2016){Farmer}, {Fields}, {Petermann}, {Dessart},
  {Cantiello}, {Paxton}, \& {Timmes}}]{Farmer2016}
{Farmer}, R., {Fields}, C.~E., {Petermann}, I., {et~al.} 2016, \apjs, 227, 22

\bibitem[{Fitzpatrick(1988)}]{fitzpatrick_properties_1988}
Fitzpatrick, E.~L. 1988, The Astrophysical Journal, 335, 703

\bibitem[{{Foley} {et~al.}(2007){Foley}, {Smith}, {Ganeshalingam}, {Li},
  {Chornock}, \& {Filippenko}}]{Foley_SN_2006jc_2007}
{Foley}, R.~J., {Smith}, N., {Ganeshalingam}, M., {et~al.} 2007, \apjl, 657,
  L105

\bibitem[{{Fragos} {et~al.}(2019){Fragos}, {Andrews}, {Ramirez-Ruiz}, {Meynet},
  {Kalogera}, {Taam}, \& {Zezas}}]{fragos_the_complete_2019}
{Fragos}, T., {Andrews}, J.~J., {Ramirez-Ruiz}, E., {et~al.} 2019, \apjl, 883,
  L45

\bibitem[{{Fremling} {et~al.}(2014){Fremling}, {Sollerman}, {Taddia}, {Ergon},
  {Valenti}, {Arcavi}, {Ben-Ami}, {Cao}, {Cenko}, {Filippenko}, {Gal-Yam}, \&
  {Howell}}]{Fremling+2014}
{Fremling}, C., {Sollerman}, J., {Taddia}, F., {et~al.} 2014, \aap, 565, A114

\bibitem[{Giacobbo \& Mapelli(2018)}]{giacobbo_progenitors_2018}
Giacobbo, N. \& Mapelli, M. 2018, Monthly Notices of the Royal Astronomical
  Society, 480, 2011

\bibitem[{Gilkis {et~al.}(2019)Gilkis, Vink, Eldridge, \&
  Tout}]{gilkis_effects_2019}
Gilkis, A., Vink, J.~S., Eldridge, J.~J., \& Tout, C.~A. 2019, Monthly Notices
  of the Royal Astronomical Society, 486, 4451

\bibitem[{{G{\"o}tberg} {et~al.}(2017){G{\"o}tberg}, {de Mink}, \&
  Groh}]{gotberg_ionizing_2017}
{G{\"o}tberg}, Y., {de Mink}, S.~E., \& Groh, J.~H. 2017, Astronomy and
  Astrophysics, 608, A11

\bibitem[{{G{\"o}tberg} {et~al.}(2018){G{\"o}tberg}, {de Mink}, {Groh},
  {Kupfer}, {Crowther}, {Zapartas}, \& {Renzo}}]{gotberg_spectral_2018}
{G{\"o}tberg}, Y., {de Mink}, S.~E., {Groh}, J.~H., {et~al.} 2018, \aap, 615,
  A78

\bibitem[{{Graur} {et~al.}(2017{\natexlab{a}}){Graur}, {Bianco}, {Huang},
  {Modjaz}, {Shivvers}, {Filippenko}, {Li}, \& {Eldridge}}]{Graur+2017a}
{Graur}, O., {Bianco}, F.~B., {Huang}, S., {et~al.} 2017{\natexlab{a}}, \apj,
  837, 120

\bibitem[{{Graur} {et~al.}(2017{\natexlab{b}}){Graur}, {Bianco}, {Modjaz},
  {Shivvers}, {Filippenko}, {Li}, \& {Smith}}]{Graur+2017}
{Graur}, O., {Bianco}, F.~B., {Modjaz}, M., {et~al.} 2017{\natexlab{b}}, \apj,
  837, 121

\bibitem[{Groh {et~al.}(2013)Groh, Meynet, Georgy, \&
  Ekström}]{groh_fundamental_2013}
Groh, J.~H., Meynet, G., Georgy, C., \& Ekström, S. 2013, Astronomy and
  Astrophysics, 558, A131

\bibitem[{Groh {et~al.}(2008)Groh, Oliveira, \& Steiner}]{groh_qwr_2008}
Groh, J.~H., Oliveira, A.~S., \& Steiner, J.~E. 2008, Astronomy and
  Astrophysics, 485, 245

\bibitem[{Habets(1986{\natexlab{a}})}]{habets_evolution_1986}
Habets, G. M. H.~J. 1986{\natexlab{a}}, Astronomy and Astrophysics, 165, 95

\bibitem[{Habets(1986{\natexlab{b}})}]{habets_evolution_1986-1}
Habets, G. M. H.~J. 1986{\natexlab{b}}, Astronomy and Astrophysics, 167, 61

\bibitem[{Han {et~al.}(1994)Han, Podsiadlowski, \&
  Eggleton}]{han_possible_1994}
Han, Z., Podsiadlowski, P., \& Eggleton, P.~P. 1994, Monthly Notices of the
  Royal Astronomical Society, 270, 121

\bibitem[{{Hosseinzadeh} {et~al.}(2019){Hosseinzadeh}, {McCully}, {Zabludoff},
  {Arcavi}, {French}, {Howell}, {Berger}, \&
  {Hiramatsu}}]{Hosseinzadeh_Type_Ibn_2019}
{Hosseinzadeh}, G., {McCully}, C., {Zabludoff}, A.~I., {et~al.} 2019, \apjl,
  871, L9

\bibitem[{{Hunter}(2007)}]{Hunter_matplotlib_2007}
{Hunter}, J.~D. 2007, Computing in Science and Engineering, 9, 90

\bibitem[{Hurley {et~al.}(2000)Hurley, Pols, \&
  Tout}]{hurley_comprehensive_2000}
Hurley, J.~R., Pols, O.~R., \& Tout, C.~A. 2000, Monthly Notices of the Royal
  Astronomical Society, 315, 543

\bibitem[{Hurley {et~al.}(2002)Hurley, Tout, \& Pols}]{hurley_evolution_2002}
Hurley, J.~R., Tout, C.~A., \& Pols, O.~R. 2002, Monthly Notices of the Royal
  Astronomical Society, 329, 897

\bibitem[{Hut(1981)}]{hut_tidal_1981}
Hut, P. 1981, Astronomy and Astrophysics, 99, 126

\bibitem[{{Iglesias} \& {Rogers}(1993)}]{iglesias_radiative_1993}
{Iglesias}, C.~A. \& {Rogers}, F.~J. 1993, \apj, 412, 752

\bibitem[{{Iglesias} \& {Rogers}(1996)}]{iglesias_updated_1996}
{Iglesias}, C.~A. \& {Rogers}, F.~J. 1996, \apj, 464, 943

\bibitem[{{Ivanova} {et~al.}(2003){Ivanova}, {Belczynski}, {Kalogera}, {Rasio},
  \& {Taam}}]{Ivanova_The_Role_2003}
{Ivanova}, N., {Belczynski}, K., {Kalogera}, V., {Rasio}, F.~A., \& {Taam},
  R.~E. 2003, \apj, 592, 475

\bibitem[{Ivanova {et~al.}(2013)Ivanova, Justham, Chen, De~Marco, Fryer,
  Gaburov, Ge, Glebbeek, Han, Li, Lu, Marsh, Podsiadlowski, Potter, Soker,
  Taam, Tauris, van~den Heuvel, \& Webbink}]{ivanova_common_2013}
Ivanova, N., Justham, S., Chen, X., {et~al.} 2013, Astronomy and Astrophysics
  Review, 21, 59

\bibitem[{{Izzard} {et~al.}(2006){Izzard}, {Dray}, {Karakas}, {Lugaro}, \&
  {Tout}}]{Izzard+2006}
{Izzard}, R.~G., {Dray}, L.~M., {Karakas}, A.~I., {Lugaro}, M., \& {Tout},
  C.~A. 2006, \aap, 460, 565

\bibitem[{{Izzard} {et~al.}(2004){Izzard}, {Tout}, {Karakas}, \&
  {Pols}}]{Izzard+2004}
{Izzard}, R.~G., {Tout}, C.~A., {Karakas}, A.~I., \& {Pols}, O.~R. 2004,
  \mnras, 350, 407

\bibitem[{Kippenhahn {et~al.}(1980)Kippenhahn, Ruschenplatt, \&
  Thomas}]{kippenhahn_time_1980}
Kippenhahn, R., Ruschenplatt, G., \& Thomas, H.-C. 1980, Astronomy and
  Astrophysics, 91, 175

\bibitem[{Kippenhahn \& Weigert(1967)}]{kippenhahn_entwicklung_1967}
Kippenhahn, R. \& Weigert, A. 1967, Zeitschrift fur Astrophysik, 65, 251

\bibitem[{Kippenhahn {et~al.}(2012)Kippenhahn, Weigert, \&
  Weiss}]{kippenhahn_stellar_2012}
Kippenhahn, R., Weigert, A., A., \& Weiss, A., A. 2012, Stellar {Structure} and
  {Evolution} - {NASA}/{ADS}, Astronomy and {Astrophysics} {Library}
  (Springer-Verlag Berlin Heidelberg)

\bibitem[{Kleiser {et~al.}(2018)Kleiser, Fuller, \&
  Kasen}]{kleiser_helium_2018}
Kleiser, I., Fuller, J., \& Kasen, D. 2018, Monthly Notices of the Royal
  Astronomical Society, 481, L141

\bibitem[{{Kleiser} \& {Kasen}(2014)}]{Kleiser+2014}
{Kleiser}, I. K.~W. \& {Kasen}, D. 2014, \mnras, 438, 318

\bibitem[{Klencki \& Nelemans(2018)}]{klencki_high_2018}
Klencki, J. \& Nelemans, G. 2018, arXiv e-prints, arXiv:1812.00012

\bibitem[{Kruckow {et~al.}(2018)Kruckow, Tauris, Langer, Kramer, \&
  Izzard}]{kruckow_progenitors_2018}
Kruckow, M.~U., Tauris, T.~M., Langer, N., Kramer, M., \& Izzard, R.~G. 2018,
  Monthly Notices of the Royal Astronomical Society, 481, 1908

\bibitem[{Maeder(1990)}]{maeder_tables_1990}
Maeder, A. 1990, Astronomy and Astrophysics Supplement Series, 84, 139

\bibitem[{{Mokiem} {et~al.}(2007){Mokiem}, {de Koter}, {Vink}, {Puls}, {Evans},
  {Smartt}, {Crowther}, {Herrero}, {Langer}, {Lennon}, {Najarro}, \&
  {Villamariz}}]{Mokiem2007}
{Mokiem}, M.~R., {de Koter}, A., {Vink}, J.~S., {et~al.} 2007, \aap, 473, 603

\bibitem[{{Neijssel} {et~al.}(2019){Neijssel}, {Vigna-G{\'o}mez}, {Stevenson},
  {Barrett}, {Gaebel}, {Broekgaarden}, {de Mink}, {Sz{\'e}csi}, {Vinciguerra},
  \& {Mandel}}]{Neijssel+2019}
{Neijssel}, C.~J., {Vigna-G{\'o}mez}, A., {Stevenson}, S., {et~al.} 2019,
  \mnras, 2457

\bibitem[{{Nomoto} {et~al.}(1993){Nomoto}, {Suzuki}, {Shigeyama}, {Kumagai},
  {Yamaoka}, \& {Saio}}]{Nomoto+_A_type_IIb_1993}
{Nomoto}, K., {Suzuki}, T., {Shigeyama}, T., {et~al.} 1993, \nat, 364, 507

\bibitem[{Nugis \& Lamers(2000)}]{nugis_mass-loss_2000}
Nugis, T. \& Lamers, H. J. G. L.~M. 2000, Astronomy and Astrophysics, 360, 227

\bibitem[{{\"{O}pik}(1924)}]{opik_notitle_1924}
{\"{O}pik}, E. 1924, Tartu Obs. Publ., 25

\bibitem[{Ouchi \& Maeda(2017)}]{ouchi_radii_2017}
Ouchi, R. \& Maeda, K. 2017, The Astrophysical Journal, 840, 90

\bibitem[{Paczyński(1971)}]{paczynski_evolutionary_1971}
Paczyński, B. 1971, Annual Review of Astronomy and Astrophysics, 9, 183

\bibitem[{{Pastorello} {et~al.}(2008){Pastorello}, {Mattila}, {Zampieri},
  {Della Valle}, {Smartt}, {Valenti}, {Agnoletto}, {Benetti}, {Benn}, {Branch},
  {Cappellaro}, {Dennefeld}, {Eldridge}, {Gal-Yam}, {Harutyunyan}, {Hunter},
  {Kjeldsen}, {Lipkin}, {Mazzali}, {Milne}, {Navasardyan}, {Ofek}, {Pian},
  {Shemmer}, {Spiro}, {Stathakis}, {Taubenberger}, {Turatto}, \&
  {Yamaoka}}]{Pastorello_Massive_2008}
{Pastorello}, A., {Mattila}, S., {Zampieri}, L., {et~al.} 2008, \mnras, 389,
  113

\bibitem[{{Pastorello} {et~al.}(2007){Pastorello}, {Smartt}, {Mattila},
  {Eldridge}, {Young}, {Itagaki}, {Yamaoka}, {Navasardyan}, {Valenti}, {Patat},
  {Agnoletto}, {Augusteijn}, {Benetti}, {Cappellaro}, {Boles}, {Bonnet-Bidaud},
  {Botticella}, {Bufano}, {Cao}, {Deng}, {Dennefeld}, {Elias-Rosa},
  {Harutyunyan}, {Keenan}, {Iijima}, {Lorenzi}, {Mazzali}, {Meng}, {Nakano},
  {Nielsen}, {Smoker}, {Stanishev}, {Turatto}, {Xu}, \&
  {Zampieri}}]{Pastorello_A_Giant_2007}
{Pastorello}, A., {Smartt}, S.~J., {Mattila}, S., {et~al.} 2007, \nat, 447, 829

\bibitem[{Paxton {et~al.}(2011)Paxton, Bildsten, Dotter, Herwig, Lesaffre, \&
  Timmes}]{paxton_modules_2011}
Paxton, B., Bildsten, L., Dotter, A., {et~al.} 2011, The Astrophysical Journal
  Supplement Series, 192, 3

\bibitem[{Paxton {et~al.}(2013)Paxton, Cantiello, Arras, Bildsten, Brown,
  Dotter, Mankovich, Montgomery, Stello, Timmes, \&
  Townsend}]{paxton_modules_2013}
Paxton, B., Cantiello, M., Arras, P., {et~al.} 2013, The Astrophysical Journal
  Supplement Series, 208, 4

\bibitem[{Paxton {et~al.}(2015)Paxton, Marchant, Schwab, Bauer, Bildsten,
  Cantiello, Dessart, Farmer, Hu, Langer, Townsend, Townsley, \&
  Timmes}]{paxton_modules_2015}
Paxton, B., Marchant, P., Schwab, J., {et~al.} 2015, The Astrophysical Journal
  Supplement Series, 220, 15

\bibitem[{Paxton {et~al.}(2018)Paxton, Schwab, Bauer, Bildsten, Blinnikov,
  Duffell, Farmer, Goldberg, Marchant, Sorokina, Thoul, Townsend, \&
  Timmes}]{paxton_modules_2018}
Paxton, B., Schwab, J., Bauer, E.~B., {et~al.} 2018, The Astrophysical Journal
  Supplement Series, 234, 34

\bibitem[{Paxton {et~al.}(2019)Paxton, Smolec, Schwab, Gautschy, Bildsten,
  Cantiello, Dotter, Farmer, Goldberg, Jermyn, Kanbur, Marchant, Thoul,
  Townsend, Wolf, Zhang, \& Timmes}]{paxton_modules_2019}
Paxton, B., Smolec, R., Schwab, J., {et~al.} 2019, The Astrophysical Journal
  Supplement Series, 243, 10

\bibitem[{{Perez} \& {Granger}(2007)}]{perez_ipython_2007}
{Perez}, F. \& {Granger}, B.~E. 2007, Computing in Science and Engineering, 9,
  21

\bibitem[{{Piro} \& {Nakar}(2013)}]{piro_what_2013}
{Piro}, A.~L. \& {Nakar}, E. 2013, \apj, 769, 67

\bibitem[{{Podsiadlowski} {et~al.}(1993){Podsiadlowski}, {Hsu}, {Joss}, \&
  {Ross}}]{Podsiadlowski+_The_Progenitor_1993}
{Podsiadlowski}, P., {Hsu}, J.~J.~L., {Joss}, P.~C., \& {Ross}, R.~R. 1993,
  \nat, 364, 509

\bibitem[{Podsiadlowski {et~al.}(1992)Podsiadlowski, Joss, \&
  Hsu}]{podsiadlowski_presupernova_1992}
Podsiadlowski, P., Joss, P.~C., \& Hsu, J. J.~L. 1992, The Astrophysical
  Journal, 391, 246

\bibitem[{Podsiadlowski {et~al.}(2004)Podsiadlowski, Langer, Poelarends,
  Rappaport, Heger, \& Pfahl}]{podsiadlowski_effects_2004}
Podsiadlowski, P., Langer, N., Poelarends, A. J.~T., {et~al.} 2004, The
  Astrophysical Journal, 612, 1044

\bibitem[{Podsiadlowski {et~al.}(2003)Podsiadlowski, Rappaport, \&
  Han}]{podsiadlowski_formation_2003}
Podsiadlowski, P., Rappaport, S., \& Han, Z. 2003, Monthly Notices of the Royal
  Astronomical Society, 341, 385

\bibitem[{Pols {et~al.}(1998)Pols, Schröder, Hurley, Tout, \&
  Eggleton}]{pols_stellar_1998}
Pols, O.~R., Schröder, K.-P., Hurley, J.~R., Tout, C.~A., \& Eggleton, P.~P.
  1998, Monthly Notices of the Royal Astronomical Society, 298, 525

\bibitem[{{Pols} {et~al.}(1995){Pols}, {Tout}, {Eggleton}, \&
  {Han}}]{Pols+1995}
{Pols}, O.~R., {Tout}, C.~A., {Eggleton}, P.~P., \& {Han}, Z. 1995, \mnras,
  274, 964

\bibitem[{Renzo {et~al.}(2017)Renzo, Ott, Shore, \&
  de~Mink}]{renzo_systematic_2017}
Renzo, M., Ott, C.~D., Shore, S.~N., \& de~Mink, S.~E. 2017, Astronomy and
  Astrophysics, 603, A118

\bibitem[{Ritter(1988)}]{ritter_turning_1988}
Ritter, H. 1988, Astronomy and Astrophysics, 202, 93

\bibitem[{Schaller {et~al.}(1992)Schaller, Schaerer, Meynet, \&
  Maeder}]{schaller_new_1992}
Schaller, G., Schaerer, D., Meynet, G., \& Maeder, A. 1992, Astronomy and
  Astrophysics Supplement Series, 96, 269

\bibitem[{{Schawinski} {et~al.}(2008){Schawinski}, {Justham}, {Wolf},
  {Podsiadlowski}, {Sullivan}, {Steenbrugge}, {Bell}, {R{\"o}ser}, {Walker},
  {Astier}, {Balam}, {Balland}, {Carlberg}, {Conley}, {Fouchez}, {Guy},
  {Hardin}, {Hook}, {Howell}, {Pain}, {Perrett}, {Pritchet}, {Regnault}, \&
  {Yi}}]{Schawinski+_Supernova_2008}
{Schawinski}, K., {Justham}, S., {Wolf}, C., {et~al.} 2008, Science, 321, 223

\bibitem[{{Schootemeijer} {et~al.}(2018){Schootemeijer}, {G{\"o}tberg}, {de
  Mink}, {Gies}, \& {Zapartas}}]{schootemeijer_clues_2018}
{Schootemeijer}, A., {G{\"o}tberg}, Y., {de Mink}, S.~E., {Gies}, D., \&
  {Zapartas}, E. 2018, \aap, 615, A30

\bibitem[{Smith(2014)}]{smith_mass_2014}
Smith, N. 2014, Annual Review of Astronomy and Astrophysics, 52, 487

\bibitem[{Spera {et~al.}(2019)Spera, Mapelli, Giacobbo, Trani, Bressan, \&
  Costa}]{spera_merging_2019}
Spera, M., Mapelli, M., Giacobbo, N., {et~al.} 2019, Monthly Notices of the
  Royal Astronomical Society, 485, 889

\bibitem[{{Sravan} {et~al.}(2019){Sravan}, {Marchant}, \&
  {Kalogera}}]{sravan_progenitors_2018}
{Sravan}, N., {Marchant}, P., \& {Kalogera}, V. 2019, \apj, 885, 130

\bibitem[{{Stevenson} {et~al.}(2017){Stevenson}, {Vigna-G{\'o}mez}, {Mandel},
  {Barrett}, {Neijssel}, {Perkins}, \& {de Mink}}]{Stevenson+2017}
{Stevenson}, S., {Vigna-G{\'o}mez}, A., {Mandel}, I., {et~al.} 2017, Nature
  Communications, 8, 14906

\bibitem[{Tauris \& Dewi(2001)}]{tauris_research_2001}
Tauris, T.~M. \& Dewi, J. D.~M. 2001, Astronomy and Astrophysics, 369, 170

\bibitem[{Tauris {et~al.}(2017)Tauris, Kramer, Freire, Wex, Janka, Langer,
  Podsiadlowski, Bozzo, Chaty, Kruckow, Heuvel, J, Antoniadis, Breton, \&
  Champion}]{tauris_formation_2017}
Tauris, T.~M., Kramer, M., Freire, P. C.~C., {et~al.} 2017, The Astrophysical
  Journal, 846, 170

\bibitem[{{Tauris} {et~al.}(2013){Tauris}, {Langer}, {Moriya}, {Podsiadlowski},
  {Yoon}, \& {Blinnikov}}]{Tauris2013}
{Tauris}, T.~M., {Langer}, N., {Moriya}, T.~J., {et~al.} 2013, \apjl, 778, L23

\bibitem[{Tauris {et~al.}(2015)Tauris, Langer, \&
  Podsiadlowski}]{tauris_ultra-stripped_2015}
Tauris, T.~M., Langer, N., \& Podsiadlowski, P. 2015, Monthly Notices of the
  Royal Astronomical Society, 451, 2123

\bibitem[{{Tout} {et~al.}(1996){Tout}, {Pols}, {Eggleton}, \&
  {Han}}]{tout_zero-age_1996}
{Tout}, C.~A., {Pols}, O.~R., {Eggleton}, P.~P., \& {Han}, Z. 1996, \mnras,
  281, 257

\bibitem[{{Trimble} \& {Paczynski}(1973)}]{trimble_helium_1973}
{Trimble}, V. \& {Paczynski}, B. 1973, \aap, 22, 9

\bibitem[{{van den Heuvel} {et~al.}(2017){van den Heuvel}, {Portegies Zwart},
  \& {de Mink}}]{van-den-Heuvel+2017}
{van den Heuvel}, E.~P.~J., {Portegies Zwart}, S.~F., \& {de Mink}, S.~E. 2017,
  \mnras, 471, 4256

\bibitem[{{van der Walt} {et~al.}(2011){van der Walt}, {Colbert}, \&
  {Varoquaux}}]{vanderWalt_numpy_2011}
{van der Walt}, S., {Colbert}, S.~C., \& {Varoquaux}, G. 2011, Computing in
  Science and Engineering, 13, 22

\bibitem[{Vigna-Gómez {et~al.}(2018)Vigna-Gómez, Neijssel, Stevenson,
  Barrett, Belczynski, Justham, de~Mink, Müller, Podsiadlowski, Renzo,
  Szécsi, \& Mandel}]{vigna-gomez_formation_2018}
Vigna-Gómez, A., Neijssel, C.~J., Stevenson, S., {et~al.} 2018, Monthly
  Notices of the Royal Astronomical Society, 481, 4009

\bibitem[{Vink(2017)}]{vink_winds_2017}
Vink, J.~S. 2017, Astronomy and Astrophysics, 607, L8

\bibitem[{{Vink} \& {de Koter}(2005)}]{Vink2005}
{Vink}, J.~S. \& {de Koter}, A. 2005, \aap, 442, 587

\bibitem[{Vink {et~al.}(2001)Vink, de~Koter, \& Lamers}]{vink_mass-loss_2001}
Vink, J.~S., de~Koter, A., \& Lamers, H. J. G. L.~M. 2001, Astronomy and
  Astrophysics, 369, 574

\bibitem[{{Webbink}(1984)}]{Webbink1984}
{Webbink}, R.~F. 1984, \apj, 277, 355

\bibitem[{Yoon(2015)}]{yoon_evolutionary_2015}
Yoon, S.-C. 2015, Publications of the Astronomical Society of Australia, 32,
  e015

\bibitem[{Yoon {et~al.}(2017)Yoon, Dessart, \& Clocchiatti}]{yoon_type_2017}
Yoon, S.-C., Dessart, L., \& Clocchiatti, A. 2017, The Astrophysical Journal,
  840, 10

\bibitem[{{Yoon} {et~al.}(2012){Yoon}, {Gr{\"a}fener}, {Vink}, {Kozyreva}, \&
  {Izzard}}]{Yoon+_detectability_2012}
{Yoon}, S.~C., {Gr{\"a}fener}, G., {Vink}, J.~S., {Kozyreva}, A., \& {Izzard},
  R.~G. 2012, \aap, 544, L11

\bibitem[{Yoon {et~al.}(2010)Yoon, Woosley, \& Langer}]{yoon_type_2010}
Yoon, S.-C., Woosley, S.~E., \& Langer, N. 2010, The Astrophysical Journal,
  725, 940

\bibitem[{Zevin {et~al.}(2019)Zevin, Kremer, Siegel, Coughlin, Tsang, Berry, \&
  Kalogera}]{zevin_can_2019}
Zevin, M., Kremer, K., Siegel, D.~M., {et~al.} 2019, arXiv:1906.11299
  [astro-ph], arXiv: 1906.11299

\end{thebibliography}

\begin{appendix}

\section{Parameters of the models}
We provide relevant parameters of our models at solar and low metallicity in Table \ref{tab:zsun} and \ref{tab:zlow}, respectively. Definitions for the symbols used in both tables are given in Table \ref{tab:def}. 

\begin{table*}[ht!]
\caption{Parameter definitions for tables \ref{tab:zsun} and \ref{tab:zlow}}
\label{tab:def}
\centering
\begin{tabular}{L{2.5cm}L{1.5cm}L{8cm}}
\hline
\hline
Parameter                          & Unit                        & Definition                                                                                                               \\
\hline
$M_{\rm{i}}$                     & $(\rm{M}_{\odot})$          & Initial total ZAMS mass                                                                                                  \\
$M_{\rm{a}}$                      & $(\rm{M}_{\odot})$          &  Total mass after the end of mass-transfer, when the nuclear helium-burning luminosity exceeds 85\% of the nuclear luminosity                   \\
$M_{\rm{b}}$                   & $(\rm{M}_{\odot})$          & Total mass at core helium depletion, when the mass fraction of helium reaches values smaller than $10^{-4}$             \\
$M_{\rm{f}}$                       & $(\rm{M}_{\odot})$          & Total mass at the end of the simulation when the mass fraction of carbon drops below $10^{-4}$                                                         \\
$M_{\rm{CO}}^{\rm{f}}$             & $(\rm{M}_{\odot})$          & Mass of the carbon/oxygen core at the end of the simulation                                                              \\
$M_{\rm{H}}^{\rm{a}}$             & $(10^{-2}\rm{M}_{\odot})$   & Total hydrogen mass after the end of mass-transfer                                                                       \\
$M_{\rm{H}}^{\rm{f}}$              & $(10^{-2}\rm{M}_{\odot})$   & Total hydrogen mass at the end of the simulation                                                                         \\
$R_{\rm{min}}$                         & $(\rm{R}_{\odot})$          & Minimum radius                                                                                                           \\
$R_{\rm{b}}$                   & $(\rm{R}_{\odot})$          & Radius at core helium depletion                                                                                          \\
$R_{\rm{p}}$      & $(\rm{R}_{\odot})$ & For low metallicity models, radius at the peak of the first radius expansion \\
$\Delta t_{\rm{p}}$  & (kyr) & For low metallicity models, difference between the final stellar age and the time of the peak of the first radius expansion\\
$R_{\rm{C\,ign.}}$                     & $(\rm{R}_{\odot})$          & Radius at core carbon ignition, when the luminosity of carbon exceeds 98\% of the nuclear luminosity                     \\
$R_{\rm{max}}$                         & $(\rm{R}_{\odot})$          & Maximum radius                                                                                                           \\
$R_{\rm{f}}$                       & $(\rm{R}_{\odot})$          & Radius at the end of the simulation                                                                                      \\
$T_{\rm{eff}}^{\rm{f}}$ &       (K)            & Effective temperature at the end of the simulation                                                                       \\
$L_{\rm{f}}$            & $(\rm{L}_{\odot})$          & Luminosity at the end of the simulation                                                                                  \\
$g_{\rm{f}}$            & $(\rm{cm}\,\rm{s}^{-2})$ & Surface gravity at the end of the simulation                                                                  \\
$\rm{S}_{\rm{f}}$                     &                             & Approximate final stellar type derived based on the effective temperature (see section \ref{sec:grid_HRD})\\
\hline
\end{tabular}
\end{table*}

\begin{sidewaystable*}[ht!]
\caption{Parameters of the solar metallicity (Z=0.0142) models.}
\label{tab:zsun}
\centering
\begin{tabular}{llllllllllllllll}
\hline
\hline
$M_{\rm{i}}/$ & $M_{\rm{a}}/$  & $M_{\rm{b}}/$ & $M_{\rm{f}}/$   & $M_{\rm{CO}}^{\rm{f}}/$ & $M_{\rm{H}}^{\rm{a}}/$ & $M_{\rm{H}}^{\rm{f}}/$ & $R_{\rm{min}}/$     & $R_{\rm{b}}/$ & $R_{\rm{C\,ign.}}/$ & $R_{\rm{max}}/$     & $R_{\rm{f}}/$   & $\log_{10}(T_{\rm{eff}}^{\rm{f}} / \rm{K})$ & $\log_{10}(L_{\rm{f}} / \rm{L}_{\odot})$ & $\log_{10}(\frac{g_{\rm{f}}}{[\rm{cm}\cdot\rm{s}^{-2}]})$ & $\rm{S}_{f}$ \\
$(\rm{M}_{\odot})$ & $(\rm{M}_{\odot})$ & $(\rm{M}_{\odot})$  & $(\rm{M}_{\odot})$ & $(\rm{M}_{\odot})$         & $(10^{-2}\rm{M}_{\odot})$  & $(10^{-2}\rm{M}_{\odot})$ & $(\rm{R}_{\odot})$ & $(\rm{R}_{\odot})$  & $(\rm{R}_{\odot})$ & $(\rm{R}_{\odot})$ & $(\rm{R}_{\odot})$ &                           &          &  &                    \\
\hline
8.86               & 2.51               & 2.39                & 2.35               & 1.40                       & 3.06                       & 0.003                     & 0.404              & 0.416               & 4.64               & 186                & 162                & 3.77                                   & 4.45                        & 0.391                       & YSG                \\
9.08               & 2.59               & 2.47                & 2.42               & 1.45                       & 3.13                       & 0.001                     & 0.401              & 0.414               & 3.74               & 176                & 169                & 3.77                                   & 4.49                        & 0.364                       & YSG                \\
9.30               & 2.68               & 2.55                & 2.50               & 1.49                       & 3.26                       & 7.98e-05                  & 0.397              & 0.410               & 3.22               & 155                & 153                & 3.79                                   & 4.50                        & 0.468                       & YSG                \\
9.52               & 2.77               & 2.63                & 2.57               & 1.53                       & 3.37                       & 4.78e-06                  & 0.389              & 0.403               & 2.91               & 135                & 133                & 3.83                                   & 4.53                        & 0.598                       & YSG                \\
9.75               & 2.87               & 2.71                & 2.65               & 1.58                       & 3.46                       & 2.79e-07                  & 0.384              & 0.399               & 2.69               & 65.9               & 65.9               & 4.00                                   & 4.59                        & 1.22                        & BSG                \\
9.99               & 2.97               & 2.80                & 2.74               & 1.63                       & 3.57                       & 8.80e-09                  & 0.380              & 0.396               & 2.51               & 31.1               & 31.1               & 4.17                                   & 4.62                        & 1.89                        & BSG                \\
10.2               & 3.07               & 2.88                & 2.83               & 1.69                       & 3.69                       & 1.42e-10                  & 0.380              & 0.397               & 2.37               & 20.7               & 20.5               & 4.27                                   & 4.65                        & 2.27                        & BSG                \\
10.5               & 3.19               & 2.98                & 2.92               & 1.75                       & 3.84                       & 4.00e-13                  & 0.384              & 0.402               & 2.26               & 15.4               & 15.4               & 4.34                                   & 4.69                        & 2.53                        & BSG                \\
10.7               & 3.30               & 3.07                & 3.02               & 1.82                       & 3.94                       & 1.45e-16                  & 0.390              & 0.409               & 2.16               & 12.7               & 12.7               & 4.39                                   & 4.72                        & 2.71                        & BSG                \\
11.0               & 3.43               & 3.17                & 3.12               & 1.88                       & 4.02                       & 6.61e-20                  & 0.396              & 0.417               & 2.07               & 10.4               & 10.4               & 4.44                                   & 4.74                        & 2.90                        & BSG                \\
11.3               & 3.55               & 3.28                & 3.22               & 1.95                       & 4.14                       & 6.05e-25                  & 0.402              & 0.425               & 2.00               & 8.87               & 8.87               & 4.48                                   & 4.76                        & 3.05                        & BSG                \\
11.5               & 3.68               & 3.38                & 3.32               & 2.02                       & 4.29                       & 1.35e-22                  & 0.408              & 0.433               & 1.92               & 7.77               & 7.64               & 4.51                                   & 4.76                        & 3.19                        & WR                 \\
11.8               & 3.81               & 3.48                & 3.43               & 2.10                       & 4.43                       & 1.52e-22                  & 0.414              & 0.441               & 1.87               & 7.19               & 7.12               & 4.54                                   & 4.81                        & 3.27                        & WR                 \\
12.1               & 3.95               & 3.59                & 3.54               & 2.19                       & 4.56                       & 2.55e-21                  & 0.419              & 0.449               & 1.80               & 6.83               & 6.80               & 4.56                                   & 4.84                        & 3.32                        & WR                 \\
12.4               & 4.10               & 3.70                & 3.65               & 2.27                       & 4.66                       & 1.16e-23                  & 0.424              & 0.457               & 1.76               & 6.31               & 6.30               & 4.58                                   & 4.86                        & 3.40                        & WR                 \\
12.7               & 4.24               & 3.82                & 3.76               & 2.36                       & 4.80                       & 1.4e-24                   & 0.429              & 0.465               & 1.69               & 5.89               & 5.83               & 4.60                                   & 4.88                        & 3.48                        & WR                 \\
13.0               & 4.39               & 3.93                & 3.88               & 2.45                       & 4.93                       & 1.37e-22                  & 0.434              & 0.473               & 1.64               & 5.46               & 5.41               & 4.62                                   & 4.90                        & 3.56                        & WR                 \\
13.3               & 4.55               & 4.05                & 4.00               & 2.55                       & 5.04                       & 1.11e-24                  & 0.439              & 0.481               & 1.59               & 5.07               & 5.07               & 4.64                                   & 4.93                        & 3.63                        & WR                 \\
13.6               & 4.71               & 4.17                & 4.12               & 2.65                       & 5.18                       & 1.74e-22                  & 0.443              & 0.489               & 1.55               & 4.62               & 4.62               & 4.67                                   & 4.95                        & 3.72                        & WR                 \\
14.0               & 4.88               & 4.29                & 4.24               & 2.75                       & 5.30                       & 1.15e-22                  & 0.448              & 0.497               & 1.51               & 4.18               & 4.18               & 4.69                                   & 4.97                        & 3.82                        & WR                 \\
14.3               & 5.05               & 4.42                & 4.37               & 2.85                       & 5.39                       & 5.09e-31                  & 0.451              & 0.505               & 1.46               & 3.97               & 3.79               & 4.72                                   & 5.00                        & 3.92                        & WR                 \\
14.6               & 5.23               & 4.54                & 4.49               & 2.97                       & 5.54                       & 9.49e-32                  & 0.455              & 0.513               & 1.41               & 4.43               & 3.44               & 4.75                                   & 5.02                        & 4.02                        & WR                 \\
15.0               & 5.41               & 4.67                & 4.62               & 3.09                       & 5.65                       & 3.3e-28                   & 0.458              & 0.521               & 1.35               & 3.26               & 3.13               & 4.77                                   & 5.04                        & 4.11                        & WR\\
\hline
\end{tabular}
\end{sidewaystable*}

\begin{sidewaystable*}[ht!]
\caption{Parameters of the low metallicity (Z=0.001) models.}
\label{tab:zlow}
\centering
\begin{tabular}{llllllllllllllllll}
\hline
\hline
$M_{\rm{i}}/$ & $M_{\rm{a}}/$  & $M_{\rm{b}}/$ & $M_{\rm{f}}/$   & $M_{\rm{CO}}^{\rm{f}}/$ & $M_{\rm{H}}^{\rm{a}}/$ & $M_{\rm{H}}^{\rm{f}}/$ & $R_{\rm{min}}/$     & $R_{\rm{b}}/$ & $R_{\rm{p}}/$    & $\Delta t_{\rm{p}}/$ & $R_{\rm{C\,ign.}}/$ & $R_{\rm{max}}/$     & $R_{\rm{f}}/$   & $\log_{10}(\frac{T_{\rm{eff}}^{\rm{f}}}{\rm{K}})$ & $\log_{10}(\frac{L_{\rm{f}}}{\rm{L}_{\odot}})$ & $\log_{10}(\frac{g_{\rm{f}}}{[\rm{cm}\cdot\rm{s}^{-2}]})$ & $\rm{S}_{f}$ \\
$(\rm{M}_{\odot})$ & $(\rm{M}_{\odot})$ & $(\rm{M}_{\odot})$  & $(\rm{M}_{\odot})$ & $(\rm{M}_{\odot})$         & $(10^{-2}\rm{M}_{\odot})$  & $(10^{-2}\rm{M}_{\odot})$ & $(\rm{R}_{\odot})$ & $(\rm{R}_{\odot})$  & $(\rm{R}_{\odot})$ & (kyr)                    & $(\rm{R}_{\odot})$ & $(\rm{R}_{\odot})$ & $(\rm{R}_{\odot})$ &                          &          &  &                    \\
\hline
8.86               & 3.11               & 3.10                 & 3.08               & 1.70                        & 13.6                       & 9.80                       & 1.38               & 2.72                & 4.79               & 68.3                   & 391                & 439                & 432                & 3.61                                   & 4.67                        & -0.345                      & RSG                \\
9.08               & 3.22               & 3.21                & 3.18               & 1.76                       & 14.2                       & 10.2                      & 1.43               & 2.83                & 5.2                & 61.9                   & 402                & 451                & 445                & 3.61                                   & 4.70                         & -0.357                      & RSG                \\
9.30                & 3.32               & 3.32                & 3.29               & 1.82                       & 14.6                       & 10.5                      & 1.47               & 2.93                & 5.58               & 56.6                   & 413                & 462                & 456                & 3.61                                   & 4.72                        & -0.364                      & RSG                \\
9.52               & 3.43               & 3.43                & 3.4                & 1.88                       & 15.1                       & 11.0                        & 1.52               & 3.06                & 6.08               & 51.5                   & 426                & 475                & 463                & 3.61                                   & 4.74                        & -0.361                      & RSG                \\
9.75               & 3.55               & 3.54                & 3.51               & 1.95                       & 15.7                       & 11.4                      & 1.57               & 3.20                 & 6.68               & 47.1                   & 437                & 488                & 471                & 3.61                                   & 4.75                        & -0.363                      & RSG                \\
9.99               & 3.66               & 3.66                & 3.63               & 2.02                       & 16.2                       & 11.7                      & 1.62               & 3.29                & 7.23               & 43.0                     & 450                & 501                & 487                & 3.61                                   & 4.78                        & -0.377                      & RSG                \\
10.2               & 3.79               & 3.78                & 3.75               & 2.10                        & 16.8                       & 12.2                      & 1.68               & 3.46                & 8.05               & 39.2                   & 462                & 515                & 507                & 3.61                                   & 4.82                        & -0.397                      & RSG                \\
10.5               & 3.91               & 3.90                 & 3.88               & 2.18                       & 17.3                       & 12.6                      & 1.73               & 3.61                & 8.95               & 36.1                   & 475                & 530                & 515                & 3.61                                   & 4.83                        & -0.397                      & RSG                \\
10.7               & 4.04               & 4.03                & 4.00                 & 2.27                       & 17.8                       & 13.0                        & 1.78               & 3.72                & 9.83               & 33.0                     & 488                & 543                & 536                & 3.61                                   & 4.87                        & -0.417                      & RSG                \\
11.0                 & 4.17               & 4.16                & 4.14               & 2.36                       & 18.5                       & 13.4                      & 1.83               & 3.87                & 11.1               & 30.3                   & 502                & 555                & 551                & 3.61                                   & 4.89                        & -0.427                      & RSG                \\
11.3               & 4.32               & 4.31                & 4.28               & 2.47                       & 19.1                       & 13.9                      & 1.90                & 4.05                & 12.8               & 27.7                   & 516                & 567                & 567                & 3.61                                   & 4.92                        & -0.439                      & RSG                \\
11.5               & 4.46               & 4.45                & 4.42               & 2.57                       & 19.8                       & 14.4                      & 1.95               & 4.21                & 14.8               & 25.5                   & 529                & 579                & 579                & 3.62                                   & 4.94                        & -0.442                      & RSG                \\
11.8               & 4.61               & 4.60                 & 4.57               & 2.68                       & 20.4                       & 14.8                      & 2.01               & 4.40                 & 17.5               & 23.4                   & 543                & 590                & 590                & 3.62                                   & 4.96                        & -0.444                      & RSG                \\
12.1               & 4.76               & 4.74                & 4.72               & 2.79                       & 21.1                       & 15.2                      & 2.07               & 4.54                & 20.4               & 21.7                   & 556                & 601                & 601                & 3.62                                   & 4.98                        & -0.446                      & RSG                \\
12.4               & 4.92               & 4.91                & 4.88               & 2.9                        & 21.9                       & 15.7                      & 2.13               & 4.74                & 25.5               & 19.9                   & 570                & 612                & 612                & 3.62                                   & 5.00                           & -0.447                      & RSG                \\
12.7               & 5.09               & 5.07                & 5.04               & 3.04                       & 22.6                       & 16.2                      & 2.20               & 4.95                & 32.7               & 18.3                   & 584                & 624                & 624                & 3.62                                   & 5.02                        & -0.449                      & RSG                \\
13.0                 & 5.25               & 5.23                & 5.20                & 3.16                       & 23.4                       & 16.7                      & 2.26               & 5.13                & 43.4               & 17.0                     & 598                & 635                & 635                & 3.62                                   & 5.04                        & -0.451                      & RSG                \\
13.3               & 5.43               & 5.41                & 5.38               & 3.3                        & 24.3                       & 17.0                        & 2.32               & 5.35                & 66.1               & 15.6                   & 610                & 645                & 645                & 3.62                                   & 5.06                        & -0.451                      & RSG                \\
13.6               & 5.61               & 5.59                & 5.56               & 3.44                       & 25.0                         & 17.4                      & 2.38               & 5.56                & 113                & 14.4                   & 623                & 656                & 656                & 3.62                                   & 5.08                        & -0.451                      & RSG                \\
14.0                 & 5.80                & 5.77                & 5.74               & 3.58                       & 25.9                       & 17.9                      & 2.46               & 5.80                 & 238                & 13.3                   & 635                & 668                & 667                & 3.62                                   & 5.10                         & -0.452                      & RSG                \\
14.3               & 5.98               & 5.95                & 5.92               & 3.73                       & 26.4                       & 18.0                        & 2.46               & 5.77                & 312                & 12.4                   & 644                & 676                & 676                & 3.63                                   & 5.12                        & -0.449                      & RSG                \\
14.6               & 6.14               & 6.11                & 6.07               & 3.87                       & 26.4                       & 17.6                      & 2.41               & 5.47                & 311                & 11.5                   & 644                & 678                & 678                & 3.63                                   & 5.14                        & -0.441                      & RSG                \\
15.0                 & 6.30                & 6.27                & 6.24               & 4.02                       & 26.0                         & 17.1                      & 2.28               & 4.98                & 232                & 10.7                   & 638                & 678                & 677                & 3.63                                   & 5.15                        & -0.428                      & RSG               \\
\hline
\end{tabular}
\end{sidewaystable*}


\section{Effects of winds on the expansion of stripped stars at low metallicity}
\label{sec:appendix_winds}
In Sections \ref{sec:ev_f_g} and \ref{sec:grid_radius} we discuss the expansion of stripped stars. At low metallicity, the models retain a hydrogen-rich layer. The mass of that layer is linked to the maximum radius each star can achieve. However, the mass of the hydrogen-rich envelope retained by a stripped star is not only determined by the binary interaction, but can also be affected by wind mass loss. The wind mass-loss rates for stripped stars are not well known \citep{yoon_evolutionary_2015, gotberg_ionizing_2017,gotberg_spectral_2018,gilkis_effects_2019}. In most stellar evolution models for stripped stars, winds are typically assumed to follow the empirically-derived prescription for Wolf-Rayet stars from \citet{nugis_mass-loss_2000}. At low metallicity, the effect of winds is expected to be limited due to the metallicity dependence of line-driven winds \citep{Vink2005, Mokiem2007}. This expectation affects predictions for the mass of the leftover hydrogen-rich layer at low metallicity, and consequently the maximum radii of stripped-star models. We discuss uncertainties introduced by this assumption below, showing comparisons for both our example model from Section \ref{sec:ev} and the most massive model from our grid.\\

\begin{figure}[ht]
\centering
\includegraphics[width=0.5\textwidth]{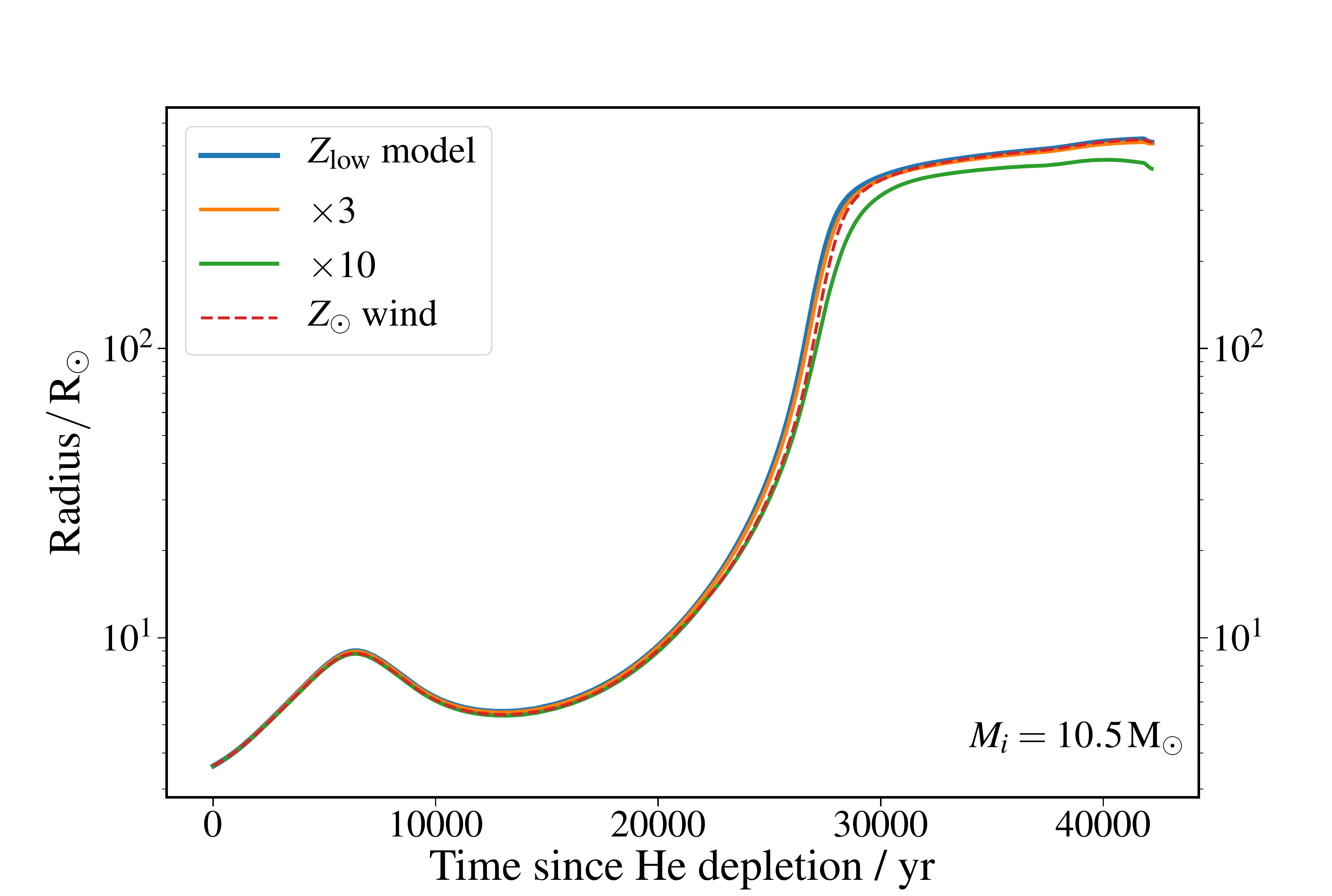}
	\caption{Radius evolution of low-metallicity stripped-star models with an initial mass of 10.5 \Msun as a function of time after core helium depletion. The blue curve represents a model computed with our default wind scheme, while the orange and green curves are for models with mass-loss rates 3 and 10 times higher than the default, respectively. The dashed red curve represents a low-metallicity model computed with the solar metallicity mass-loss scheme.}
	\label{fig:wind_test_10_radius_ev}
\end{figure}
\begin{figure}[ht]
\centering
\includegraphics[width=0.5\textwidth]{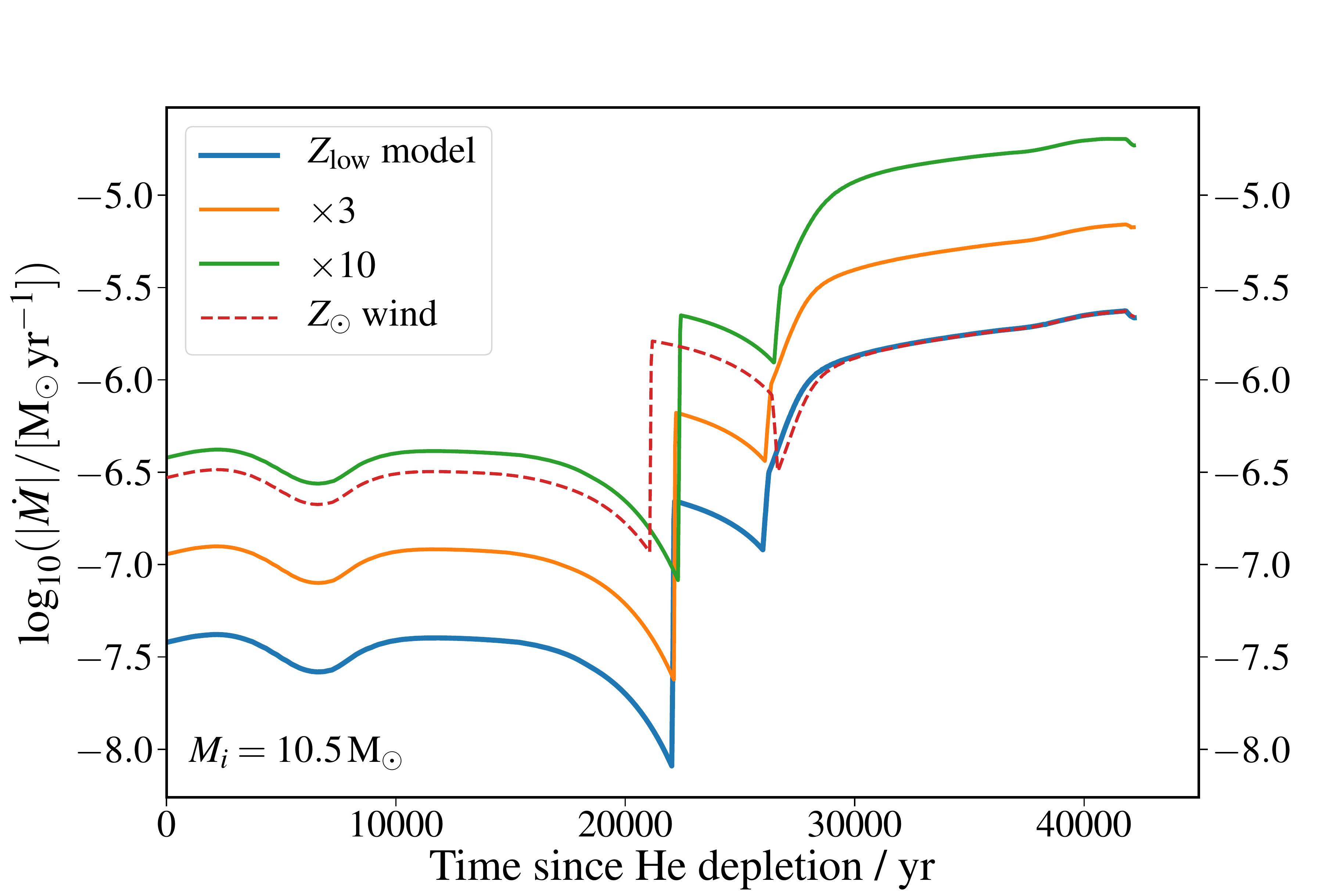}
	\caption{Evolution of the wind-mass loss rate for the low-metallicity 10.5 \Msun stripped-star model as a function of time after core helium depletion. The line styles are associated with the same models as in Fig. \ref{fig:wind_test_10_radius_ev}.}
	\label{fig:wind_test_10_mdot_ev}
\end{figure}

Figures \ref{fig:wind_test_10_radius_ev} and \ref{fig:wind_test_10_mdot_ev} demonstrate the effects of increasing our assumed wind-loss rates on our exemplary model, with an initial mass of 10.5 \Msun.  Figure \ref{fig:wind_test_10_radius_ev} shows the evolution of the stellar radius after core helium burning as a function of time. We compare our default model to two models for which the mass-loss rate is increased by constant factors of three and ten. We also present a model with the mass-loss scheme behaving as if the star was at solar metallicity. For the first 20,000 years of the radius evolution, no significant impact of the wind mass-loss rate can be observed. Differences appear towards the end of the evolution. The model with a factor of ten higher mass-loss rate ends its evolution with a smaller radius, of 417\Rsun, compared to the 515\Rsun of the model with our default wind mass-loss scheme.  The model with the solar-metallicity wind mass-loss scheme ends its life with a radius of 516\Rsun, i.e., very similar to the default model. 

Figure \ref{fig:wind_test_10_mdot_ev} shows the evolution of the wind mass-loss rates of the models in Fig. \ref{fig:wind_test_10_radius_ev}. For all models, the dominant mass loss is late in the evolution, when the stars have become giants. The model with a solar-metallicity wind mass-loss rate closely follows the ten times higher mass-loss rate for the first 20,000 yr. This is due to the metallicity dependence of the \citet{vink_mass-loss_2001} mass-loss rate, which is $(Z)^{0.85}$, so approximately ten times higher for solar metallicity. The sharp changes in the mass-loss rates can be attributed to the bistability jumps of the winds. Because the temperature at which these take place is metallicity-dependent \citep{vink_mass-loss_2001}, they occur earlier for the solar metallicity model. Towards the end of the evolution, the model assuming solar-metallicity winds follows the same mass-loss rate as our default model, because we assume these cool-star winds to be metallicity-independent. However, even if our model predictions had been close to the green curve our conclusions would have remained qualitatively unchanged. 

\begin{figure}[ht]
\centering
\includegraphics[width=0.5\textwidth]{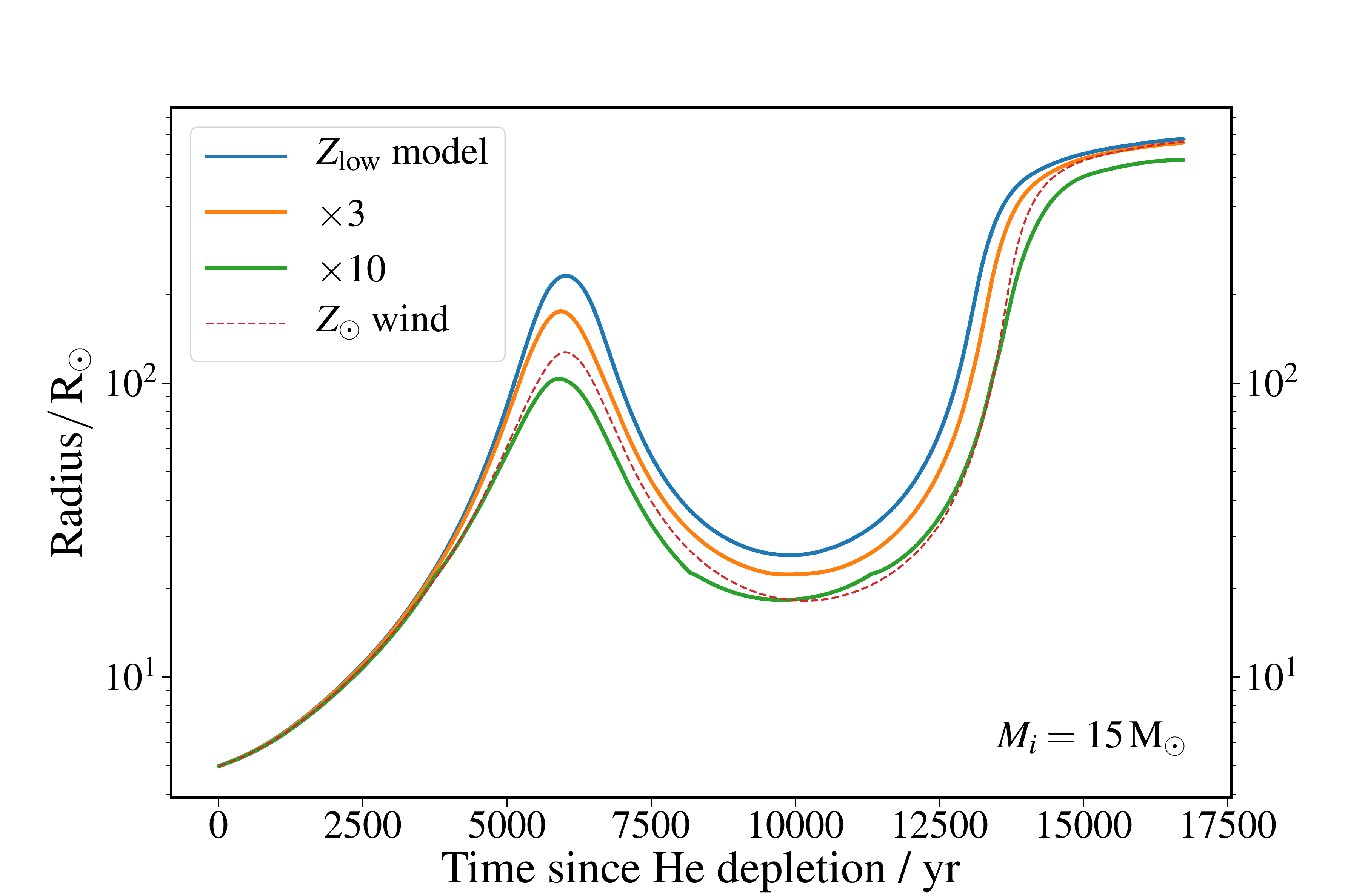}
	\caption{Same as Fig. \ref{fig:wind_test_10_radius_ev} for the low-metallicity model with an initial mass of 15\Msun.}
	\label{fig:wind_test_15_radius_ev}
\end{figure}

Since the effect of wind-mass loss is mass-dependent, we also investigate the impact of increased wind mass-loss rates on the highest-mass model in our grid, which has an initial mass of 15\Msun. We present the results of this test in Fig. \ref{fig:wind_test_15_radius_ev}. The impact of changing these assumptions on the radius evolution is more pronounced throughout the evolution than for the 10.5\Msun model, but the differences in final radii are still not enough to affect our qualitative conclusions.   The largest relative differences can be observed at the peak of the first radius expansion phase, with radii of 232, 175, 104\Rsun for the default wind assumptions, three-times higher, and ten-times higher mass-loss rates, respectively. However, the predicted radius at the end of the evolution is less affected. The model with a ten-times higher wind mass-loss rate has a final radius only 15\% lower than the default model (and only 3\% lower than the default model for the three-times higher wind mass-loss rate variation). Again we also show the radius evolution of a model assuming solar-metallicity winds. The majority of the evolution again follows the same trend as the model with a ten-times higher wind mass-loss rate. At the peak of the first radius expansion, the solar metallicity wind model has a radius of 127 \Rsun. However the final radius is only 2\% smaller than that of the default mass loss model. 

From these tests we conclude that the effect of reasonable uncertainties in wind mass loss on the maximum radius these stars are predicted to reach is very small.

\section{Effects of the orbital period on the expansion of stripped stars at low metallicity}
\label{sec:appendix_porb}
In Appendix \ref{sec:appendix_winds}, we demonstrate that plausible ranges of stellar winds have only a very small impact on the expansion of stars stripped in binaries at low metallicity. The binary interaction is mainly responsible for determining the mass of the remaining hydrogen layer at low metallicity. \citet{yoon_type_2017} explored how the choice of orbital separation has a large impact on the final effective temperature and radius of stripped stars at low metallicity, for stars that retain a hydrogen layer with a mass that exceeds 0.15\Msun. In this section we discuss how our choice of the orbital separation at low metallicity affects our results. We first show results for models with one initial stellar mass, which are stripped in a binary computed at multiple orbital separations. We then compare the maximum radii obtained for the grid of models presented in the main text with two grids at alternative orbital separations.

\begin{figure}[ht]
\centering
\includegraphics[width=0.5\textwidth]{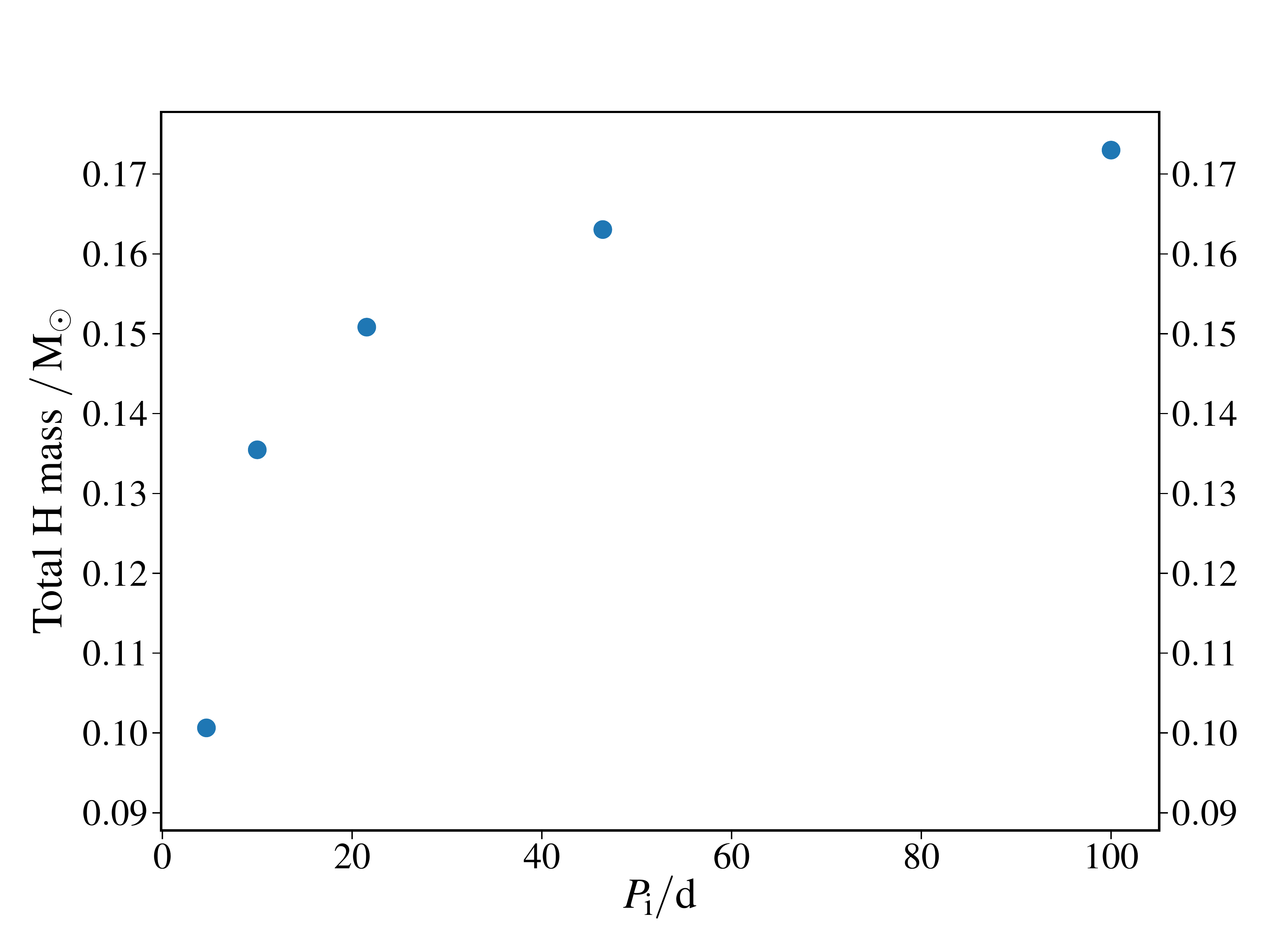}
	\caption{Total hydrogen mass at the moment of core helium depletion for low-metallicity models with an initial mass of 11.3\Msun, shown as a function of initial orbital period, $P_{i}$.}
	\label{fig:porb_test_11_total_h_mass}
\end{figure}

\begin{figure}[ht]
\centering
\includegraphics[width=0.5\textwidth]{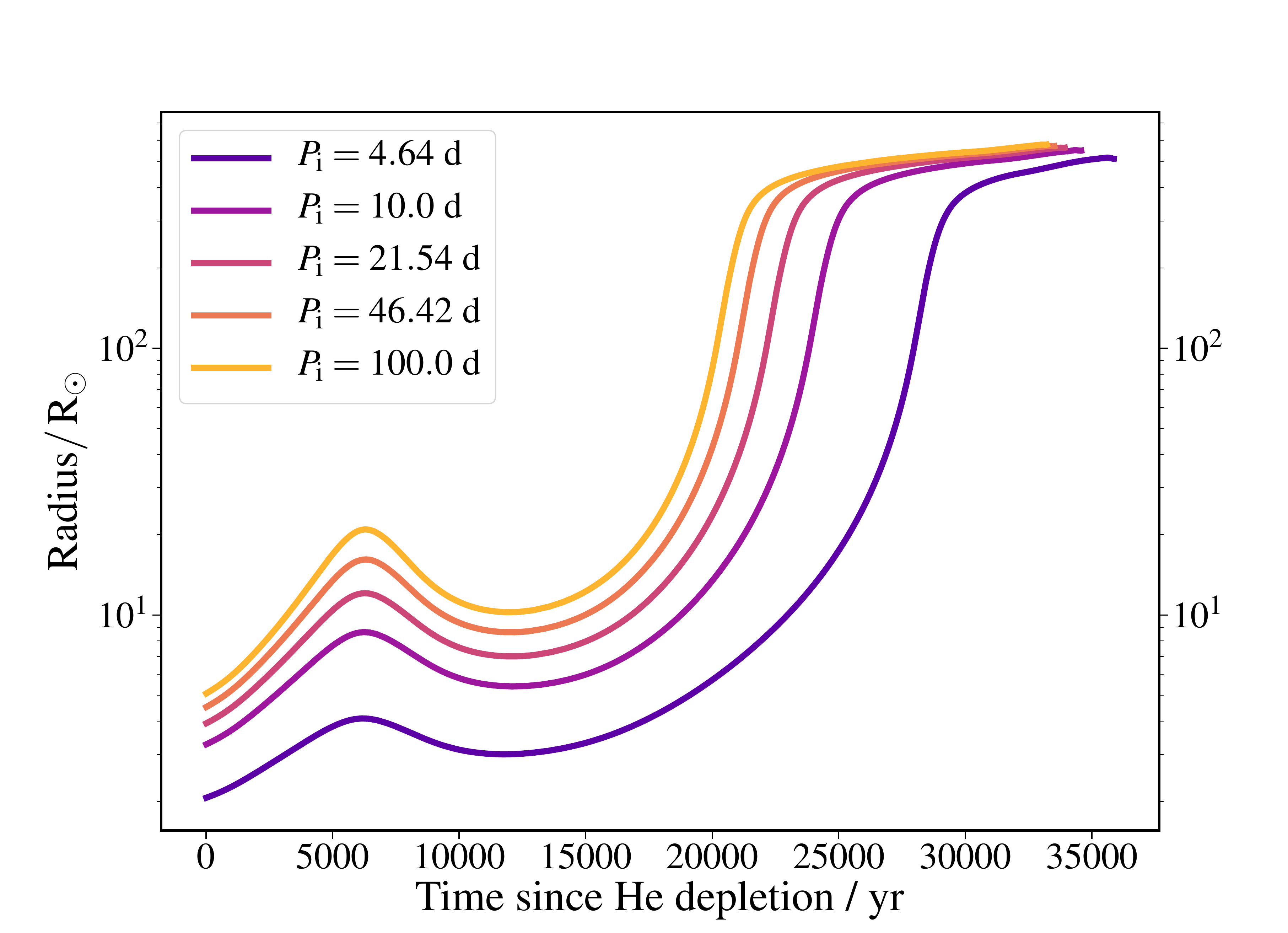}
	\caption{Radius evolution of low-metallicity stripped star models with an initial mass of 11.3\Msun as a function of time after core helium depletion. Colors indicate the initial orbital period as specified in the legend.}
	\label{fig:porb_test_11_radius_ev}
\end{figure}

\begin{figure}[ht]
\centering
\includegraphics[width=0.5\textwidth]{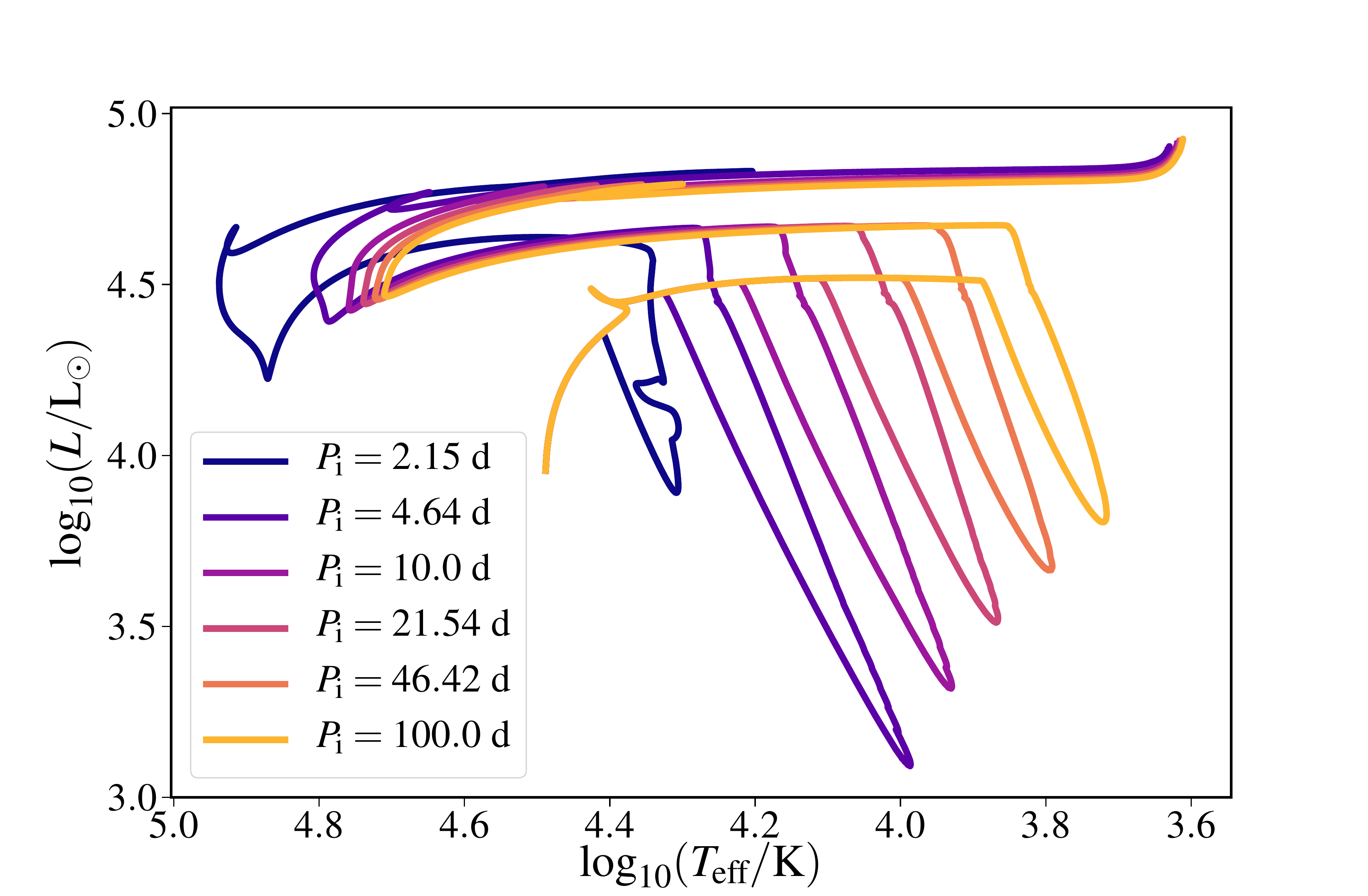}
	\caption{Full evolution of models with an initial mass of 11.3\Msun with varying initial orbital periods on the Hertzsprung-Russell diagram. Colors indicate the initial orbital period as specified in the legend. All models with an initial orbital period longer than three days interact with their companions after leaving the main-sequence (case B mass-transfer).}
	\label{fig:porb_test_11_hrd}
\end{figure}

Figs.\,\ref{fig:porb_test_11_total_h_mass} and \ref{fig:porb_test_11_radius_ev} show the impact of the initial orbital separation on the remaining hydrogen mass and radius evolution, for stellar models with the same initial mass of 11.3\Msun. These results are from a set of calculations with logarithmically-spaced orbital periods ranging from 1 d to 100 d. We did not find converged solutions for the model with an initial orbital period of 1 d. We plot the full evolution of all models on the Hertzsprung-Russell diagram in Fig. \ref{fig:porb_test_11_hrd}. The model with the shortest orbital period already interacts with its companion during the main-sequence (case A mass transfer), which is why it has a distinct evolutionary track. We focus on the case B mass-transfer models and thus do not discuss this model below. In Fig. \ref{fig:porb_test_11_total_h_mass}, we display the total hydrogen mass at core helium depletion as a function of the initial orbital period. As expected, the mass of the hydrogen-rich layer increases for longer orbital periods, but it is still within a factor of two.

Fig. \ref{fig:porb_test_11_radius_ev} demonstrates the consequent effect of the remaining hydrogen-rich layer on the radius evolution of these low-metallicity stripped stars after core helium depletion (cf.\  Fig.\ \ref{fig:Rad_ev}). The models display a parallel radius evolution, where models with longer orbital periods have larger radii overall. However, all models reach similar final radii (between 514 and 580 \Rsun).

\begin{figure}[ht]
\centering
\includegraphics[width=0.5\textwidth]{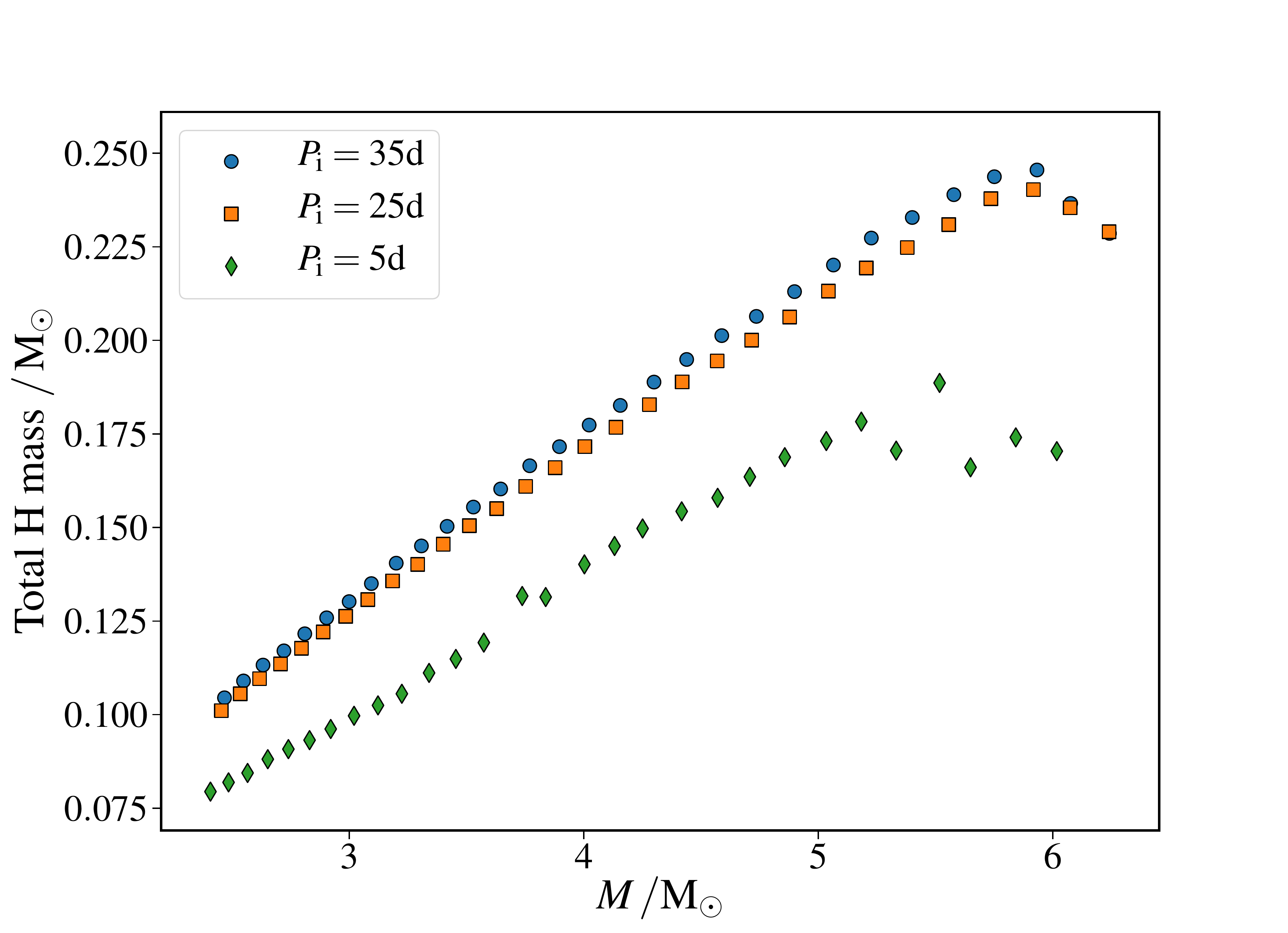}
	\caption{Total hydrogen mass at the moment of core helium depletion as a function of stellar mass. Green diamonds, orange squares, and blue circles indicate grids computed with fixed orbital periods of 5, 25, and 35 d, respectively.}
	\label{fig:porb_test_grids_radius_ev}
\end{figure}

\begin{figure}[ht]
\centering
\includegraphics[width=0.5\textwidth]{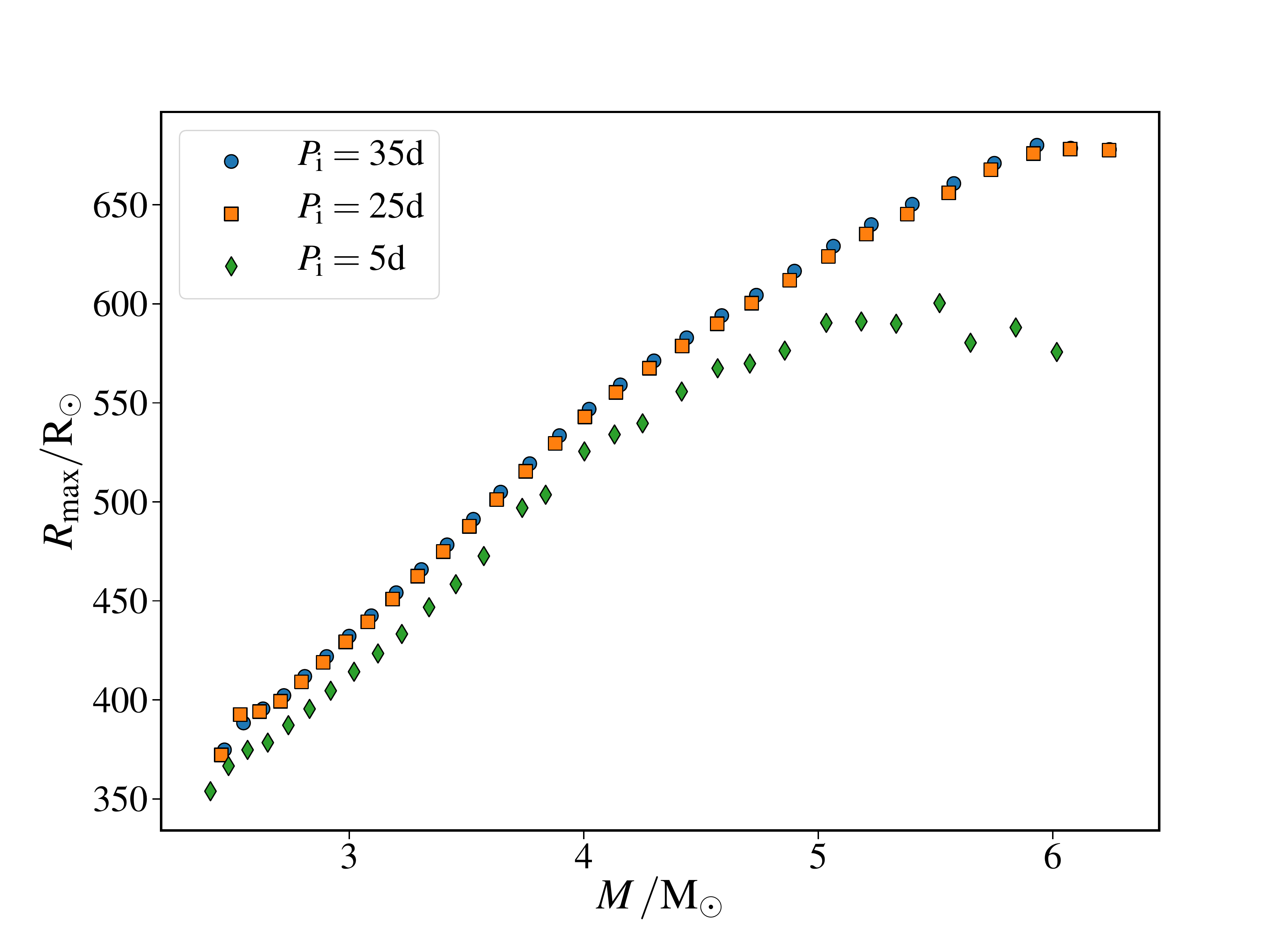}
	\caption{Maximum radius as a function of mass for the same models as in Fig. \ref{fig:porb_test_grids_total_h_mass}.}
	\label{fig:porb_test_grids_total_h_mass}
\end{figure}

Because the radius evolution also depends on the total stellar mass, we compare the maximum radii of our default model grid at low metallicity with two model grids computed with initial orbital periods of 5 d and 35 d, respectively. 
Results from these grids are displayed in Figs.\,\ref{fig:porb_test_grids_radius_ev} and \ref{fig:porb_test_grids_total_h_mass}. (For comparison with Figs.\,\ref{fig:porb_test_11_total_h_mass} and \ref{fig:porb_test_11_radius_ev}, we note that the post-stripping masses from the models shown in those figures ranges from 4 to 4.35\Msun.)  Fig.\,\ref{fig:porb_test_grids_total_h_mass} presents the total hydrogen mass at the moment of core helium depletion as a function of mass. All models display a linear trend of increasing hydrogen mass as a function of mass, except for the highest mass models, which have slightly lower hydrogen mass that can be attributed to the increased effect of stellar winds. As discussed in appendix \ref{sec:appendix_winds}, the impact of stellar winds on the evolution is very small at this low metallicity. However, since the hydrogen layer has such a low mass (below 0.3 \Msun), even a small effect can have large consequences. The total hydrogen masses for the models with initial orbital periods of 25 and 35 d are very similar, and the models with a shorter orbital periods have only slightly smaller total hydrogen masses. The total hydrogen masses are notably smaller for the models with an orbital period of 5d. We demonstrate the impact of the orbital period on the maximum radius in Fig. \ref{fig:porb_test_grids_radius_ev}, where we find very similar trends as in the total hydrogen mass. Overall, models with the same initial mass and for these different orbital periods reach very similar final radii.

From a population perspective, assuming the initial orbital periods are distributed uniformly in log space, only a minority of the stars in this mass range stripped by stable mass transfer in the Hertzsprung Gap would have initial orbital periods that lead to final radii which are marginally (within 100 \Rsun) different from the predictions of the grid shown in the main text. This effect does not impact our main finding, i.e., that the radii of such stripped stars are severely underestimated in population synthesis models, especially at low metallicity. We conclude that for case B mass-transfer, our choice of the orbital period has only a small effect on the maximum radius of stripped stars and is representative for the population. 
For stripped stars created by other channels (e.g., case A mass-transfer or common envelope evolution), the maximum radius at low metallicity could be significantly different because the mass of the remaining hydrogen-rich layer could be much smaller.

\section{Analytic fitting functions for the radius of stripped stars}
\label{sec:appendix_fits}

For convenience we provide simple analytic fits to the minimum and maximum stellar radii from our stripped-star models. These fits accurately represent our models to within a few percent.

In the following formulae, $M$ is the total mass of the stripped star at the relevant time, i.e., at the minimum radius following post-stripping contraction or at the maximum radius towards the end of the nuclear burning. The difference between those masses is small for these relatively low-luminosity models and our adopted winds.

\begin{figure}[]
	\resizebox{\hsize}{!}{\includegraphics[width=\textwidth]{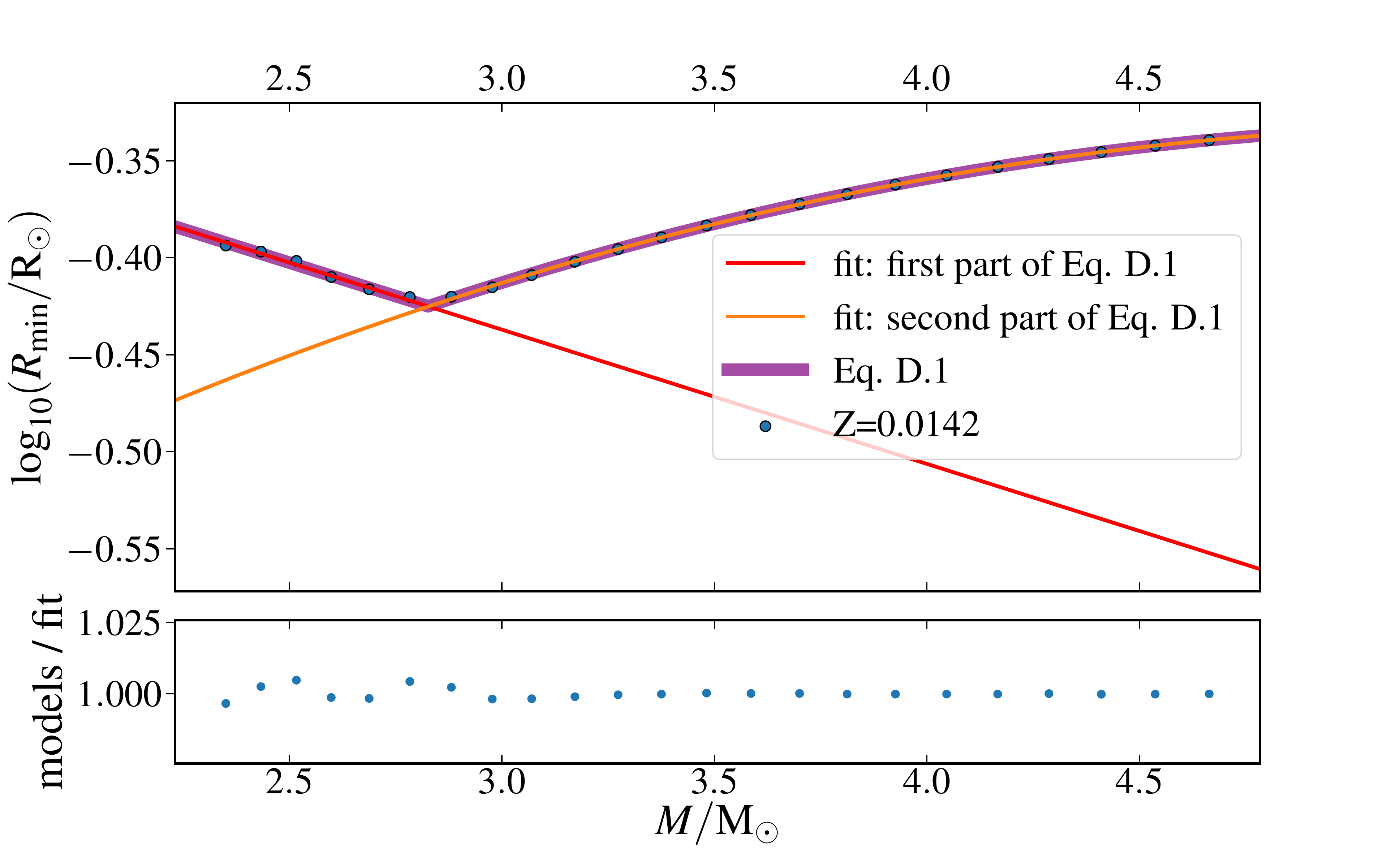}}
	\caption{Minimum radii of stripped-envelope stars as a function of their total masses at solar metallicity. The best fit obtained (Eq.\,\ref{eq:Rmin_Zsun}) is shown in purple and letters indicate the best fit parameters obtained. The lower panel shows the residuals of the fit, which are defined as fractional (i.e., dimensionless).}
	\label{fig:fit_Rmin_Zsun_all}
\end{figure}

At solar metallicity, the minimum radius, $R_{\rm{min}}$, for each model stripped star is well-described by:
\begin{equation}
\label{eq:Rmin_Zsun}
\begin{aligned}
    \log{ \left( \frac{R_{\rm{min}}}{ \Rsun} \right)} =  
    \max \bigg\{ & -0.069 \left( \frac{M}{\Msun} \right) - 0.23, \\
    & -0.0142 \left( \frac{M}{\Msun} \right) ^2 + 0.153 \left( \frac{M}{\Msun} \right) - 0.744 \bigg\}  \, \textrm{.}
\end{aligned}
\end{equation}
\noindent This fit is shown in Fig.\, \ref{fig:fit_Rmin_Zsun_all}, together with the 
ratio of the model to the fit (in all cases these ratios are for linear quantities, e.g., computed model radius in \Rsun over fitted radius in \Rsun).

\begin{figure}[]
	\resizebox{\hsize}{!}{\includegraphics[width=\textwidth]{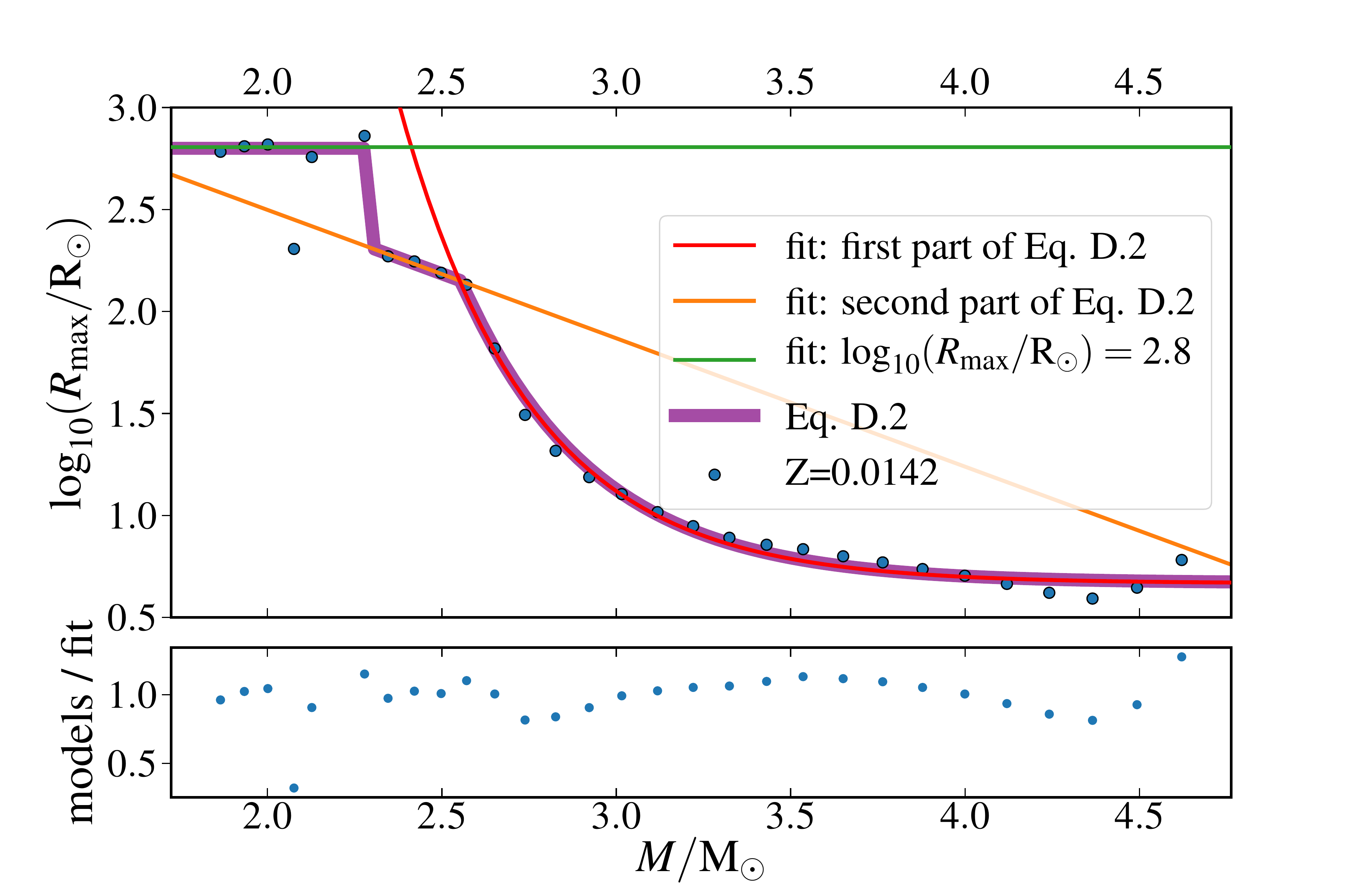}}
	\caption{Maximum radii of stripped-envelope stars as a function of their total masses at solar metallicity, extended to lower mass models. The fit is shown in purple. The lower panel shows the fractional residuals of the fit.}
	\label{fig:fit_Rmax_Zsun_all}
\end{figure}

The maximum radii, $R_{\rm{max}}$, of the solar-metallicity stripped-star models above $2.28 M_{\odot}$ can be expressed by:
\begin{equation}
\label{eq:Rmax_Zsun}
\begin{aligned}
    \log{\left( \frac{R_{\rm{max}}}{\Rsun} \right)} = \min \bigg\{ \, & 1258\exp{ \left( -2.64 \left( \frac{M}{\Msun} \right) \right) }  + 0.67, \\ 
    & -0.63 \left( \frac{M}{\Msun} \right) + 3.76 \bigg\} \textrm{.}
    \end{aligned}
\end{equation}
This fit is presented in Fig.\,\ref{fig:fit_Rmax_Zsun_all}. The maximum radius of the most massive progenitor was excluded from this fit due to numerical uncertainties in this model; otherwise, this fit has a maximum deviation from our models of about $2\,\Rsun$. For solar-metallicity stripped stars in our grid of $2.28 \Msun$ and below, we find setting $\log{(R_{\rm{max}}/\Rsun)} = 2.8$ is reasonable. The model with a radius of $\log(R/\Rsun) = 2.3$  did not reach core carbon ignition, and so was excluded from the fit.

\begin{figure}[]
	\resizebox{\hsize}{!}{\includegraphics[width=\textwidth]{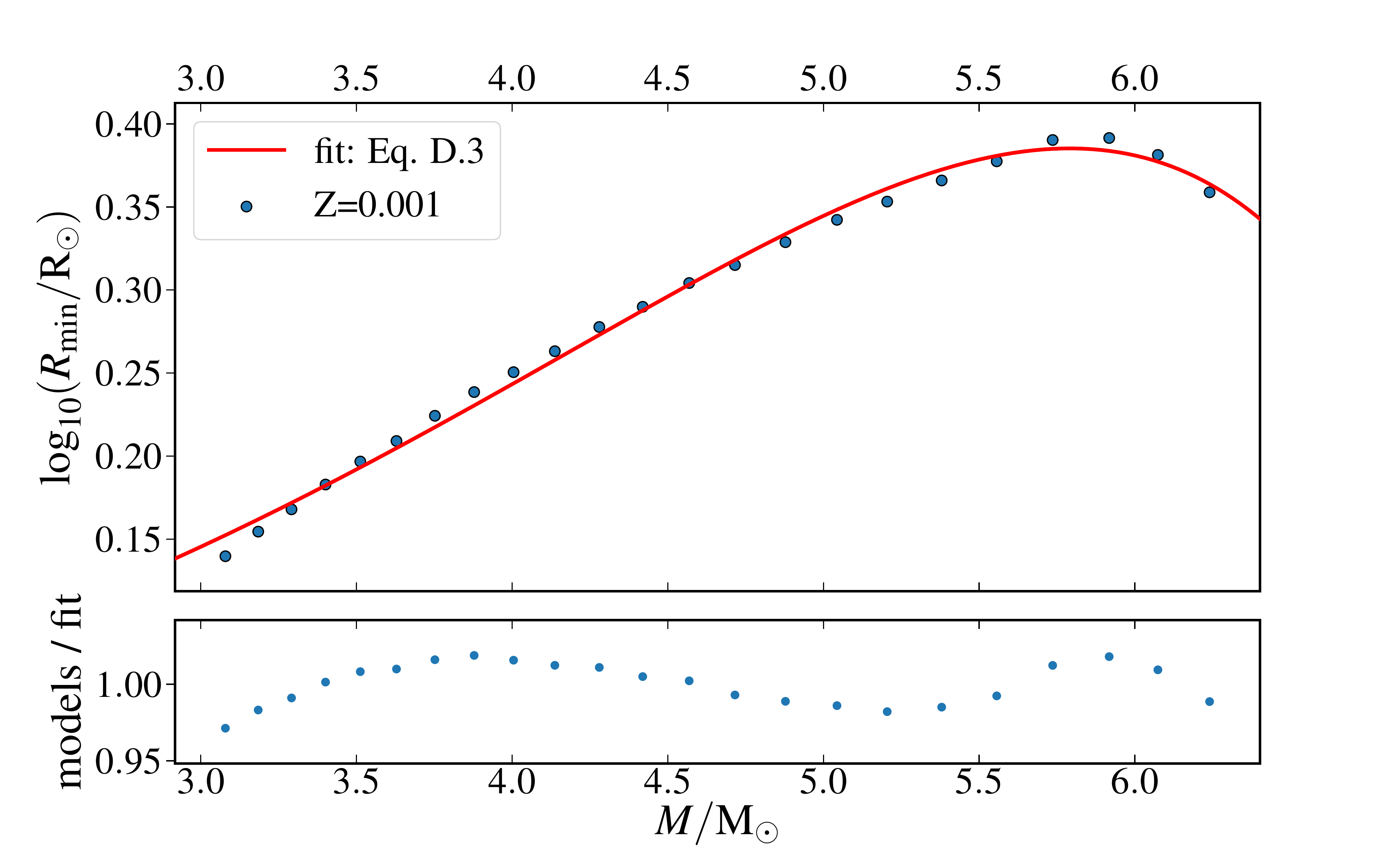}}
	\caption{Minimum radii of stripped-envelope stars as a function of their total masses at low metallicity. The best fit obtained (eq. \ref{eq:Rmin_Zlow}) is show in red. The lower panel shows the residuals of the fit.}
	\label{fig:fit_Rmin_Zlow}
\end{figure}
\begin{figure}[]
 \hspace{0.3cm}
	\resizebox{\hsize}{!}{\includegraphics[width=\textwidth]{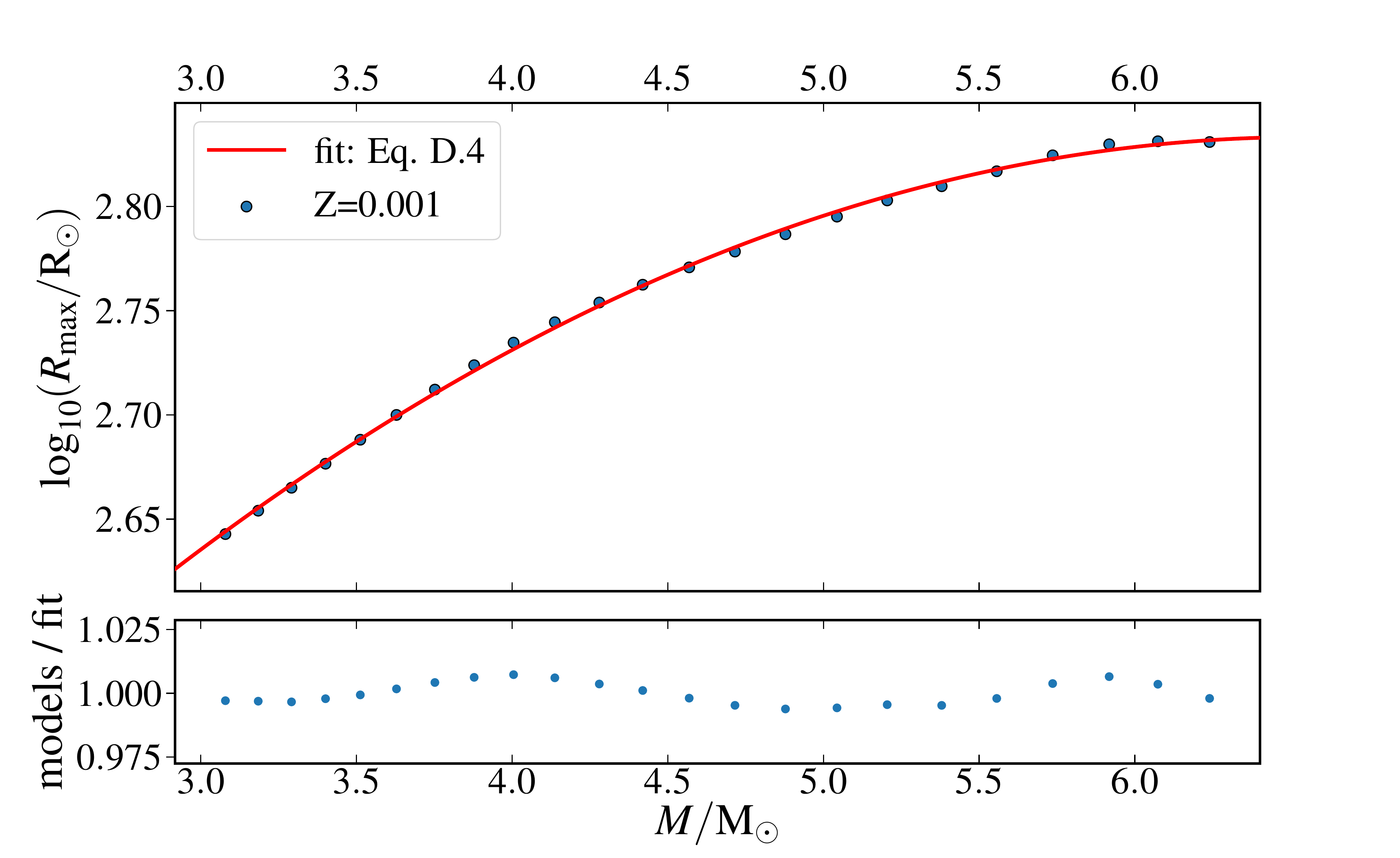}}
	\caption{Maximum radii of stripped-envelope stars as a function of their total mass at low metallicity. The best fit obtained (eq. \ref{eq:Rmax_Zlow}) is show in red. The lower panel shows the residuals of the fit.}
	\label{fig:fit_Rmin_Zsun}
\end{figure}

For our lower-metallicity models, we fit the minimum and the maximum radii for each model with: 
\begin{equation}
\label{eq:Rmin_Zlow}
    \log{ \left( \frac{R_{\rm{min}}}{\Rsun} \right) } = -9.8\times10^{-8} \left( \frac{M}{ \Msun} \right) ^8 + 0.015 \left( \frac{M}{\Msun} \right) ^2 + 0.013 \textrm{,}
\end{equation}
\begin{equation}
\label{eq:Rmax_Zlow}
    \log{ \left( \frac{R_{\rm{max}}}{\Rsun} \right) } = -0.016 \left( \frac{M}{\Msun} \right) ^2 + 0.21 \left( \frac{M}{\Msun} \right) + 2.2\textrm{.}
\end{equation}
This fit to $R_{\rm{min}}$ is shown in Fig.\,\ref{fig:fit_Rmin_Zlow}. The fit to $R_{\rm{max}}$ is in Fig.\,\ref{fig:fit_Rmin_Zsun}, and matches our low-metallicity models to within approximately 1\Rsun.

\begin{figure}[]
	\resizebox{\hsize}{!}{\includegraphics[width=\textwidth]{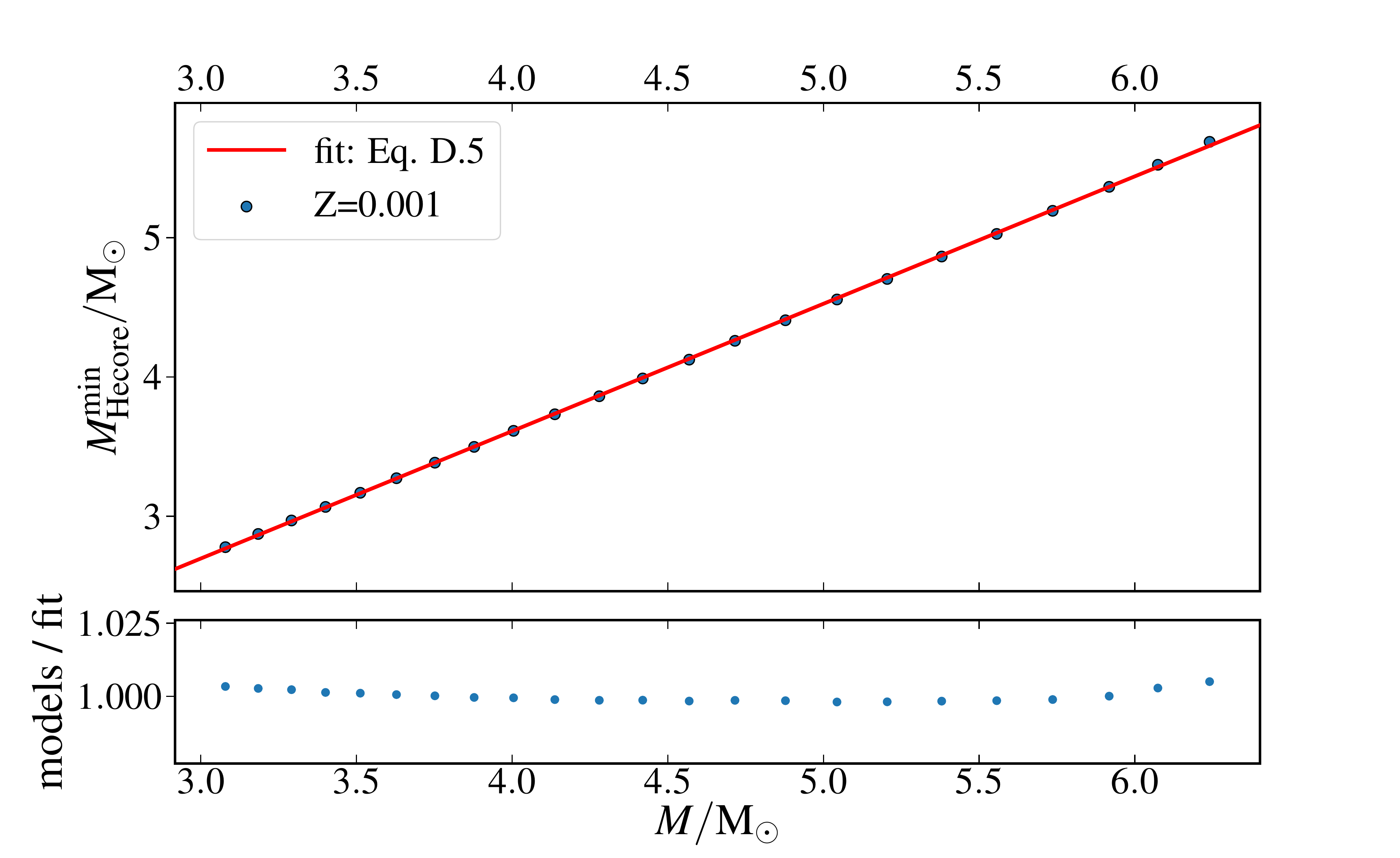}}
	\caption{Helium core mass at the minimum radius as a function of their total masses at low metallicity. The best fit obtained (eq. \ref{eq:MHe_min_Zlow}) is show in red. The lower panel shows the residuals of the fit defined as the difference between the fit and the models in solar masses.}
	\label{fig:fit_minHecoremass_Zlow}
\end{figure}

\begin{figure}[]
	\resizebox{\hsize}{!}{\includegraphics[width=\textwidth]{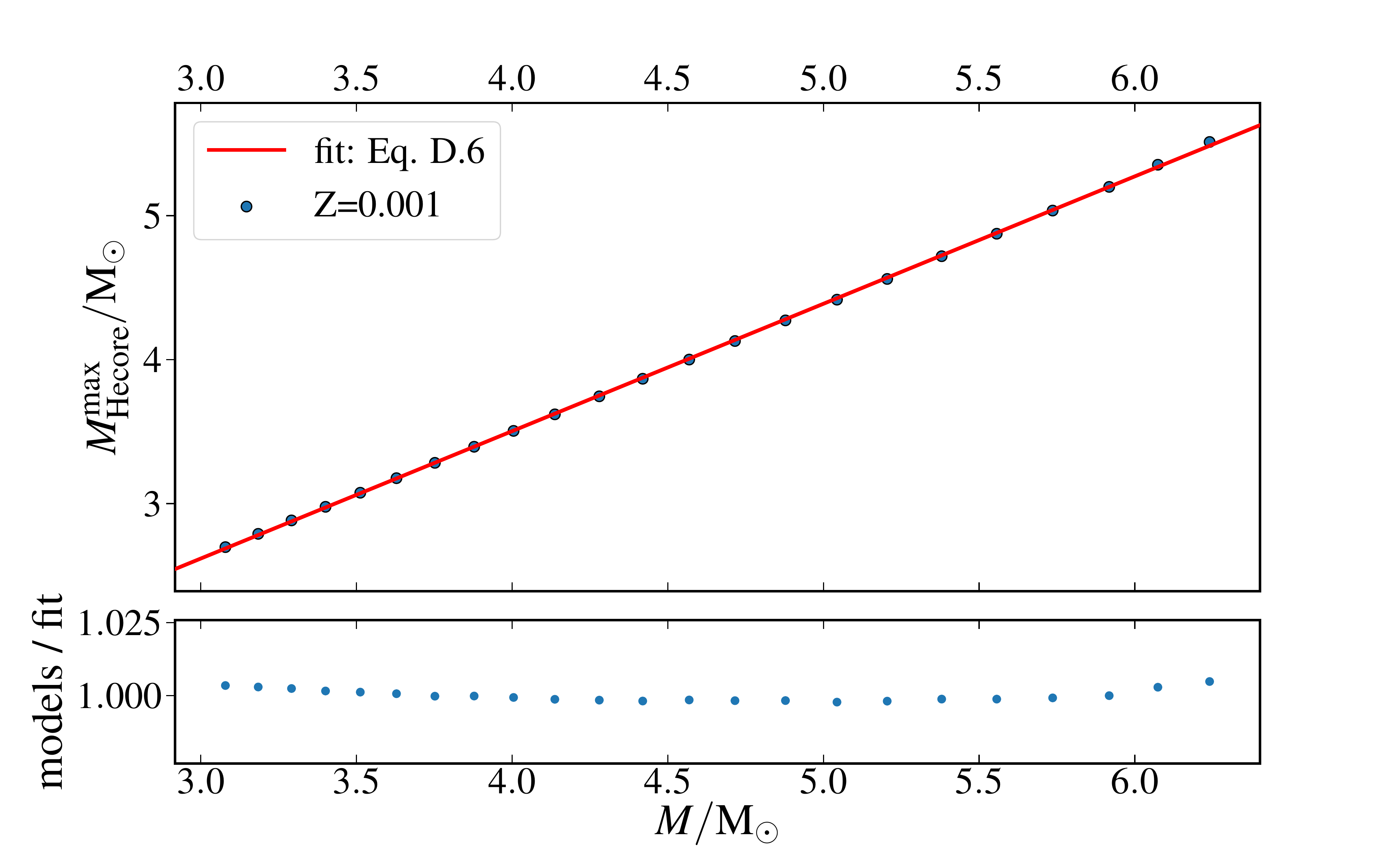}}
	\caption{Helium core mass at the maximum radius as a function of the total mass at low metallicity. The best fit obtained (eq. \ref{eq:MHe_max_Zlow}) is show in red. The lower panel shows the residuals of the fit defined as the difference between the fit and the models in solar masses.}
	\label{fig:fit_maxHecoremass_Zlow}
\end{figure}

For these low-metallicity models, it would be mistaken to take the initial stripped-star mass as a helium-core mass since the helium core grows as a result of the helium produced by hydrogen shell burning. So for these low-metallicity models, we also provide a fit for the initial $M_{\rm{He core}}^{\rm{min}}$ and final $M_{\rm{He core}}^{\rm{max}}$ helium core masses, as: 
\begin{equation}
\label{eq:MHe_min_Zlow}
    \left( \frac{M_{\rm{He core}}^{\rm{min}}}{\Msun} \right) = 0.91 \left( \frac{M}{\Msun} \right) -0.047\textrm{,}
\end{equation}
\begin{equation}
\label{eq:MHe_max_Zlow}
    \left( \frac{M_{\rm{He core}}^{\rm{max}}}{\Msun} \right) = 0.89\left( \frac{M}{\Msun} \right) -0.037\textrm{.}
\end{equation}
These fits are shown in Fig. \ref{fig:fit_minHecoremass_Zlow} and \ref{fig:fit_maxHecoremass_Zlow}, respectively.

We intend our fits to be used as a first improvement to current population synthesis calculations, especially focusing on predictions for double neutron-star systems. However, we do not provide a full way to integrate with the Hurley prescriptions. Our detailed calculations show that stars at low metallicity are partially stripped and exhibit intermediate behavior, which does not fit into the powerful but relatively simple scheme provided by Hurley and collaborators. Although one could add an extra parameter to keep track of the mass of the remaining envelope, this would still not be sufficient to, for example, predict the response of such stars to mass loss. More extended grids of detailed models would be needed to encompass multiple variations of the orbital period, the metallicity, and the mass-transfer efficiency, and for an even wider mass range than we present here. The use of such dense grids of stellar models for population studies would be preferable to current population synthesis models given the approximations we point out in Section \ref{sec:comparison}.

\section{Binding energy of the envelope}
\label{sec:appendix_eb}
In Section \ref{sec:grid_E_B} we present the envelope binding energies computed from our models.  However, the magnitude of the envelope binding energy is sensitive to the location of the boundary between core and envelope, and on the whether or not internal energy terms are incorporated in the calculation \citep[see, e.g., ][]{ivanova_common_2013}, which we discuss below. 

\subsection{Binding energy with and without internal energy}
\begin{figure}[]
    \resizebox{\hsize}{!}{\includegraphics[width=\textwidth]{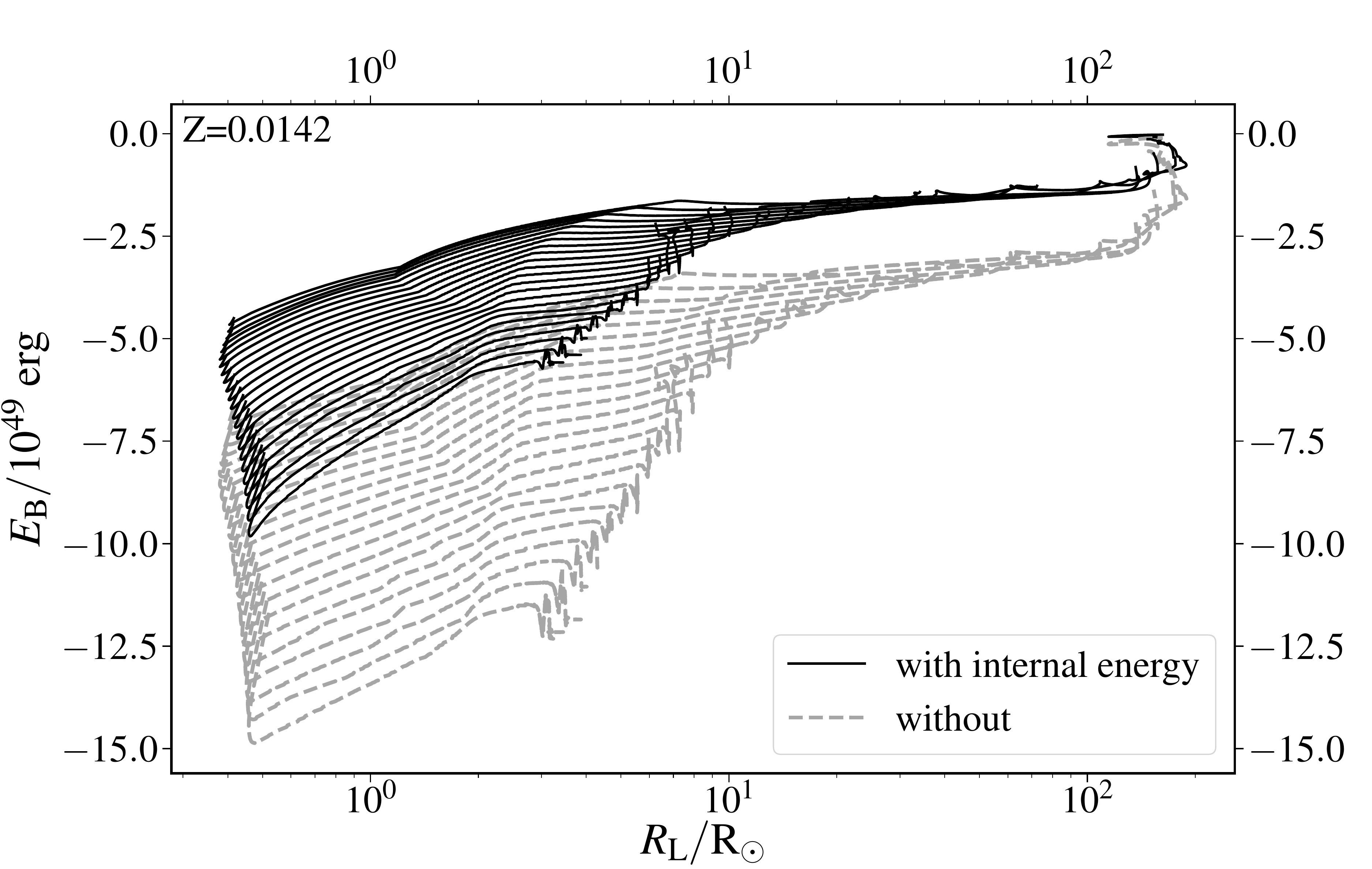}}
	\resizebox{\hsize}{!}{\includegraphics[width=\textwidth]{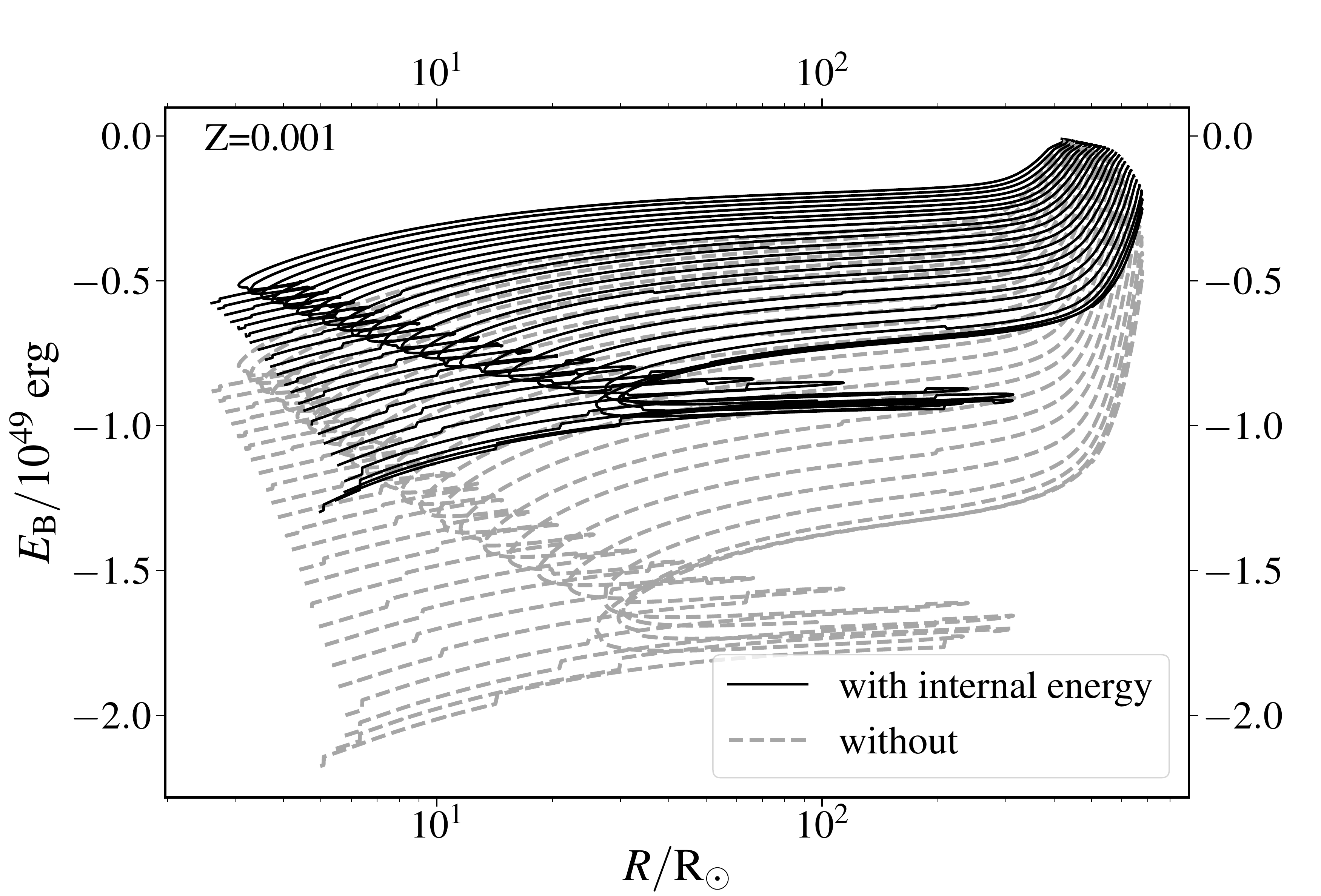}}
	\caption{Binding energy of the hydrogen-rich envelope as a function of the radius after core helium depletion at solar (top) and lower (bottom) metallicity. Here, we compare the binding energy computed with and without the internal energy terms. On the top axis, we indicate the orbital separation at which a neutron star companion would be expected if the star would fill its Roche lobe.}
	\label{fig:E_B_comp_eps_no_eps}
\end{figure}

When considering common-envelope evolution, the term "binding energy" is inconsistently used in the literature.  Sometimes the term is used purely for the gravitational binding energy, without including either the thermal internal energy of the envelope material or the electrostatic potential energy of ionised and dissociated matter. In this context it is common to refer to the thermal energy and recombination energy terms (including molecular dissociation energy) collectively simply as "internal energy", as we do.  The difference between binding energies which ignore or include these terms can be substantial, and can qualitatively affect common-envelope outcomes \citep[e.g.,][and references therein]{han_possible_1994, podsiadlowski_formation_2003, ivanova_common_2013}. In Fig. \ref{fig:E_B_comp_eps_no_eps}, we compare the effect of including or excluding the internal energy terms for the calculation of the binding energy for our grids at high and low metallicity. Models for which the binding energy was computed with the internal energy terms have a lower magnitude of binding energy -- i.e., are less bound -- and span a smaller absolute range of binding energies at the beginning of the evolution.\\

\begin{figure}[]
    \resizebox{\hsize}{!}{\includegraphics[width=0.9\textwidth]{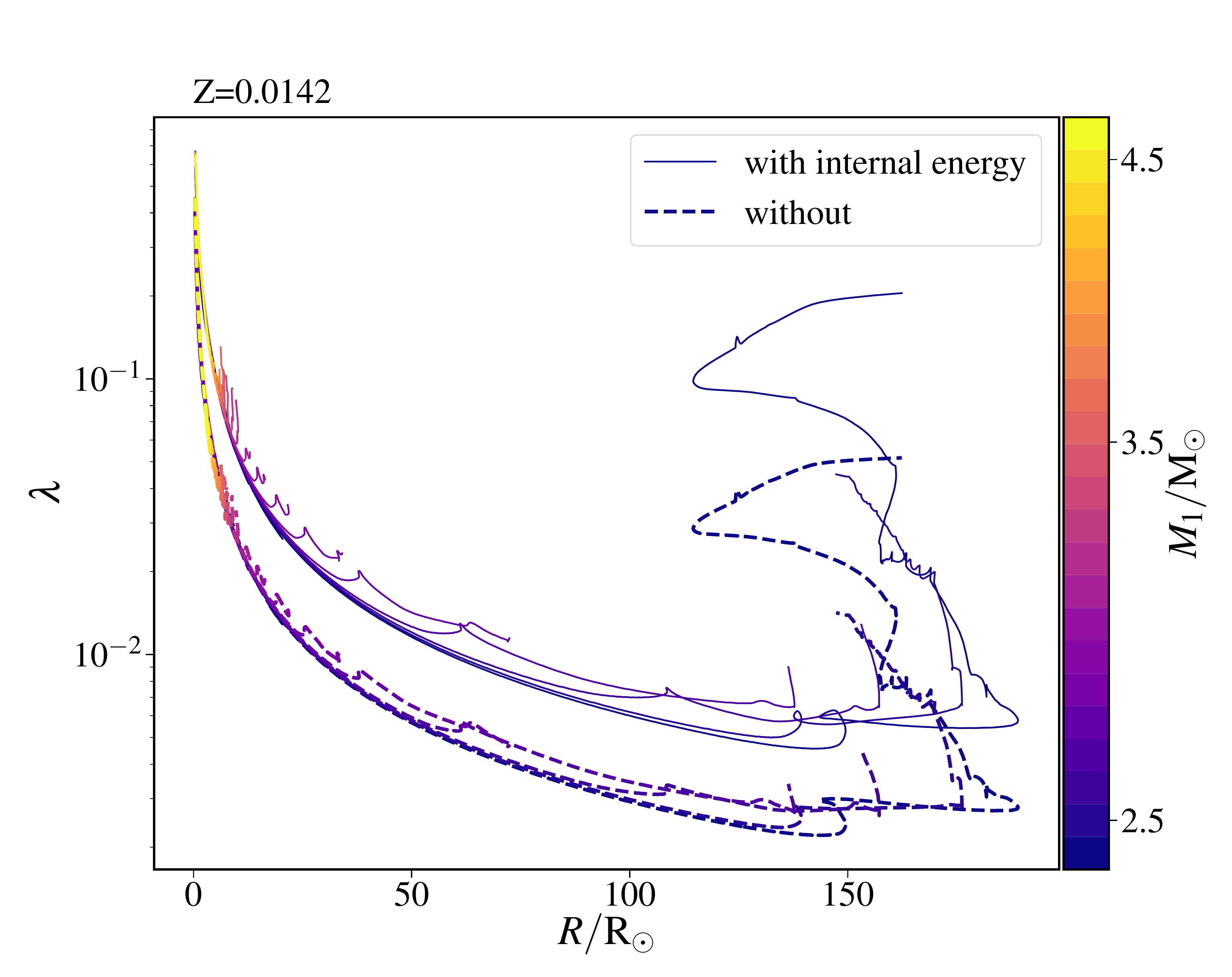}}
	\resizebox{\hsize}{!}{\includegraphics[width=0.9\textwidth]{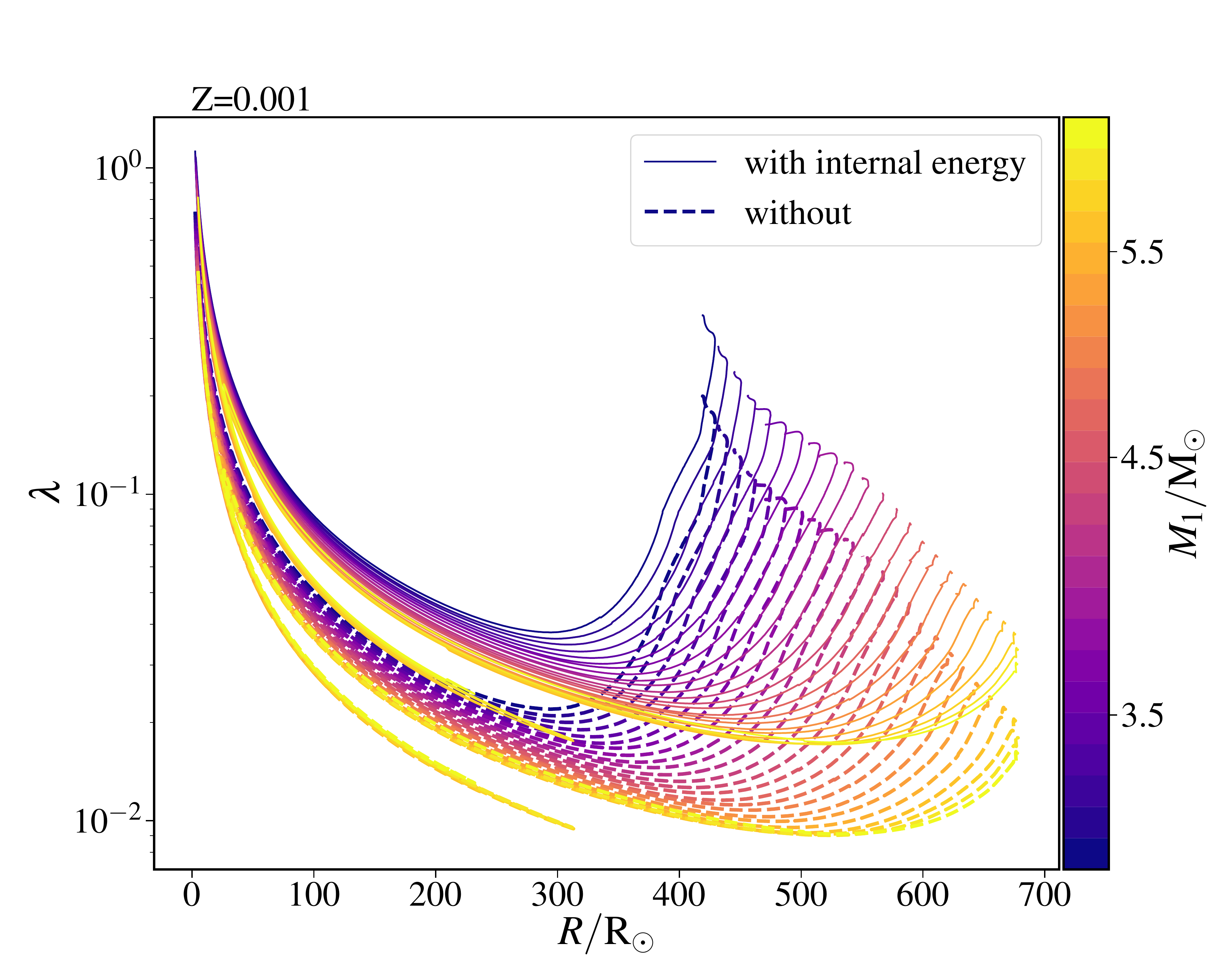}}
	\caption{Envelope structure parameter $\lambda$ as a function of the radius after core helium depletion at solar (top) and lower (bottom) metallicity. The dashed line indicates $\lambda$ parameters computed using only the gravitational energy term for the binding energy.}
	\label{fig:lambda}
\end{figure}

The dimensionless $\lambda$ parameter is commonly employed to encode how the structure of an envelope affects its binding energy. This was introduced by \citet{de_kool_common_1990} for calculating the outcome of common-envelope evolution with the "alpha prescription". It is defined as 
\begin{equation}
    \centering
    \lambda = \frac{G M_1 M_{1, \rm{env}}}{E_{\rm{B}} R_1} \,\rm{,}
\end{equation}
where $M_1$ is the mass of the primary star transferring mass, $M_{1, \rm{env}}$ the mass of its envelope, and $R_1$ its radius.
We show the $\lambda$ parameter at high and low metallicity for both definitions of the binding energy in Fig. \ref{fig:lambda}.

\subsection{Choice of core/envelope mass boundary}
Subtle differences in the definition of the core and envelope boundary mass can also significantly affect the value of the binding energy \citep[see, e.g.,][]{han_possible_1994, tauris_research_2001, podsiadlowski_formation_2003, ivanova_common_2013}. We do not investigate the effect of choosing such different definitions for the core-envelope boundaries on our results.

However, for these stripped stars we do investigate the differences between choosing a hydrogen-rich or helium-rich boundary. We compare the binding energy computed using Eq.~\ref{eq:E_b} for the hydrogen-rich (H) and the helium-rich (H + He) envelope in Fig.~\ref{fig:Eb_time} as a function of time after core helium depletion at high and low metallicities. For both grids, the binding energy of the H-envelope is an order of magnitude lower than that of the the H + He envelope. This is unsurprising, given that the potential well is deeper and steeper closer to the core of the star. For both regions, the magnitude of the total absolute binding energy increases with increasing initial mass.\\
At solar metallicity, only the lowest-mass models have a hydrogen envelope shortly after core helium depletion, before it disappears due to wind mass loss after about 30 kyr.\\
In contrast, at low metallicity, all models retain a hydrogen envelope. 
At the moment of the first radius expansion, the binding energy of the H-envelope drops before increasing shortly, and decreasing again, while that of the H + He envelope increases, and later decreases. This is consistent with the double-mirror effect. 
\end{appendix}
\end{document}